\documentstyle[12pt,epsf]{article}

\renewcommand{\theequation}{\arabic{section}.\arabic{equation}}

\newcommand{\quark}{\langle\bar q q\rangle}
\newcommand{\squark}{\langle\bar s s\rangle}
\newcommand{\gluon}{\left\langle\,\frac{\alpha_s}{\pi}\,G^2\right\rangle}
\newcommand{\mixed}{\langle\bar q \sigma gG q\rangle}
\newcommand{\smixed}{\langle\bar s \sigma gG s\rangle}
\newcommand{\ms}{\overline{m}_s}
\newcommand{\fperp}{f_\rho^T}

\newcommand{\hs}{h_\parallel^{(s)}}
\newcommand{\hsWW}{h_\parallel^{(s)WW}}
\newcommand{\hsg}{h_\parallel^{(s)g}}
\newcommand{\hsm}{h_\parallel^{(s)m}}
\newcommand{\htt}{h_\parallel^{(t)}}
\newcommand{\httWW}{h_\parallel^{(t)WW}}
\newcommand{\httg}{h_\parallel^{(t)g}}
\newcommand{\httm}{h_\parallel^{(t)m}}

\newcommand{\fA}{f_{3\rho}^A}
\newcommand{\fV}{f_{3\rho}^V}
\newcommand{\fT}{f_{3\rho}^T}

\makeatletter
\def\slash#1{{\mathpalette\c@ncel{#1}}} 
\makeatother
\newcommand{\deriv}{\stackrel{\leftrightarrow}{D}}
\newcommand{\derleft}{\stackrel{\leftarrow}{D}}
\newcommand{\derright}{\stackrel{\rightarrow}{D}}

\newcommand{\ub}{\bar u}

\newcommand\beq{\begin{eqnarray}}
\newcommand\eeq{\end{eqnarray}}

\newcommand\la{\langle}
\newcommand\ra{\rangle}

\def\xslash{\rlap/{\mkern-1mu x}}

\def\ubar{\overline{u}}

\def\Gtilde{\tilde{G}}

\newcommand{\wperp}{\Omega_n^\perp}
\newcommand{\cu}{{\cal U}}
\newcommand{\opdot}{M_{2\cdot} + i M_{1\cdot}}
\newcommand{\opstar}{M_{2*} + i M_{1*}}

\begin{document}


\begin{titlepage}
\begin{flushright}
\begin{tabular}{l}
FERMILAB--PUB--98/028--T\\
NORDITA--98--6--P\\
JUPD-9811\\
hep-ph/9802299
\end{tabular}
\end{flushright}
\vskip0.5cm
\begin{center}
  {\Large \bf 
             Higher Twist Distribution Amplitudes of Vector Mesons in
             QCD: Formalism and Twist Three  Distributions
  \\}

\vspace{1cm}
{\sc Patricia~Ball}${}^1$, {\sc V.M.~Braun}${}^{2}$,
 {\sc Y.~Koike}${}^3$ and {\sc K.~Tanaka}${}^4$
\\[0.3cm]
\vspace*{0.1cm} ${}^1$ {\it 
Fermi National Accelerator Laboratory,
  P.O.\ Box 500, Batavia, IL 60510, USA} \\[0.3cm]
\vspace*{0.1cm} ${}^2${\it NORDITA, Blegdamsvej 17, DK-2100 Copenhagen,
Denmark}
\\[0.3cm]
\vspace*{0.1cm} ${}^3$ {\it Dept. of Physics, Niigata University,
Niigata 950--2181, Japan}
\\[0.3cm]
\vspace*{0.1cm} ${}^4$ {\it Dept. of Physics, Juntendo University,
Inba-gun, Chiba 270-1606, Japan}
\\[1cm]


  \vskip1.8cm
  {\large\bf Abstract:\\[10pt]} \parbox[t]{\textwidth}{ 
  We present a systematic study of  twist three light-cone distribution
  amplitudes of vector mesons in QCD, which is based on conformal expansion and
  takes into account meson and quark mass corrections. 
  A complete set of distribution amplitudes is constructed  
  for $\rho$, $\omega$, $K^*$ and $\phi$ mesons, which satisfies all (exact) 
  equations of motion and constraints from conformal expansion. 
  Nonperturbative input parameters
  are calculated from QCD sum rules, including an update of  
  SU(3) breaking corrections in the leading twist distributions.
}
  \vskip1cm 
{\em Submitted to Nuclear Physics B}
\end{center}
\end{titlepage}

\section{Introduction}
\setcounter{equation}{0}

The success of QCD as fundamental theory of strong interactions is
intimately tied to its ability to describe hard inclusive reactions,
which has been tested in numerous experiments. 
Progress in the theory and phenomenology of hard exclusive processes
\cite{BLreport} has been more modest for several reasons.
First of all, exclusive processes 
are in general more difficult to study experimentally as they
constitute only a small fraction of the inclusive rates at comparable
momentum transfers. 
In addition, there is growing understanding that, although
the quark counting rules appear to start working at small momentum
transfers, QCD factorization in its standard form \cite{exclusive}
may  be valid only at very large momentum transfers. This is 
in contrast to inclusive processes like deep-inelastic scattering, 
where the leading twist factorization approximation is adequate already 
at $Q\sim 1$~GeV. Evidently, the situation calls for 
a systematic study of preasymtotic corrections to hard 
exclusive amplitudes.

Not much is known yet about these preasymptotic corrections.
The task is complicated by the fact that it actually comprises two
different problems. First, it is not known how to 
combine higher twist contributions to hadron form factors with the  
so-called ``soft'' or ``end-point'' contributions which are of the
same order of magnitude.
Second, in order to be able to calculate higher twist corrections, 
one needs to know {\em both} the dependence of the leading Fock state
wave function -- with a minimal number of (valence) partons -- 
on transverse
momentum {\em and} the distribution amplitudes with a non-minimal
parton configuration with additional gluons and/or quark-antiquark
pairs. 
These two effects are physically different,
but related to each other by the (exact) QCD equations of motion; 
taking into account one effect and neglecting the other is inconsistent
with QCD, unless kinematic suppression or enhancement of  a particular
mechanism can be established. 
One has to find a basis of 
independent distributions and to develop a 
meaningful approximation to describe them by a minimal number 
of parameters. 

In this paper we address the second problem only, leaving aside the 
questions of ``end-point'' contributions and how to generalize the
factorization
formalism beyond leading twist. The main goal of our study is to find out 
whether higher twist components in hadrons have an economic description
in QCD.
Light-cone distributions beyond leading twist were previously
addressed in Refs.~\cite{CZreport,ZZC85,BBK89,G89,BF90,H97}. 
The existing results are, however, far from being complete and sometimes
even contradictory. The aim of this paper is to 
develop a systematic formalism for  constructing a basis
of independent higher twist distributions in such a form that it
is possible to model them while automatically including
all QCD constraints. 
Our approach is an extension of earlier work \cite{BF90}
with the basic idea that the equations of motion 
can be solved order by order in the expansion in conformal spin.
These relations are exact in perturbation theory 
although conformal
symmetry is broken beyond one loop. Taking into account 
a few low-order terms in the
conformal expansion, one obtains a consistent set of distribution amplitudes 
which involve  a minimal number of 
 independent nonperturbative parameters that can be estimated from QCD
sum rules (or, eventually, calculated on the lattice). 
Although in principle this can be done for arbitrary twist, we
concentrate on twist three distribution amplitudes of vector mesons
in this work.
A detailed treatment of twist four  distributions
will be presented elsewhere.
 
In Ref.~\cite{BF90}, this program 
was  realized  for the pion for which a complete 
set of distributions of twist three and twist four is now available.
Vector mesons bring in the complication of polarization 
and are more difficult to treat as the meson mass cannot be
neglected. This requires a generalization
of the techniques of \cite{BF90}, which we work out in the present paper.
In addition, we take into account SU(3) 
flavour violation effects induced by quark 
masses and construct a  complete set of twist three distribution
amplitudes for $\rho$, $\omega$, 
$K^*$ and $\phi$ mesons, which is our main result.
All necessary nonperturbative constants are calculated from QCD sum rules
and the  scale-dependence is worked out in leading
logarithmic approximation. 

Apart from providing the leading corrections to hard exclusive amplitudes, 
 twist three distributions are
of special interest as they are free from renormalon 
ambiguities (power divergences of the corresponding operators, in a different
language) and their evolution with $Q^2$ is simple in the limit of
 a large number of colours,  
$N_c\to\infty$, as will be clarified in this work. 
One may hope that these distributions are accessible 
experimentally.
Some immediate applications of our results, which we do not pursue in this
paper, are to exclusive 
semileptonic and radiative $B$ decays and to hard electroproduction of 
vector mesons at HERA. 

The paper is organized as follows. Section 2 is mainly introductory. 
We collect necessary definitions and explain basic ideas.
Section 3 contains a detailed study on chiral-odd distribution 
amplitudes, including solution of the equations of motion, conformal expansion
and renormalization. A similar program is carried 
out for chiral-even distributions in Sec.~4. 
Section 5 contains explicit models for the $\rho$, $K^*$ and 
$\phi$ meson distribution amplitudes up to twist three,  
which involve a minimal number of nonperturbative parameters and 
satisfy all QCD constraints.
The final Sec.~6 is reserved for a summary and conclusions. 
The paper contains three appendices: in App.~A we collect some useful 
formulae about orthogonal polynomials,  
in App.~B we elaborate on the structure of conformal expansion for the 
so-called Wandzura-Wilczek contributions, 
and App.~C contains QCD sum rules for the  nonperturbative 
expansion coefficients used in Sec.~5.

\section{General Framework}
\setcounter{equation}{0}
\subsection{Kinematics and notations}
Amplitudes of light-cone dominated processes involving vector mesons
can be expressed in terms of matrix elements of gauge-invariant
nonlocal operators sandwiched between the vacuum and the vector meson state,
\begin{equation}
    \langle 0|\bar u(x) \Gamma [x,-x] d(-x)|\rho^-(P,\lambda)\rangle,
\label{eq:1}
\end{equation}
where $\Gamma$ is a generic Dirac matrix structure
and where we use the notation 
$[x,y]$ for the path-ordered gauge factor along the straight line 
connecting the points $x$ and $y$:
\begin{equation}
[x,y] =\mbox{\rm Pexp}[ig\!\!\int_0^1\!\! dt\,(x-y)_\mu A^\mu(tx+(1-t)y)].
\label{Pexp}
\end{equation}
To simplify notations,  we will explicitly consider  
charged $\rho$ mesons; the distribution amplitudes 
 of $\rho^0$ and of $K^*$ and $\phi$ mesons can  
be obtained by choosing appropriate SU(3) currents.
In order to be able to study SU(3) breaking effects, we
keep all quark mass terms.

The asymptotic expansion of exclusive amplitudes in powers of 
large momentum transfer is governed by contributions 
from small transverse separations between the constituents, which are
obtained by expanding amplitudes like (\ref{eq:1})
in powers of the deviation from the light-cone $x^2= 0$.    
To implement the light-cone expansion in a systematic way, it is
convenient to use light-like vectors.
Let $P_\mu$ be the $\rho$ meson momentum and $e^{(\lambda)}_\mu$ its 
polarization vector, so that 
\begin{equation}
  P^2=m_\rho^2, \qquad e^{(\lambda)}\cdot e^{(\lambda)} =-1, \qquad
 P\cdot e^{(\lambda)} =0.
\end{equation}
We introduce light-like vectors $p$ and $z$ with
$$ p^2=0, \qquad z^2=0,$$
such that $p\to P$ in the limit $m^2_\rho\to 0$ and 
$z\to x$ for $x^2\to 0$. {}From this it follows that
\begin{eqnarray}
  z_\mu &=& 
x_\mu-P_\mu\frac{1}{m_\rho^2}\left[xP -\sqrt{(xP)^2-x^2m^2_\rho}\,\right]
\nonumber\\
&=& x_\mu -\frac{1}{2}P_\mu\frac{x^2}{zp}+ \mbox{\cal O}(x^4)
\nonumber\\
&=& x_\mu\left[1-\frac{x^2m_{\rho}^2}{4(zp)^2}\right]
-\frac{1}{2}p_\mu\frac{x^2}{zp}
    + \mbox{\cal O}(x^4),
\nonumber\\
p_\mu &=& P_\mu-\frac{1}{2}z_\mu \frac{m^2_\rho}{pz}.
\label{lc-variables}
\end{eqnarray}
Useful scalar products are
\begin{eqnarray}
  zP = zp &=& \sqrt{(xP)^2-x^2m_\rho^2},
\nonumber\\
  p\cdot e^{(\lambda)} &=& -\frac{m^2_\rho}{2 pz} z\cdot e^{(\lambda)}.
\end{eqnarray}
The polarizaton vector $e^{(\lambda)}$ can be decomposed into projections 
onto the two light-like vectors and the orthogonal plane:
\begin{eqnarray}
 e^{(\lambda)}_\mu &=& \frac{(e^{(\lambda)}\cdot z)}{pz} p_\mu +
                     \frac{(e^{(\lambda)}\cdot p)}{pz} z_\mu +
                     e^{(\lambda)}_{\perp\mu}
\nonumber\\
&=&  \frac{(e^{(\lambda)}\cdot z)}{pz}\left( p_\mu -\frac{m^2_\rho}{2pz} z_\mu
                                            \right)+e^{(\lambda)}_{\perp\mu}. 
\label{polv}
\end{eqnarray}
Note that 
\begin{equation}
  (e^{(\lambda)}\cdot z) = (e^{(\lambda)}\cdot x).
\end{equation}
In terms of the original variables one has
\begin{equation}
  e^{(\lambda)}_\mu = (e^{(\lambda)}\cdot x)
 \frac{P_\mu (xP)-x_\mu m_\rho^2}{(xP)^2-x^2m_\rho^2}+e^{(\lambda)}_{\perp\mu}.
\end{equation}
We also need the projector onto the directions orthogonal to $p$ and $z$,
\begin{equation}
       g^\perp_{\mu\nu} = g_{\mu\nu} -\frac{1}{pz}(p_\mu z_\nu+ p_\nu z_\mu), 
\end{equation}
and will often use the notations
\begin{equation}
    a.\equiv a_\mu z^\mu, \qquad a_\ast \equiv a_\mu p^\mu/(pz),
\end{equation}
for an arbitrary Lorentz vector $a_\mu$.

We use the standard Bjorken-Drell 
convention \cite{BD65} for the metric and the Dirac matrices; in particular
$\gamma_{5} = i \gamma^{0} \gamma^{1} \gamma^{2} \gamma^{3}$,
and the Levi-Civita tensor $\epsilon_{\mu \nu \lambda \sigma}$
is defined as the totally antisymmetric tensor with $\epsilon_{0123} = 1$.
The covariant derivative is defined as 
$D_{\mu} \equiv \overrightarrow{D}_\mu= \partial_{\mu} - igA_{\mu}$, 
which is consistent with
the gauge phase factor (\ref{Pexp}), and we also use the notation
$\overleftarrow{D}_{\mu} = \overleftarrow\partial_{\mu} 
+ig A_{\mu}(x)$ in later sections. The dual gluon field strength
tensor is defined as $\widetilde{G}_{\mu\nu} =
\frac{1}{2}\epsilon_{\mu\nu \rho\sigma} G^{\rho\sigma}$.
Finally, we use a covariant normalization for  one-particle states,
i.e.\ 
$\langle \rho^-(P,\lambda) |\rho^-(P',\lambda')\rangle
= (2 \pi)^{3} 2P^{0} \delta^{(3)}(P - P') \delta_{\lambda \lambda'}$.

\subsection{Classification of  two-particle distribution amplitudes}

By definition, light-cone distribution amplitudes are obtained from
 Bethe-Salpeter wave functions at (almost) zero 
transverse separations of the constituents,
\begin{equation}
   \phi(x) \sim \int^{k_\perp^2<\mu^2}\! d^2k_\perp\, \phi(x,k_\perp)\,,
\end{equation}
and are given by vacuum-to-meson matrix elements of nonlocal 
operators on the light-cone:
\begin{equation}
\langle 0|\bar u(z) \Gamma [z,-z] d(-z)|\rho^-(P,\lambda)\rangle.
\end{equation}
Note that unlike in Eq.~(\ref{eq:1}) the separation between the 
quark and the antiquark is strictly light-like. The expansion of  
(\ref{eq:1}) near the light-cone in terms of operators with
light-like separations is the subject of the operator
product expansion and considered at length in Ref.~\cite{BB88}.
 
It turns out that all quark-antiquark 
distribution amplitudes of vector mesons can be classified 
in the same way as the more familiar nucleon structure functions
(parton distributions) which
correspond to the independent tensor structures in matrix elements of type
$\langle N(P, S)|\bar \psi(z) \Gamma [z,- z] 
\psi(-z)|N(P,S)\rangle$
over nucleon states $|N(P,S)\rangle$ with 
momentum $P$ and spin $S$.

Jaffe and Ji \cite{JJ92} find nine independent quark distributions
whose spin, twist and chiral classifications are shown in 
Tab.~\ref{tab:1}.
\begin{table}
\begin{center}
\renewcommand{\arraystretch}{1.3}
\begin{tabular}{|c|ccc|}
\hline
Twist    & 2 & 3 & 4 \\
    & $O(1)$  & $O(1/Q)$& $O(1/Q^{2})$ \\ \hline
spin ave.& $f_{1}$ & \underline{$e$} & $f_{4}$ \\
$S_{\parallel}$ & $g_{1}$ & \underline{$h_{L}$} & $g_{3}$ \\
$S_{\perp}$ & \underline{$h_{1}$} & $g_{T}$ & \underline{$h_{3}$}\\[2pt]
\hline
\end{tabular}
\renewcommand{\arraystretch}{1}
\end{center}
\caption{Spin, twist and chiral classification of the 
nucleon structure functions.}
\label{tab:1}
\end{table}%
The parton distributions in 
the first row are spin-independent, those in the
second and third rows describe
longitudinally ($S_{\parallel}$) and transversely ($S_{\perp}$)
polarized nucleons.
Each column refers to twist: a distribution of twist $t$
contributes to inclusive cross sections with
coefficients
which contain $t-2$ or more powers of $1/Q$.
The underlined distributions are referred to as 
chiral-odd, because they correspond to 
chirality-violating Dirac matrix structures 
$\Gamma = \{\sigma_{\mu \nu} i \gamma_{5},\, 1\}$. The other distributions
are termed chiral-even, because they are obtained for the 
chirality-conserving structures
$\Gamma = \{\gamma_{\mu},\, \gamma_{\mu}\gamma_{5}\}$.

The nucleon parton distributions are defined by
\begin{eqnarray}
\lefteqn{\hspace*{-1.5cm}\langle N(P, S)|\bar \psi(z) 
 \sigma_{\mu \nu} i \gamma_{5}[z,-z] 
\psi(-z)|N(P,S)\rangle } \nonumber \\
&=& 2\Bigg[ (S_{\perp \mu}p_\nu -
S_{\perp \nu}p_\mu )
\int_{-1}^{1} \!dx\, e^{2i x p \cdot z} h_{1}(x, \mu^{2}) 
\nonumber \\
& &{}+ (p_\mu z_\nu - p_\nu z_\mu )
\frac{S \cdot z}{(p \cdot z)^{2}}
M^{2} 
\int_{-1}^{1} \!dx\, e^{2i x p \cdot z} h_{L}(x, \mu^{2}) 
\nonumber \\
& &{}+ 
(S_{\perp \mu} z_\nu -S_{\perp \nu} z_\mu) 
\frac{M^{2}}{p \cdot z} 
\int_{-1}^{1} \!dx\, e^{2i x p \cdot z} h_{3}(x, \mu^{2}) \Bigg],
\label{eq:ftda}
\end{eqnarray}
\begin{equation}
\langle N(P, S)|\bar \psi(z) [z,-z] 
\psi(-z)|N(P,S)\rangle
= 2M \int_{-1}^{1} \!dx\, e^{2i x p \cdot z} e(x, \mu^{2}),
\label{eq:fsda}
\end{equation}
\begin{eqnarray}
\lefteqn{\langle N(P, S)|\bar \psi(z) \gamma_\mu [z,-z] \psi(-z) |N(P,S)\rangle
\ }\makebox[2cm]{\ }\nonumber\\
& = & 2 \left[ p_\mu \int_{-1}^{1} \!dx\, e^{2i x p \cdot z}
f_1(x,\mu^2) + z_\mu\, \frac{M^2}{p\cdot z}  \int_{-1}^{1} \!dx\, e^{2i
x p \cdot z} f_4(x,\mu^2)\right],
\label{eq:favda}\\
\lefteqn{\langle N(P, S)|\bar \psi(z) \gamma_{\mu} \gamma_{5}[z,-z] 
\psi(-z)|N(P,S)\rangle  \hfill}\makebox[1.4cm]{\ } \nonumber \\
&=& 2M\left[ p_{\mu}
\frac{S \cdot z}{p \cdot z}
\int_{-1}^{1} \!dx\, e^{2i x p \cdot z} g_{1}(x, \mu^{2}) \right. 
+ S_{\perp \mu}
\int_{-1}^{1} \!du\, e^{2i x p \cdot z} g_{T}(x, \mu^{2}) 
\nonumber \\
& & \left.{}+ z_{\mu}
\frac{S\cdot z }{(p \cdot z)^{2}} M^{2}
\int_{-1}^{1} \!dx\, e^{2i x p \cdot z} g_{3}(x, \mu^{2}) \right],
\label{eq:fvda}
\end{eqnarray}
where $P_{\mu}$ and $S_{\mu}$ are decomposed similarly
to (\ref{lc-variables}) and (\ref{polv}), respectively,
with $m_{\rho}$ replaced by the nucleon mass $M$;
$x$ is the Bjorken scaling variable. 
The nucleon spin vector is normalized as $S\cdot S= -1$,
which differs from the definition used in \cite{JJ92} by a factor $M^2$.

The analysis of vector meson distribution amplitudes
reveals an analogous pattern, which is no surprise as 
the operator structures are the same 
and the $\rho$ meson polarization vector formally substitutes 
the nucleon spin vector in the Lorentz structures.
We find eight independent two-particle distributions whose
classification with respect to spin, twist and chirality is summarized
in Tab.~\ref{tab:2}.
\begin{table}
\begin{center}
\renewcommand{\arraystretch}{1.3}
\begin{tabular}{|c|ccc|}
\hline
Twist    & 2 & 3 & 4 \\
    & $O(1)$  & $O(1/Q)$& $O(1/Q^{2})$ \\ \hline
$e_{\parallel}$ & $\phi_{\parallel}$ & \underline{$\htt$}, \underline{$\hs$}& 
$g_{3}$ \\
$e_{\perp}$ & \underline{$\phi_{\perp}$} & $g_{\perp}^{(v)}$,
$g_{\perp}^{(a)}$ & \underline{$h_{3}$}\\[2pt] \hline
\end{tabular}
\renewcommand{\arraystretch}{1}
\end{center}
\caption{Spin, twist and chiral classification of the $\rho$ meson
distribution 
amplitudes.}
\label{tab:2}
\end{table}%
One distribution amplitude is obtained 
for longitudinally ($e_{\parallel}$) and transversely ($e_{\perp}$)
polarized $\rho$ mesons of twist~2 and twist~4, respectively.
On the other hand, the number of  twist~3
distribution
amplitudes is doubled for each polarization.
Due to this analogous structure
we take over the notations
from Tab.~\ref{tab:1} for some quantities. Again, 
the higher twist distribution amplitudes
contribute to a hard exclusive amplitude with
additional powers of $1/Q$ compared to the leading
twist~2 ones.
The underlined distribution amplitudes 
are chiral-odd, the others chiral-even.
Because the matrix element
$\langle 0|\bar u(z) \Gamma [z,-z] d(-z)|\rho^-(P,\lambda)\rangle$
depends on the polarization vector $e_{\mu}^{(\lambda)}$
linearly, there is no spin-independent
distribution amplitude. This is in contrast to the nucleon
parton distributions
of Tab.~\ref{tab:1}, where the dependence on the polarization
vector $S_{\mu}$ is determined by
the density matrix $(1 + \gamma_{5} \rlap/{\mkern-1mu S})/2$ 
which contains a spin-independent part.
One more difference is that $e_{\mu}^{(\lambda)}$
for the $\rho$ meson is a vector, while $S_{\mu}$ for the nucleon
is a pseudovector.
Thus, an insertion of additional $i\gamma_5$ is necessary in order that
 matrix elements of relevant operators have the analogous Lorentz
 decomposition.

The explicit definitions of the chiral-odd $\rho$ distributions are:
\begin{eqnarray}
\lefteqn{\hspace*{-1.5cm}\langle 0|\bar u(z) \sigma_{\mu \nu} [z,-z] 
d(-z)|\rho^-(P,\lambda)\rangle =} \nonumber \\
&=& i f_{\rho}^{T} \left[ ( e^{(\lambda)}_{\perp \mu}p_\nu -
e^{(\lambda)}_{\perp \nu}p_\mu )
\int_{0}^{1} \!du\, e^{i \xi p \cdot z} \phi_{\perp}(u, \mu^{2}) \right. 
\nonumber \\
& &{}+ (p_\mu z_\nu - p_\nu z_\mu )
\frac{e^{(\lambda)} \cdot z}{(p \cdot z)^{2}}
m_{\rho}^{2} 
\int_{0}^{1} \!du\, e^{i \xi p \cdot z} \htt (u, \mu^{2}) 
\nonumber \\
& & \left.{}+ \frac{1}{2}
(e^{(\lambda)}_{\perp \mu} z_\nu -e^{(\lambda)}_{\perp \nu} z_\mu) 
\frac{m_{\rho}^{2}}{p \cdot z} 
\int_{0}^{1} \!du\, e^{i \xi p \cdot z} h_{3}(u, \mu^{2}) \right],
\label{eq:tda}
\end{eqnarray}
\begin{equation}
\langle 0|\bar u(z) [z,-z] 
d(-z)|\rho^-(P,\lambda)\rangle
= -i \left(f_{\rho}^{T} - f_{\rho}\frac{m_{u} + m_{d}}{m_{\rho}}
\right)(e^{(\lambda)}\cdot z) m_{\rho}^{2}
\int_{0}^{1} \!du\, e^{i \xi p \cdot z} \hs(u, \mu^{2}),
\label{eq:sda}
\end{equation}
while the chiral-even distributions are defined as (cf.~\cite{BBrho})
\begin{eqnarray}
\lefteqn{\langle 0|\bar u(z) \gamma_{\mu} [z,-z] 
d(-z)|\rho^-(P,\lambda)\rangle = }\makebox[2cm]{\ } \nonumber \\
&=& f_{\rho} m_{\rho} \left[ p_{\mu}
\frac{e^{(\lambda)}\cdot z}{p \cdot z}
\int_{0}^{1} \!du\, e^{i \xi p \cdot z} \phi_{\parallel}(u, \mu^{2}) \right. 
+ e^{(\lambda)}_{\perp \mu}
\int_{0}^{1} \!du\, e^{i \xi p \cdot z} g_{\perp}^{(v)}(u, \mu^{2}) 
\nonumber \\
& & \left.{}- \frac{1}{2}z_{\mu}
\frac{e^{(\lambda)}\cdot z }{(p \cdot z)^{2}} m_{\rho}^{2}
\int_{0}^{1} \!du\, e^{i \xi p \cdot z} g_{3}(u, \mu^{2}) \right]
\label{eq:vda}
\end{eqnarray}
and 
\begin{eqnarray}
\lefteqn{\langle 0|\bar u(z) \gamma_{\mu} \gamma_{5}[z,-z] 
d(-z)|\rho^-(P,\lambda)\rangle = }\makebox[2cm]{\ } \nonumber \\
&=& \frac{1}{2}\left(f_{\rho} - f_{\rho}^{T}
\frac{m_{u} + m_{d}}{m_{\rho}}\right)
m_{\rho} \epsilon_{\mu}^{\phantom{\mu}\nu \alpha \beta}
e^{(\lambda)}_{\perp \nu} p_{\alpha} z_{\beta}
\int_{0}^{1} \!du\, e^{i \xi p \cdot z} g^{(a)}_{\perp}(u, \mu^{2}).
\label{eq:avda}
\end{eqnarray}
Here and below we use the shorthand notation
$$\xi = u - (1-u) = 2u-1.$$ 
The distribution amplitudes are dimensionless functions of $u$ and 
describe
the probability amplitudes to find the $\rho$ in a state with minimal
number of constituents ---
 quark and antiquark --- which carry
momentum fractions $u$ (quark) and $1-u$ (antiquark), respectively, and
have a small transverse separation 
of order $1/\mu$.
The nonlocal operators on the left-hand side are
 renormalized at scale $\mu$, so that
 the distribution amplitudes depend on $\mu$ as well. This
dependence can be calculated in perturbative QCD and will be considered
below in Secs.~3 and 4.

The vector and tensor  decay constants $f_\rho$ and $f_\rho^T$ are defined 
as usually as
\begin{eqnarray}
\langle 0|\bar u(0) \gamma_{\mu}
d(0)|\rho^-(P,\lambda)\rangle & = & f_{\rho}m_{\rho}
e^{(\lambda)}_{\mu},
\label{eq:fr}\\
\langle 0|\bar u(0) \sigma_{\mu \nu} 
d(0)|\rho^-(P,\lambda)\rangle &=& i f_{\rho}^{T}
(e_{\mu}^{(\lambda)}P_{\nu} - e_{\nu}^{(\lambda)}P_{\mu}),
\label{eq:frp}
\end{eqnarray}
and their numerical values are discussed in Sec.~5.

All eight distributions $\phi=\{\phi_\parallel, \phi_\perp,
g_\perp^{(v)},g_\perp^{(a)},\htt,\hs,h_3,g_3\}$ are normalized as
\begin{equation}
\int_0^1\!du\, \phi(u) =1,
\label{eq:norm}
\end{equation}
which can be checked by comparing both sides
of the  defining equations in the limit $z_\mu\to 0$ and using the
equations of motion.
The rationale for  keeping  the (tiny) corrections proportional 
to the $u$ and $d$ quark masses $m_u$ and $m_d$ is that it will allow 
us to calculate the SU(3) breaking corrections for $K^*$ and $\phi$ 
mesons.

Note that
the meson-to-vacuum matrix element vanishes for $\Gamma = i \gamma_{5}$,
because it is not possible to construct a pseudoscalar quantity
from $p_{\mu}, z_{\mu}$, and $e_{\mu}^{(\lambda)}$.
On the other hand,
(\ref{eq:avda}) would correspond to
(\ref{eq:favda}) which defines the spin-averaged nucleon 
distributions and is the only exception to the complete analogy
between the nucleon distribution functions 
and the $\rho$ meson distribution amplitudes. In this case it is
the difference in parity between $e_{\mu}^{(\lambda)}$ and $S_{\mu}$,
which leads to a completely different decomposition of the matrix elements.

Because of the analogous structure, the twist classification of the
various distributions does not require a separate study and 
can be inferred directly from \cite{JJ92}, see Tab.~\ref{tab:2}.
Its physical interpretation, however,  deserves a discussion.

One convenient way to understand the twist classification
of distribution amplitudes directly from their definitions
is to go over to the infinite momentum frame.
{}For definiteness, we assume for the moment that
$\rho$ is moving in the positive $\hat{\bf e}_3$ direction
and $p^+$ and $z^-$ are the only nonzero component of $p$
and $z$, respectively.  Then the infinite momentum frame 
can be understood as 
$p^+ \sim Q \rightarrow \infty$ with a fixed  $p\cdot z \sim 1$
in (\ref{eq:tda})--(\ref{eq:avda}). 
{}From (\ref{polv}) it follows that 
in this frame 
$(e^{(\lambda)} \cdot z) \sim 1$ and $e^{(\lambda)}_{\perp} \sim 1$.
This determines  the power counting in $Q$ 
of all terms in the right-hand sides of (\ref{eq:tda})--(\ref{eq:avda}).
The first, second, and third terms in 
(\ref{eq:tda}) and (\ref{eq:vda})
behave as $O(Q)$, $O(1)$ and
$O(1/Q)$, respectively, and the right hand sides of
(\ref{eq:sda}) and (\ref{eq:avda}) are of $O(1)$.

A mathematically similar, but conceptually different approach to
twist counting is based on the light-cone quantization 
formalism \cite{KS,BLreport,JJ92}. In this approach
 quark fields are decomposed into  ``good'' 
and ``bad'' components,
so that $\psi = \psi_{+} + \psi_{-}$ with 
$\psi_{+} = (1/2)\gamma_*\gamma_. \psi$ and
$\psi_{-} = (1/2)\gamma_.\gamma_*\psi$.
As discussed in \cite{JJ92}, a ``bad'' component $\psi_{-}$
introduces one unit
of twist. Therefore, a quark-antiquark operator of type $\bar ud$
contains twist~2 ($\bar u_{+} d_{+}$), twist~3 ($\bar u_{+} d_{-}$, 
$\bar u_{-} d_{+}$), and twist~4 ($\bar u_{-} d_{-}$) contributions.
This explains why the number of twist~3 distribution amplitudes is
doubled compared with the twist~2 and twist~4 ones 
(see Tab.~\ref{tab:2}).

The physical content of this classification is that 
a ``good'' component $\psi_{+}$ represents an independent
degree of freedom corresponding to the particle content of the
``Fock state''.
On the other hand, the ``bad'' components are not dynamically independent,
but can be expressed in terms of the higher
components in the Fock wave function with a larger number of constituents,
in particular  corresponding to a coherent quark-gluon pair.
Only the twist~2 distribution amplitudes correspond to the 
valence quark-antiquark
component in the $\rho$ meson wave function, while the 
higher twist amplitudes involve contributions of multi-particle states.
This point will be discussed in detail in Secs.~3 and 4.

One important comment is in order. The definition of twist based 
on power counting in the infinite momentum frame 
is convenient, because it is directly related
to the power of $1/Q$, with which the corresponding distributions
appear in the physical scattering amplitudes,
and  hence is frequently employed in recent works \cite{JJ92}.
On the other hand, this definition is not Lorentz invariant
and does not match the conventional and more consistent definition of
twist as ``dimension minus spin'' of the relevant operators.
For example, the nucleon structure function $g_T$ which is identified 
as  twist 3 by power counting in fact contains contributions
of both operators of twist~2 and twist~3. Similarly, the distribution
amplitudes of vector mesons that were identified as twist~3 above,
actually contain contributions of twist~2 operators as well.
In Secs.~3 and 4 we will study in detail the operator structure 
of twist~3 distribution amplitudes based on the operator product expansion. 
In this context the conventional ``operator'' definition of twist will
be more adequate.
The mismatch of different definitions of twist has to be kept in mind, but 
hopefully will not yield confusion.    

To summarize, Eqs.~(\ref{eq:tda})--(\ref{eq:avda}) define a complete
set of valence 
light-cone distribution amplitudes and provide full information
on the quark-antiquark component of the Fock wave function
of the $\rho$ meson at zero transverse separation. As mentioned
above, not all of these distributions 
 are independent. In the following sections we will
derive exact relations between the twist~3
quark-antiquark distribution amplitudes and those
involving one additional gluon, which are introduced below.

\subsection{Three-particle distribution amplitudes of twist three}

Higher Fock components of the meson wave function are described by 
multi-particle distribution amplitudes. In this paper 
we will explicitly deal with three-particle twist~3 quark-antiquark-gluon
distributions, defined as
\begin{eqnarray}
\langle 0|\bar u(z) \gamma_\alpha [z,vz]gG_{\mu\nu}(vz)[vz,-z] 
         d(-z)|\rho^-(P,\lambda)\rangle & = &
      ip_\alpha[p_\mu e^{(\lambda)}_{\perp\nu}-p_\nu e^{(\lambda)}_{\perp\mu}]
      f_{3\rho}^V{\cal V}(v,pz)+\ldots\nonumber\\[-10pt]
\label{eq:V3}\\[-5pt]
\langle 0|\bar u(z) \gamma_\alpha \gamma_5[z,vz]
         g\widetilde G_{\mu\nu}(vz)[vz,-z] 
         d(-z)|\rho^-(P,\lambda)\rangle & = &  
 p_\alpha[p_\nu e^{(\lambda)}_{\perp\mu}-p_\mu e^{(\lambda)}_{\perp\nu}]
      f_{3\rho}^A{\cal A}(v,pz)+\ldots\nonumber\\[-10pt]
\label{eq:A3}\\[-5pt]
\langle 0|\bar u(z) \sigma_{\alpha\beta} [z,vz]
         gG_{\mu\nu}(vz)[vz,-z] 
         d(-z)|\rho^-(P,\lambda)\rangle & = & \nonumber \\
\lefteqn{=\ \frac{e^{(\lambda)}\cdot z }{2 (p \cdot z)}
    [ p_\alpha p_\mu g^\perp_{\beta\nu} 
     -p_\beta p_\mu g^\perp_{\alpha\nu} 
     -p_\alpha p_\nu g^\perp_{\beta\mu} 
     +p_\beta p_\nu g^\perp_{\alpha\mu} ] 
     f_{3\rho}^T m_{\rho} {\cal T}(v,pz)+\ldots,}\makebox[7cm]{\ }
\label{eq:T3}
\end{eqnarray}
where the ellipses stand for Lorentz structures
 of twist higher than three  and where we used the following shorthand
notation for the integrals defining three-particle distribution amplitudes:
\begin{equation}
   {\cal F}(v,pz) \equiv  \int {\cal D}\underline{\alpha}
   \,e^{-ipz(\alpha_u-\alpha_d+v\alpha_g)}{\cal F}(\alpha_d,\alpha_u,\alpha_g).
\label{eq:short} 
\end{equation}
Here ${\cal F} = \{{\cal V,A,T}\}$ refers in an obvious way to the vector,
axial-vector and tensor distributions, 
$\underline{\alpha}$ is the set of three 
momentum fractions: $\alpha_d$ ($d$ quark), $\alpha_u$ ($u$ quark)
and $\alpha_g$ (gluon), and the integration measure is defined as
\begin{equation}
 \int {\cal D}\underline{\alpha} \equiv \int_0^1 d\alpha_d
  \int_0^1 d\alpha_u\int_0^1 d\alpha_g \,\delta(1-\sum \alpha_i).
\label{eq:measure}
\end{equation}
The normalization constants $f_{3\rho}^V, f_{3\rho}^A, f_{3\rho}^T$ 
are defined in such a way that 
\begin{eqnarray}
 \int\! {\cal D}\underline{\alpha}\, (\alpha_d-\alpha_u)\,{\cal V}
    (\alpha_d,\alpha_u,\alpha_g) &=&1,
\nonumber\\
\int\! {\cal D}\underline{\alpha}\,{\cal A} (\alpha_d,\alpha_u,\alpha_g) &=&1,
\nonumber\\
 \int\! {\cal D}\underline{\alpha} \,(\alpha_d-\alpha_u)\,{\cal T}
    (\alpha_d,\alpha_u,\alpha_g) &=&1.
\label{eq:normalize}
\end{eqnarray} 
Choosing the normalization in this way  we anticipate 
that the function ${\cal A}$ is symmetric and the functions ${\cal V}$
and ${\cal T}$
are antisymmetric under the interchange $\alpha_u \leftrightarrow \alpha_d$
in the SU(3)  limit (cf.\ \cite{CZreport}), which follows from the
behaviour of the corresponding matrix elements under G-parity transformations.
Numerical values of all constants are discussed in Sec.~5.

With these major definitions at hand, we now proceed to a systematic 
study of the twist~3 distribution amplitudes.

\section{Chiral-odd Distribution Amplitudes}
\setcounter{equation}{0}

This section is devoted to the general discussion of chiral-odd 
distributions of twist~3. We  demonstrate that the 
two-particle distribution amplitudes $\htt(u, \mu^{2})$ 
and $\hs(u, \mu^{2})$ can be eliminated in favour of independent dynamical 
degrees of freedom and expressed in terms of 
leading twist two-- and three-particle 
distributions. The corresponding relations
are worked out in detail and solved explicitly. A similar relation
between the nucleon structure function $g_1$ and the twist~2 part
of $g_T$ is known as Wandzura-Wilczek relation \cite{WW77}. 
We also investigate the expansion of twist~3 distributions in terms of
matrix elements of conformal operators.
We demonstrate  that the equations of motion are satisfied order 
by order in the conformal expansion which provides, for this reason, a
systematic approach to the construction of models of distribution 
amplitudes, consistent with QCD constraints.
The renormalization of all distributions is worked out 
in the leading logarithmic approximation.

\subsection{Equations of motion}

The basis of twist~3 distributions defined in Sec.~2 is overcomplete.
Due to the QCD equations of motion, the number of independent 
degrees of freedom is less than the number of independent Lorentz
structures, and our first task will be to reveal the corresponding 
constraints.

The standard technique for this purpose is to derive relations between
towers of local operators which arise in the Taylor expansion of
the nonlocal operators in Eqs.~(\ref{eq:tda}), (\ref{eq:sda}) and 
whose matrix elements are just moments of the distribution amplitudes.
A more elegant and economic approach is to use exact operator 
identities between the nonlocal operators 
\cite{BBK89,BF90} (see also \cite{BB88}).
In the present context, we need the identities
\begin{eqnarray}
\lefteqn{\frac{\partial}{\partial x_{\mu}}
\left\{ \ub(x) \sigma_{\mu \nu} x^{\nu} [x, -x]
d (-x) \right\} =}\makebox[1cm]{\ }
\nonumber\\
&=&
i \int_{-1}^{1}\! dv\, v \; \ub(x) x^{\alpha}
\sigma_{\alpha \beta} [x, vx]x^{\mu}gG_{\mu \beta}(vx)
[vx, -x] d(-x)
\nonumber \\
&&{}- i x^{\beta}\partial_{\beta} \left\{ \ub(x)
[x, -x] d(-x) \right\} - (m_{u} - m_{d}) \ub(x)
\!\not\!x \: [x, -x]d(-x),
\label{eq:3id1} \\
\lefteqn{\ub(x)[x, -x]d(-x) - \ub(0) d(0) =}\makebox[1cm]{\ }
\nonumber\\&=&
\int_{0}^{1} dt \int_{-t}^{t}dv\,
\ub(tx) x^{\alpha} \sigma_{\alpha \beta} [tx, vx]
x^{\mu} gG_{\mu \beta} (vx) [vx, -tx] d(-tx)
\nonumber\\
&&{}+ i \int_{0}^{1}\!dt\, \partial^{\alpha}\left\{
\ub(tx) \sigma_{\alpha \beta}x^{\beta}[tx, -tx]
d(-tx) \right\}
\nonumber\\&&
+ i (m_{u} + m_{d}) \int_{0}^{1} \!dt \,\ub(tx)
\!\not\! x \: [tx, -tx] d(-tx).
\label{eq:3id2}
\end{eqnarray}
Here we introduced a shorthand notation
for the derivative over the total translation:
\begin{equation}
\partial_{\alpha}\left\{ \ub(tx)
\Gamma [tx, -tx] d(-tx) \right\} \equiv
\left. \frac{\partial}{\partial y^{\alpha}}
\left\{ \ub(tx + y) \Gamma
[tx + y, -tx + y] d(-tx + y)\right\} \right|_{y \rightarrow 0},
\label{eq:3tdrv}
\end{equation}
with the generic Dirac matrix structure $\Gamma$.

In the light-cone limit $x^2\to 0$  matrix elements of the operators 
on both sides of Eqs.~(\ref{eq:3id1}) and (\ref{eq:3id2}), 
sandwiched between the vacuum 
and the $\rho$ meson state, can be expressed in terms of the distribution
amplitudes defined in Sec.~2:
\begin{eqnarray}
\lefteqn{
\langle 0|\ub(x) \sigma_{\mu \nu} x^{\nu} [x, -x]
d (-x) |\rho^-(P,\lambda)\rangle =}\makebox[1.6cm]{\ }
\nonumber\\
&=& 
 i f_{\rho}^{T} \left\{ \left( e^{(\lambda)}_{\mu} -
\frac{(e^{(\lambda)}\cdot x)}{Px}P_\mu \right) (Px)
\int_{0}^{1} \!du\, e^{i \xi P \cdot x} 
\Big[\phi_{\perp}(u, \mu^{2})+O(x^2)\Big] \right.
\nonumber\\
&&\left. -m^2_\rho \frac{(e^{(\lambda)}\cdot x)}{Px} 
\left(x_\mu-\frac{x^2}{Px}P_\mu \right)
\int_{0}^{1} \!du\, e^{i \xi P \cdot x}\Big[\htt (u,\mu^{2})
-\phi_\perp(u,\mu^{2})\Big]\right\},
\label{eq:OPE1}
\end{eqnarray}
\begin{eqnarray}
\lefteqn{\langle 0|\bar u(x) [x,-x] 
d(-x)|\rho^-(P,\lambda)\rangle
=}\makebox[1.6cm]{\ }
\nonumber\\&=& 
-i \left(f_{\rho}^{T} - f_{\rho}\frac{m_{u} + m_{d}}{m_{\rho}}
\right)(e^{(\lambda)}\cdot x) m_{\rho}^{2}
\int_{0}^{1} \!du\, e^{i \xi P \cdot x} \Big[\hs(u, \mu^{2})+O(x^2)\Big],
\label{eq:OPE2}
\end{eqnarray}
and likewise for three-particle distributions.

The matrix elements of Eqs.~(\ref{eq:3id1}) and (\ref{eq:3id2}) yield
a system of integral equations between two- and 
three-particle light-cone distribution 
amplitudes:\footnote{The suppressed corrections $O(x^2)$ drop out;
the $O(x^2)$ term in  the Lorentz structure in the second line of
(\ref{eq:OPE1}),
however, does give a contribution to the left-hand side of (\ref{eq:3id1}) 
after taking the  derivative with respect to $x_{\mu}$.}
\begin{eqnarray}
\lefteqn{-i pz \int_{0}^{1}\!du\, e^{i \xi pz}
\xi\, \htt (u) - 2 \int_{0}^{1}\!du\, e^{i \xi pz}
\left( \htt (u) - \phi_{\perp}(u) \right)=} \nonumber \\
&=& \zeta_{3\rho}^{T}(pz)^{2}
\!\int_{-1}^{1}\!dv\, v {\cal T}(v, pz)
+ \left(1 - \delta_{+}\right) (pz)^{2} \!\int_{0}^{1} 
e^{i\xi pz}\hs(u) 
+ i \delta_{-}pz \!\int_{0}^{1}\!du\, e^{i \xi pz} 
\phi_{\parallel}(u),
\label{eq:3cie1} 
\end{eqnarray}
\begin{eqnarray}
\left(1 - \delta_{+} \right)\! \int_{0}^{1}\! du\, e^{i \xi pz} \hs(u)
&=& i 
\zeta_{3\rho}^{T} pz\!
\int_{0}^{1}tdt \int_{-1}^{1}dv\, {\cal T}(v, tpz)
+ \int_{0}^{1}\!dt \int_{0}^{1}\!du \,e^{i \xi t pz} \htt (u)
\nonumber \\
& & {}-\delta_{+} \int_{0}^{1}\!dt \int_{0}^{1}\!du\, e^{i \xi t pz} 
\phi_{\parallel}(u),
\label{eq:3cie2}
\end{eqnarray}
where we have discarded all corrections of order $x^2$, set $x_\mu=z_\mu$
and introduced the notations
\begin{equation}
\delta_{\pm} = \frac{f_{\rho}}{f_{\rho}^{T}}
\frac{m_{u} \pm m_{d}}{m_{\rho}}, \qquad
\zeta_{3\rho}^{T} = \frac{f_{3 \rho}^{T}}{f_{\rho}^{T}m_\rho}.
\label{eq:3pm}
\end{equation}
Eqs.~(\ref{eq:3cie1}) and (\ref{eq:3cie2}) are 
exact in QCD. Note the terms with total derivatives in 
(\ref{eq:3id1}) and (\ref{eq:3id2}),  which induce mixing between
$\hs(u)$ and $\htt (u)$. Such contributions 
are specific for exclusive processes and have no
analogue in  deep-inelastic scattering. 
Note also that quark mass corrections 
bring in  the leading twist chiral-even distribution 
$\phi_{\parallel}(u)$.

We can solve Eqs.~(\ref{eq:3cie1}) and (\ref{eq:3cie2})
for $\hs(u)$ and $\htt (u)$ in terms of the other distributions.
To simplify the algebra, it is convenient to consider moments in
 an intermediate step. Defining
\begin{equation}
M_{n}^{\parallel,\perp} = \int_{0}^{1}\! du\, \xi^{n} 
\phi_{\parallel,\perp}(u),\quad M_{n}^{(s),(t)} = \int_{0}^{1}\! du\,
\xi^{n} h_\parallel^{(s),(t)}(u)
\label{eq:3mom1}
\end{equation}
and
\begin{equation}
{\cal T}_{n}(v) = \left. (-i)^{n} \frac{\partial^{n}}{\partial \tau^{n}}
{\cal T}(v, \tau)\right|_{\tau = 0} = \int\!{\cal D}\underline{\alpha}\,
(\alpha_{d} - \alpha_{u} - v \alpha_{g})^{n} {\cal T}
(\alpha_{d}, \alpha_{u}, \alpha_{g})
\label{eq:3mom2}
\end{equation}
and expanding (\ref{eq:3cie1}) and (\ref{eq:3cie2})
in powers of $(pz)$, we obtain
\begin{equation}
M_{n}^{(t) } - \frac{2}{n+2}\, M_{n}^{\perp}
- (1 - \delta_{+})\frac{(n-1)n}{n+2} \,M_{n-2}^{(s)}
+ \delta_{-}
\frac{n}{n+2} 
\,M_{n-1}^{\parallel}
-\zeta_{3\rho}^{T} \frac{(n-1)n}{n+2}
\int_{-1}^{1}\!\!\!dv\, v {\cal T}_{n-2}(v) = 0,
\label{eq:3mome1} 
\end{equation}
\begin{equation}
(1 - \delta_{+})\,M_{n}^{(s)} - \frac{1}{n+1}\,M_{n}^{(t) }
+ \delta_{+}\frac{1}{n+1}\,M_{n}^{\parallel}
- \zeta_{3 \rho}^{T}\frac{n}{n+1} \int_{-1}^{1}\!dv \,{\cal T}_{n-1}(v)
= 0.
\label{eq:3mome}
\end{equation}
Combining these two equations, one gets the following 
recurrence relations for $\htt $ and $\hs$:
\begin{eqnarray}
(n+2)\, M_{n}^{(t) } - n\, M_{n-2}^{(t) }
& = &2 M_{n}^{\perp} +\zeta_{3 \rho}^{T}
\int_{-1}^{1}\!dv \left\{(n-1)n \,v{\cal T}_{n-2}(v) 
+(n-2)n\, {\cal T}_{n-3}(v)\right\}
\nonumber \\
& &- \delta_{+} \: n \,M_{n-2}^{\parallel}
- \delta_{-} \: n\, M_{n-1}^{\parallel},
\label{eq:3rec1}
\end{eqnarray}
\begin{eqnarray}
\lefteqn{\hspace*{-2.0cm}(1 - \delta_{+})\left\{
(n+1)(n+2) M_{n}^{(s)} - (n-1)n M_{n-2}^{(s)}\right\}~=}
\nonumber \\
&=&2 M_{n}^{\perp}+\zeta_{3 \rho}^{T}
\int_{-1}^{1}dv \left\{n(n+2)\, {\cal T}_{n-1}(v)
+ (n-1) n\, v \,{\cal T}_{n-2}(v) \right\}
\nonumber\\&&{}
- \delta_{+} \: (n+2) \,M_{n}^{\parallel}
- \delta_{-} \: n\, M_{n-1}^{\parallel}.
\label{eq:3rec2}
\end{eqnarray}
Recurrence relations of this type are easily solved by  
transforming them into differential equations. For instance, 
for the distribution amplitude $\hs(u)$ one finds a second order equation:
\begin{equation}
(1-\delta_+) \,u\bar u \,(\hs)''(u) = - \, \Phi(u)
\end{equation} 
with 
\begin{eqnarray}
\Phi(u) & = & 2\phi_\perp(u) - \delta_+ \left( 
\phi_\parallel(u) - \frac{1}{2}\,\xi\phi'_\parallel(u)\right) +
\frac{1}{2}\,\delta_- \phi_\parallel'(u)\nonumber\\
& & {}+\zeta_{3 \rho}^{T}
\frac{d}{du}\,\int\limits_0^u\!\!d\alpha_d\int\limits_0^{\bar
u}\!\! d\alpha_u
\,\frac{1}{1-\alpha_u-\alpha_d}\left(\alpha_d\,\frac{d}{d\alpha_d} +
\alpha_u\,\frac{d}{d\alpha_u}\, - 1\right) {\cal T}(\underline{\alpha}).
\label{eq:Phi}
\end{eqnarray}
 Here and below
we use the shorthand notation $\bar u =1-u$.
The solution of this equation with boundary conditions specified by the
values of the first two moments reads
\begin{equation}
(1-\delta_+)\, \hs(u)  =  \bar u \int\limits_0^u\!\! dv\, 
\frac{1}{\bar v}\,\Phi(v)
+ u \int\limits_u^1\!\! dv\, \frac{1}{v}\,\Phi(v)\,.
\label{eq:e_solution}
\end{equation} 
The solution for $\htt(u)$ can be obtained in a similar manner and reads:
\begin{eqnarray}
\htt (u) & = & \frac{1}{2}\,\xi \left(\int\limits_0^u\!\!
dv\frac{1}{\bar v}\,\Phi(v) - \int\limits_u^1\!\!
dv\frac{1}{v}\,\Phi(v)\right) + \delta_+\phi_\parallel(u)
\nonumber\\
& & 
{} + \zeta_{3 \rho}^{T}\frac{d}{du}\,\int\limits_0^u\!\!d\alpha_d
\int\limits_0^{\bar u}\!\! d\alpha_u
\,\frac{1}{1-\alpha_u-\alpha_d}\,{\cal T}(\underline{\alpha}).
\label{eq:hL_solution}
\end{eqnarray}
According to the various ``source'' terms 
on the right-hand side of (\ref{eq:3rec1}) and (\ref{eq:3rec2}),
one can decompose the solution in an obvious way into three
pieces as
\begin{eqnarray}
\htt (u) &=& \httWW(u) + \httg(u) + \httm(u),
\label{eq:3solh}\\
\hs(u) &=& \hsWW(u) + \hsg(u) + \hsm(u),
\label{eq:3sole}
\end{eqnarray}
where $\httWW(u)$ and $\hsWW(u)$ denote
the ``Wandzura-Wilczek'' type contributions of twist~2 operators,
$\httg(u)$ and $\hsg(u)$ stand for contributions of three-particle
distributions and $\httm(u)$ and $\hsm(u)$ 
are due to the quark mass corrections.
In particular, we get 
\begin{eqnarray}
\httWW(u) &=& \xi \left( \int_{0}^{u} dv 
\frac{\phi_{\perp}(v)}{\bar v}
- \int_{u}^{1} dv \frac{\phi_{\perp}(v)}{v} \right),
\label{eq:3hww} \\
\hsWW(u) &=& 2 \left( \ub \int_{0}^{u} dv 
\frac{\phi_{\perp}(v)}{\bar v} + u \int_{u}^{1} dv 
\frac{\phi_{\perp}(v)}{v} \right).
\label{eq:3eww}
\end{eqnarray}
These are the analogues of the Wandzura-Wilczek
contributions to the nucleon structure functions
$g_{T}(x, Q^{2})$ \cite{WW77} and $h_{L}(x, Q^{2})$ \cite{JJ92}.

The relations Eqs.~(\ref{eq:e_solution}) and (\ref{eq:hL_solution}) 
are the principal results of this section: chiral-odd two-particle
distribution amplitudes of twist~3 are expressed in terms of 
the leading twist amplitudes and the three-particle twist~3
distribution. In the next subsection we will discuss how  
to proceed further with this rather complicated 
formal solution, concentrating on the massless quark limit.

\subsection{Conformal expansion}
\label{sec32}

The conformal expansion of light-cone distribution
amplitudes is analogous to the partial wave expansion
of  wave functions in standard quantum mechanics.   
In conformal expansion, 
the invariance of massless QCD under conformal transformations
substitutes the rotational symmetry in quantum mechanics.
In quantum mechanics, the purpose of partial wave 
decomposition is to separate angular degrees of freedom 
from radial ones 
(for spherically symmetric potentials). All dependence on the angular 
coordinates is included in spherical harmonics which form an irreducible
representation of the group O(3), and the dependence on the single 
remaining radial coordinate is governed by a one-dimensional 
Schr\"{o}dinger equation. Similarly, the conformal expansion of distribution 
amplitudes in QCD aims to separate longitudinal degrees of freedom 
from  transverse ones. All dependence on the longitudinal momentum fractions 
is included in terms of functions (orthogonal polynomials) 
forming irreducible representations of the so-called collinear subgroup 
of the conformal group, SL(2,R), describing M\"obius transformations on the 
light-cone. The transverse-momentum dependence
(the scale-dependence) is governed by simple renormalization group equations:
the different partial waves,
labelled by different ``conformal spins'',
behave independently and do not mix with each other.
Since the conformal invariance of QCD is 
broken by quantum corrections, mixing of different terms of the 
conformal expansion is only absent to leading logarithmic accuracy.
Still, conformal spin is a good quantum number in hard processes,
up to small corrections of order $\alpha_{s}^{2}$. 
Application of conformal symmetry to the studies of exclusive processes
in leading twist have received a lot of attention in the literature,
see e.g.~\cite{B+,Makeenko,Mueller}.

Despite certain complications, 
the conformal expansion presents a natural approach 
to the study of higher twist distributions, which has even  more power
than in leading twist. The reason is that conformal transformations commute
with the exact QCD equations of motion since the latter are not 
renormalized\footnote{More precisely, one can regularize the theory in 
such a way as to preserve the equations of motion.}. Thus, the equations of 
motion can be solved order by order in the conformal expansion. 
In this section, we use the approach of \cite{O82,BF90}
to work out the explicit form 
of the conformal expansion for the chiral-odd distributions
$\phi_{\perp}(u), \htt (u), \hs(u)$ and ${\cal T}(\underline{\alpha})$,
and solve the constraints (\ref{eq:3solh}) and (\ref{eq:3sole})
order by order in conformal spin.

Since quark mass terms break the conformal symmetry of the QCD Lagrangian
explicitly, one might expect difficulties to incorporate  
SU(3) breaking corrections in the formalism.
In fact, the inclusion of quark mass corrections turns out to be 
straightforward
and produces two types of effects. First, matrix elements of conformal 
operators are modified and in general do not have the symmetry of the
massless theory. This is not a ``problem'', since the conformal expansion
is designed to simplify the transverse momentum  
dependence of the wave functions by relating it to the scale dependence 
of the relevant operators. This dependence is given by operator anomalous
dimensions which are not affected by quark masses, provided they are
smaller than the scales involved. Second, new higher twist operators arise, 
in which quark masses multiply operators of lower twist, see the previous
section. These additional operators, again, do not pose a ``problem''
and can be expanded systematically in conformal partial waves, leaving 
the quark masses as multiplicative factors. Explicit examples are 
considered later in Sec.~5, while in this section
we neglect operators proportional to the quark masses and
set $\delta_{\pm} = 0$ in the formulae obtained 
in Sec.~3.1. 

The conformal expansion of distribution amplitudes is especially simple
when each constituent field has fixed (Lorentz) spin projection onto the 
light-cone.
Such constituent fields correspond to the so-called primary fields 
in conformal field theories, and their conformal spin equals
\begin{equation}
  j=\frac{1}{2}\,(l+s),
\label{eq:cspin}
\end{equation}
where $l$ is the canonical dimension  and $s$ the 
(Lorentz) spin projection.
Multi-particle states built of primary fields can be expanded in increasing 
conformal spin: the lowest possible spin equals to the sum of spins of 
constituents, and its ``wave function'' is given by the product of one-particle
states. This state is nondegenerate and cannot mix with other states 
because of conformal symmetry. Its evolution is given, therefore, by a 
simple renormalization group equation and one can check (see Sec.~3.3) that 
the corresponding anomalous dimension is the lowest one in the whole spectrum.
Therefore, this state  is the only one which
survives in the formal limit $Q^2\to \infty$; following established tradition 
we will refer to the multi-particle state with the minimal conformal spin
as ``asymptotic distribution amplitude''.

An explicit expression for the asymptotic distribution amplitude 
of a multi-particle state built of primary fields was obtained in 
Refs.~\cite{O82,BF90}:
\begin{equation}
\phi_{as}(\alpha_1,\alpha_2,\ldots,\alpha_m) = 
\frac{\Gamma[2j_1+\ldots +2j_m]}{\Gamma[2j_1]\ldots \Gamma[2j_m]}
\alpha_1^{2j_1-1}\alpha_2^{2j_2-1}\ldots \alpha_m^{2j_m-1}.
\label{eq:asymptotic}
\end{equation}
Here the $j_k$ are the conformal spins of the constituent fields 
(quark or gluons
with fixed spin projections). This distribution has conformal spin 
$j=j_1+\ldots+j_m$.
Multi-particle irreducible representations with higher spin $j+n,n=1,2,\ldots$ 
are given by  orthogonal polynomials of $m$ variables (with the constraint 
$\sum_{k=1}^m \alpha_k=1$ ) with the weight function (\ref{eq:asymptotic}). 
  
A classical example is the leading twist quark-antiquark 
distribution amplitude. The distribution amplitude
$\phi_{\perp}(u)$, defined in (\ref{eq:tda}), has
the expansion
\begin{equation}
\phi_{\perp}(u) = 6u\ub \sum_{n = 0}^{\infty}
a_{n}^{\perp} C_{n}^{3/2}(\xi) ,
\label{eq:3ceph}
\end{equation}
where $C_{n}^{3/2}(\xi)$ are Gegenbauer
polynomials (see e.g.\ \cite{ER53}). 
The dimension of  quark fields is $l=3/2$ and 
the leading twist distribution corresponds to positive spin
projection $s=+1/2$ for both the quark and the antiquark. Thus, according to
(\ref{eq:cspin}), the conformal spin of each field is $j_q=j_{\bar q} = 1$;
the asymptotic distribution amplitude (\ref{eq:asymptotic}) 
equals $\phi_{as}(\alpha_q,\alpha_{\bar q}) = 6 \alpha_q \alpha_{\bar q}$
and has conformal spin $j=2$.
Taking into account $ \alpha_q +\alpha_{\bar q} =1$ and denoting $u=\alpha_q$  
we arrive at the first term in the expansion (\ref{eq:3ceph}). The 
Gegenbauer polynomials correspond to contributions with higher conformal 
spin $j+n$ and are orthogonal over the weight function $6u\bar u$.

Note that $a^{\perp}_{0} = 1$ due to the normalization condition 
(\ref{eq:norm}). In the strict massless limit only the terms with 
even  $n$ survive in Eq.~(\ref{eq:3ceph}) because of G-parity 
invariance.
The conformal expansion, however, 
can be performed at the operator level and is 
disconnected from particular symmetries of states such as G-parity. The 
following discussion is, therefore,  valid for arbitrary $n$. We keep
terms with $n=2k+1$ for the later discussion of SU(3) breaking corrections.

The conformal expansion of the twist~3 two-particle
distribution amplitudes $\htt (u)$ and $\hs(u)$ is 
less immediate. As a first step, one has to
decompose them into components built of primary fields ---
with fixed spin projections. 
To this end, we define a
set of auxiliary amplitudes $h^{\uparrow \downarrow}(u)$
and $h^{\downarrow \uparrow}(u)$ using the spin projection operators
$P_+ = (1/2) \gamma_*\gamma_.$ and $P_- = (1/2) \gamma_.\gamma_*$ 
to single out quark states with $s=+1/2$ and $s=-1/2$, respectively, 
(see \cite{BF90,ABS}):
\begin{eqnarray}
\langle 0| \ub(z) \gamma. \gamma_{\ast}[z,-z] 
d(-z)|\rho^-(P,\lambda)\rangle
&=& f_{\rho}^{T}m_{\rho}^{2}\,
\frac{e^{(\lambda)}\cdot z}{p\cdot z} 
\int_{0}^{1} du\, e^{i\xi pz} h^{\uparrow \downarrow} (u), 
\label{eq:hud} \\
\langle 0| \ub(z) \gamma_{\ast} \gamma.[z,-z] 
d(-z)|\rho^-(P,\lambda)\rangle
&=& f_{\rho}^{T}m_{\rho}^{2}\,
\frac{e^{(\lambda)}\cdot z}{p\cdot z} 
\int_{0}^{1} du\, e^{i\xi pz} h^{\downarrow \uparrow} (u), 
\label{eq:hdu}
\end{eqnarray}
which are related to $\htt (u)$ and $\hs(u)$ by 
(see (\ref{eq:tda}), (\ref{eq:sda}))
\begin{eqnarray}
h^{\uparrow \downarrow}(u)
&=& 
\htt (u) + \frac{1}{2}\frac{d\hs(u)}{du},
\label{eq:3hlud}\\
h^{\downarrow \uparrow}(u)
&=& -\htt (u)
+ \frac{1}{2}\frac{d\hs(u)}{du}.
\label{eq:3eud}
\end{eqnarray}
The conformal expansion of $h^{\uparrow \downarrow}(u)$
and $h^{\downarrow \uparrow}(u)$ is straightforward
and is given by
\begin{eqnarray}
h^{\uparrow \downarrow} (u) &=& 2\ub \sum_{n=0}^{\infty}
h_{n}^{\uparrow \downarrow} P_{n}^{(1,0)}(\xi),
\label{eq:3cehud}\\
h^{\downarrow \uparrow} (u) &=& 2u\sum_{n=0}^{\infty}
h_{n}^{\downarrow \uparrow} P_{n}^{(0,1)}(\xi),
\label{eq:3cehdu}
\end{eqnarray}
where $ P_{n}^{(0,1)}(\xi)$ are Jacobi polynomials (see e.g.\ \cite{ER53}) and
the $n$-th term corresponds to conformal spin $j= n+3/2$.
Substituting these expansions in
(\ref{eq:3hlud}) and (\ref{eq:3eud}) and  using the identities 
(\ref{eq:jg1}), (\ref{eq:jg2}) and (\ref{eq:jg3}) 
in App.~\ref{app:a}, we obtain
\begin{eqnarray}
\htt (u) &=& \sum_{n=0, 2, 4, \ldots} \left(H_{n} - H_{n-1}\right)
C_{n}^{1/2}(\xi)
+ \sum_{n=1, 3, 5, \ldots} \left(h_{n} - 
h_{n-1}\right)
C_{n}^{1/2}(\xi),
\label{eq:3ceh}\\
\hs(u) &=& 4u\ub \left(\sum_{n=0,2,4, \ldots} 
\frac{H_{n} - H_{n+1}}{(n+1)(n+2)}
C_{n}^{3/2}(\xi)
+\sum_{n=1,3,5, \ldots} 
\frac{h_{n} - h_{n+1}}{(n+1)(n+2)}
C_{n}^{3/2}(\xi)
\right),\makebox[1cm]{\ }
\label{eq:3cee}
\end{eqnarray}
where $H_{-1}=h_{-1}=0$ is implied, and 
\begin{eqnarray}
H_{n} &\equiv& \frac{ h_{n}^{\uparrow \downarrow} -
(-1)^{n}h_{n}^{\downarrow\uparrow}}{2},
\nonumber \\
h_{n} &\equiv& \frac{ h_{n}^{\uparrow \downarrow} +
(-1)^{n}h_{n}^{\downarrow\uparrow}}{2},
\label{eq:3coeff}
\end{eqnarray}
for $n= 0, 1, 2, \ldots$ correspond to 
G-parity conserving and G-parity violating 
contributions, respectively.
Note that the coefficient in front of each orthogonal polynomial
in (\ref{eq:3ceh}) and (\ref{eq:3cee}) does not correspond to a definite 
conformal spin; in contrast to (\ref{eq:3ceph}), (\ref{eq:3cehud}),
and (\ref{eq:3cehdu}), it is rather given by difference 
between the contributions of two successive conformal spins.

The conformal expansion of the twist~3 three-particle distribution
amplitude gives yet another example for the general 
expression Eq.~(\ref{eq:asymptotic}). The expansion reads
\begin{equation}
{\cal T}(\alpha_{d}, \alpha_{u}, 1-\alpha_{d}-\alpha_{u})
= 360 \alpha_{d}\alpha_{u}(1-\alpha_{d}-\alpha_{u})^{2}
\sum_{k,l = 0}^{\infty} \omega^{T}_{k,l}J_{k,l}(\alpha_{d}, \alpha_{u}),
\label{eq:3cet}
\end{equation}
where 
$J_{k,l}(\alpha_{d}, \alpha_{u}) \equiv 
J_{k,l}(6, 2, 2, \alpha_{d}, \alpha_{u})$ are   particular Appell  
polynomials of two variables (see p269 of \cite{ER53}).
The conformal spin of a generic term in this expression equals
$j = k + l + 7/2$ and is the same for all contributions 
with equal sum $n=k+l$. This illustrates that three-particle 
conformal representations are degenerate; the number of independent 
operators with the same spin in fact increases with the spin.
Conformal symmetry does not allow mixing between contributions 
with different $j=n+7/2$; it does allow, however,  mixing with each other of 
different states with the same value of $j$. Therefore, the mixing matrix 
for higher twist operators becomes only block-diagonal in the conformal basis,
instead of being diagonalized like in leading twist.

Next, we are going to demonstrate that conformal expansion is fully consistent
with the equations of motion. To this end we need to show that the conformal 
expansion coefficients for two-particle twist~3 distributions can be 
expressed in terms of the expansion coefficients for 
three-particle distributions with the same conformal spin, and 
we need to separate
the Wandzura-Wilczek contributions.
The calculation is straightforward, although somewhat tedious. 

We decompose
\begin{equation}
H_{n} =H_{n}^{WW} + H_{n}^{g}, \;\;\;\;\;\;\;
h_{n} = h_{n}^{WW} + h_{n}^{g},
\label{eq:3solc}
\end{equation}
and start with the Wandzura-Wilczek contributions 
(\ref{eq:3hww}) and (\ref{eq:3eww}) which give rise to 
auxiliary amplitudes ${h^{\uparrow \downarrow}}^{WW}(u)$ with
\begin{eqnarray}
{h^{\uparrow \downarrow}}^{WW}(u) 
&=& 2\ub
\left(- \int_{0}^{u}dv\frac{\phi_{\perp}(v)}{\bar v}
+ \int_{u}^{1}dv\frac{\phi_{\perp}(v)}{v}\right),
\label{eq:3udww}\\
{h^{\downarrow \uparrow}}^{WW}(u) 
&=& 2u
\left(- \int_{0}^{u}dv\frac{\phi_{\perp}(v)}{\bar v}
+ \int_{u}^{1}dv\frac{\phi_{\perp}(v)}{v}\right).
\label{eq:3duww}
\end{eqnarray} 
The integrals on the right-hand side of (\ref{eq:3udww}) and (\ref{eq:3duww}) 
can easily be done using (\ref{eq:jd1}) and (\ref{eq:jg4}):
\begin{equation}
- \int_{0}^{u}dv\frac{\phi_{\perp}(v)}{\bar v}
+ \int_{u}^{1}dv\frac{\phi_{\perp}(v)}{v}
= -3 \sum_{n = 0}^{\infty} a_{n}^{\perp} P_{n+1}^{(0,0)}(\xi).
\label{eq:3intph}
\end{equation}
Substituting the recurrence relations for  
Jacobi polynomials, (\ref{eq:jrec2}),
into this result, one immediately obtains
\begin{eqnarray}
H_{n}^{WW} &=& \frac{3(n+1)}{2n+3} a_{n}^{\perp};
\;\;\;\;\;
h_{n}^{WW}= - \frac{3(n+1)}{2n+1} a_{n-1}^{\perp}
\;\;\;\;\;\;\;\;(n = 0, 2, 4, \ldots),
\nonumber \\
H_{n}^{WW}&=&- \frac{3(n+1)}{2n+1} a_{n-1}^{\perp};
\;\;\;\;\;
h_{n}^{WW}= \frac{3(n+1)}{2n+3} a_{n}^{\perp}
\;\;\;\;\;\;\;\;(n = 1, 3, 5, \ldots).
\label{eq:3hnww}
\end{eqnarray}
For even $n$, we
find that $a_n^\perp$ which corresponds to the
conformal spin $n+2$ in the expansion for
the twist~2 distribution amplitude gives rise to $H^{WW}_n$ and $H^{WW}_{n+1}$
which corresponds to the conformal spin $n+3/2$ and
$n+5/2$, respectively.  Likewise for odd $n$, $a_n^\perp$
gives rise to $h_n^{WW}$ and $h_{n+1}^{WW}$.  
These values of the conformal spin do not match the expansion 
in Eqs.~(\ref{eq:3cehud}) and (\ref{eq:3cehdu}). This is, however, not a 
contradiction since
Wandzura-Wilczek terms are in fact not intrinsic twist~3 distributions,
but correspond to matrix elements of twist~2 operators 
over $\rho$ mesons with different (longitudinal) polarization.
To relate matrix elements of conformal operators over longitudinal and 
transverse $\rho$ mesons, one has to perform a spin rotation 
(in the $\rho$ meson rest frame) which does not
commute with the generators of collinear conformal group. 
As shown in App.~\ref{app:K}, this rotation gives rise to the shift 
in conformal spin and exactly explains the mismatch appearing in 
Eq.~(\ref{eq:3hnww}). Therefore, the conformal symmetry is realized 
in Wandzura-Wilczek contributions as well, but to see this one has to 
supplement the conformal classification of operators by conformal 
transformation properties of the meson states.

The three-particle contributions $H_{n}^{g}$ 
and $h_{n}^{g}$ of (\ref{eq:3solc})
can be treated similarly.
{}From the solutions  for $\httg(u)$ and $\hsg(u)$ 
in Sec.~3.1 we obtain the corresponding auxiliary amplitudes:
\begin{eqnarray}
{h^{\uparrow \downarrow}}^{g} &=&
\zeta_{3 \rho}^{T} \left[\ub
\left( - \int_{0}^{u}dv \frac{K(v)}{\bar v}
+ \int_{u}^{1}dv \frac{K(v)}{v}\right)
+ \frac{d}{du}\int_{0}^{u}d\alpha_{d}
\int_{0}^{\ub}d\alpha_{u} \frac{1}{\alpha_{g}}
{\cal T}(\underline{\alpha})\right],
\label{eq:3udg}\\
{h^{\downarrow \uparrow}}^{g} &=&
\zeta_{3 \rho}^{T} \left[ u
\left( - \int_{0}^{u}dv \frac{K(v)}{\bar v}
+ \int_{u}^{1}dv \frac{K(v)}{v}\right)
- \frac{d}{du}\int_{0}^{u}d\alpha_{d}
\int_{0}^{\ub}d\alpha_{u} \frac{1}{\alpha_{g}}
{\cal T}(\underline{\alpha})\right],
\label{eq:3dug}
\end{eqnarray}
where $\alpha_{g} = 1 - \alpha_{d}- \alpha_{u}$, and
\begin{equation}
K(u) = \frac{d}{du}\int_{0}^{1}d\alpha_{d} \int_{0}^{\ub}
d\alpha_{u} \frac{1}{\alpha_{g}}
\left( \alpha_{d}\frac{d}{d \alpha_{d}}
+ \alpha_{u}\frac{d}{d \alpha_{u}} -1
\right) {\cal T}(\underline{\alpha}).
\label{eq:3k}
\end{equation}
Substitution of  the conformal expansion (\ref{eq:3cet})
into (\ref{eq:3k}) yields
\begin{equation}
K(u) = 180\,u\ub \sum_{k,l=0}^{\infty}
\omega_{k,l}^{T} \frac{k!\,l!\,(-1)^{k}}{(k+l+2)!}
\left( \frac{k-l}{(k+l+3)}P_{k+l+2}^{(1,1)}(\xi)
+ P_{k+l+1}^{(1,1)}(\xi) \right),
\label{eq:3k2}
\end{equation}
where we have used Eq.~(\ref{eq:apint}) to perform the integration.
The final integration involving $K(v)$ on the right-hand side
of (\ref{eq:3udg}) and (\ref{eq:3dug})
can be done similarly to (\ref{eq:3intph}) by using (\ref{eq:jd1}):
\begin{equation}
- \int_{0}^{u}\!\!dv \frac{K(v)}{\bar v}
+ \int_{u}^{1}\!\!dv \frac{K(v)}{\bar v} =
180 \sum_{k,l=0}^{\infty}\omega_{k,l}^{T}
\frac{k!\,l!\,(-1)^{k+1}}{(k+l+3)!}\left( \frac{k-l}{k+l+4}
P^{(0,0)}_{k+l+3}(\xi)+P_{k+l+2}^{(0,0)}(\xi)\right)\!,
\label{eq:3intk}
\end{equation}
and the last term of (\ref{eq:3udg}) and (\ref{eq:3dug})
can be integrated  using (\ref{eq:apint2}):
\begin{equation}
\frac{d}{du}\int_{0}^{u}\!\!d\alpha_{d}
\int_{0}^{\ub}\!\!d\alpha_{u} \frac{1}{\alpha_{g}}
{\cal T}(\underline{\alpha})
= 180u\ub\sum_{k,l=0}^{\infty} 
\omega_{k,l}^{T}\frac{k!\,l!(-1)^{k}}{(k+l+3)!}
\left(\frac{k-l}{k+l+3}P_{k+l+2}^{(1,1)}(\xi) - P_{k+l+1}^{(1,1)}(\xi)
\right)\!.
\label{eq:3intt}
\end{equation}
Substituting (\ref{eq:3intk}) and (\ref{eq:3intt})
into (\ref{eq:3udg}), (\ref{eq:3dug}),
and using the identities (\ref{eq:jrec12})--(\ref{eq:jrec2}),
we find
\begin{eqnarray}
H_{n}^{g} &=& 180 \zeta_{3 \rho}^{T}
\sum_{k=0}^{n-2} \frac{k!(n-k-2)! (k+2)}{(n+2)!}
(-1)^{n-k} \omega^{T}_{[k,n-k-2]}
\;\;\;\;\;\;\;\;\;(n= 3, 4, 5, \ldots),
\label{eq:3hng}\\
h_{n}^{g} &=& -180 \zeta_{3 \rho}^{T}
\sum_{k=0}^{n-2} \frac{k!(n-k-2)! (k+2)}{(n+2)!}
(-1)^{n-k} \omega^{T}_{\{k,n-k-2\}}
\;\;\;\;\;\;(n= 2, 3, 4, \ldots),
\label{eq:3hng2}
\end{eqnarray}
while $H_{0}^{g} = H_{1}^{g} =H_{2}^{g}=h_{0}^{g}
=h_{1}^{g} = 0$.
Here we introduced the following quantities:
\begin{eqnarray}
\omega^{T}_{[k,l]} &\equiv& \frac{\omega_{k,l}^{T} - \omega_{l,k}^{T}}{2},
\label{eq:3oma}\\
\omega^{T}_{\{k,l\}} &\equiv& \frac{\omega_{k,l}^{T} + \omega_{l,k}^{T}}{2}.
\label{eq:3oms}
\end{eqnarray}
We find that the coefficients $\omega^{T}_{k,l}$ with fixed 
$k+l = n-2$, which correspond to the conformal spin $j=n+3/2$, all 
contribute to $H_{n}^{g}$ and $h_{n}^{g}$ corresponding to
the same conformal spin $j=n+3/2$, as anticipated.
For later convenience, we give the lowest order coefficients
 (\ref{eq:3hng}) and (\ref{eq:3hng2}) explicitly:
\begin{eqnarray}
H_{3}^{g} &=& \frac{15}{2}\zeta_{3\rho}^{T} \omega_{[1,0]}^{T} =
-70 \zeta_{3\rho}^{T}, \;\;\; H_{4}^{g} = \zeta_{3\rho}^{T} 
\omega_{[2,0]}^{T}, \;\;\; H_{5}^{g} = \zeta_{3\rho}^{T}
\left( \frac{3}{2} \omega_{[3,0]}^{T} - \frac{1}{2}\omega_{[2,1]}^{T}
\right), \ldots,
\label{eq:3exa1}\\
h_{2}^{g} &=& - 15 \zeta_{3\rho}^{T}\omega_{\{0,0\}}^{T},\;\;\;
h_{3}^{g} = - \frac{3}{2}\zeta_{3\rho}^{T}\omega_{\{1,0\}}^{T}, \;\;\;
h_{4}^{g} = \zeta_{3\rho}^{T}\left(\frac{3}{4}\omega_{\{1,1\}}^{T}
- 3 \omega_{\{2,0\}}^{T} \right), \ldots, 
\label{eq:3exa2}
\end{eqnarray}
where we substitute $\omega^{T}_{[0,1]} = 28/3$, which
follows from the normalization condition (\ref{eq:normalize}).

{}From (\ref{eq:3solc}), (\ref{eq:3hnww}), and
(\ref{eq:3hng}) it follows that the two lowest coefficients 
$H_{0}$ and $H_{1}$ are completely determined by the value of 
$a_{0}^{\perp}=1$, which results in $H_{0}=1$ and $H_{1} = -2$.
It is easy to see that these values for $H_{0}$ and $H_{1}$
ensure $\int_{0}^{1}du \,\htt (u) = \int_{0}^{1}du \,\hs(u) = 1$ and
therefore, the normalization condition 
for $\phi_{\perp}(u)$ ensures  correct normalization
of $\htt (u)$ and $\hs(u)$.

To summarize, we have demonstrated that the equations of motion 
that relate different twist~3 operators can be solved order by order 
in the conformal expansion. In other words, 
equations of motion impose ``horizontal''  relations between 
operators of the same conformal spin and do not involve other spins.
This picture is somewhat complicated by the Wandzura-Wilczek contributions
of the operators of lower (leading) twist which have  a more 
peculiar structure.
The explicit relations derived above can be made somewhat more compact
by assuming
G-parity invariance. In this case 
$a_{n}^{\perp} =0$ for odd $n$, 
$\omega_{k,l}^{T} = - \omega_{l,k}^{T}$, 
$H_{n} = h_{n}^{\uparrow \downarrow}$,
and $h_{n} = 0$. As mentioned above, the G-parity violating terms
are only relevant for SU(3) breaking corrections in the distribution 
amplitudes of $K^*$ mesons.

\subsection{Renormalization and scale-dependence}

The scale-dependence of the chiral-odd distribution
amplitudes $\phi_{\perp}(u, \mu^{2})$, $\htt (u, \mu^{2})$
and $\hs(u, \mu^{2})$ is governed by the renormalization group (RG)
equation for the relevant nonlocal light-cone operators appearing
in the definitions (\ref{eq:tda}), (\ref{eq:sda}), and (\ref{eq:T3}).
Unlike inclusive processes, operators involving total derivatives
have to be taken into account since they contribute to nonforward
matrix elements, and this leads to
additional operator mixing.
{}For example, consider the  set of local operators  
$\ub(0) (\overleftarrow{D}.)^{n-k} \sigma_{\perp}.
(\overrightarrow{D}.)^{k} d(0)$ ($k= 0, 1, 2, \ldots$),
which contribute to the $n$-th moment of the twist~2 distribution amplitude
$\phi_{\perp}(u, \mu^{2})$.
These operators differ by total derivatives and all
mix with each other under renormalization;
to calculate the scale-dependence, one has to find the eigenvalues
and eigenvectors of the corresponding anomalous dimension matrix.

As is well known \cite{B+,Makeenko,O82}, 
 conformal expansion provides the solution to this problem.
Analysis based on the anomalous Ward identities
for the dilatation and special conformal
transformation (which are members of the conformal group)
shows \cite{O82} that, to leading logarithmic accuracy,
the conformal operators with different conformal spin do not mix with 
each other and thus diagonalize
the anomalous dimension matrix.
As a result, by employing a conformal operator basis
we do not encounter any additional operator mixing compared 
to inclusive processes.
The relevant anomalous dimensions,
which correspond to the eigenvalues of the anomalous dimension matrix, 
can be extracted directly from the results for renormalization
 of the corresponding parton distribution functions. 
In particular, the one-loop anomalous dimensions 
for $\htt (u, \mu^{2})$ and $\hs(u, \mu^{2})$ are the same as
for the chiral-odd parton distribution functions \cite{KT95,KN97,BM97,AM90}
as will be shown in the following. 

Our main task in this section is to reveal the explicit operator 
content of the conformal operators, corresponding to particular
coefficients in the
conformal expansions (\ref{eq:3ceh}), (\ref{eq:3cee}),
(\ref{eq:3ceph}), and (\ref{eq:3cet}).
We give a one-to-one correspondence between the conformal basis
and the basis used in the inclusive case.
This allows us to determine the anomalous dimensions
of the conformal operators and to find the evolution of the 
distribution amplitudes $\htt (u, \mu^{2})$ and $\hs(u, \mu^{2})$
through the conformal expansion.
We will work out this program for arbitrary conformal spin.

One complication is that the conformal
operator basis for the higher twist operators 
is degenerate (see (\ref{eq:3cet})) and the mixing matrix becomes 
only block-diagonal instead of being fully diagonalized like in leading twist.
Consequently, the conformal expansion for three-particle contributions
to $\htt (u, \mu^{2})$ and $\hs(u, \mu^{2})$
does not resolve possible mixing between components with the same conformal
spin. This is similar to mixing of the many quark-gluon correlation
operators for the corresponding twist~3 parton distribution 
functions \cite{KT95,KN97,BM97}. It has been shown recently 
\cite{BBKT96,ABH91}, however, that an important simplification occurs
in the limit of a large number of colors or of large
spin (moment of parton distribution function).
In these limits all complicated mixing disappears and 
the twist~3 parton distribution functions obey simple DGLAP-type 
evolution.
We will demonstrate that the twist~3 distribution amplitudes obey 
a similar pattern.

Let us start with the Wandzura-Wilczek terms $\httWW(u, \mu^{2})$
and $\hsWW(u, \mu^{2})$ (see (\ref{eq:3ceh}), (\ref{eq:3cee}),
and (\ref{eq:3solc})). The coefficients $H_{n}^{WW}$ and $h_{n}^{WW}$
in their conformal expansions have been expressed by $a_{k}^{\perp}$ 
of (\ref{eq:3ceph}) as shown in (\ref{eq:3hnww}).
Therefore, the scale-dependence of $a_{k}^{\perp}$,
i.e.\ of the twist~2 amplitude $\phi_{\perp}(u, \mu^{2})$, 
completely determines
that of $\httWW(u, \mu^{2})$ and $\hsWW(u, \mu^{2})$.
{}From (\ref{eq:3ceph}) and orthogonality relations of Gegenbauer
polynomials \cite{ER53}, we obtain
\begin{equation}
a_{n}^{\perp}(\mu^{2}) = \frac{2(2n+3)}{3(n+1)(n+2)}
\int_{0}^{1}du\, C_{n}^{3/2}(\xi) \phi_{\perp}(u, \mu^{2}).
\label{eq:3anp}
\end{equation}
Substituting (\ref{eq:tda}) into the right-hand side
of (\ref{eq:3anp}) gives 
\begin{equation}
\left( f_{\rho}^{T} a_{n}^{\perp} \right)
(\mu^{2}) =
i \frac{2(2n+3)}{3(n+1)(n+2)} \frac{1}{(p\cdot z)^{n+1}}
\langle 0| \Omega_{n}^{\perp}(0; \mu^{2})
| \rho^{-} (P, \lambda) \rangle
\label{eq:3anp2}
\end{equation}
with
\begin{equation}
\Omega_{n}^{\perp}(x; \mu^{2})=
\left(i\partial. \right)^{n} \ub(x)
e_{\perp}^{(\lambda)\nu}\sigma_{\nu}. C_{n}^{3/2}
\left(\frac{\deriv\!.}
{\partial.}\right) d(x),
\label{eq:3omg}
\end{equation}
where the local operator on the right-hand side
is renormalized at $\mu^{2}$,
$\deriv = \overrightarrow{D}-\overleftarrow{D}$,
and $\partial_{\mu}$ is the total derivative (\ref{eq:3tdrv}).
$\Omega_{n}^{\perp}(x, \mu^{2})$ is the conformal
operator of conformal spin $j=n+2$ \cite{Makeenko,O82}\footnote{
In principle, one can construct
a tower of conformal operators 
$(\partial.)^{m} \Omega_{n}^{\perp}$ ($m= 0, 1, \ldots$)
with the same conformal spin, but with the different
``third component'' of it.}.
Therefore, it is RG covariant to leading logarithmic
accuracy and satisfies the RG equation:
\begin{equation}
\left( \mu \frac{\partial}{\partial \mu} + \beta(g) 
\frac{\partial}{\partial g} + \frac{\alpha_{s}}{2 \pi}
\gamma_{n}^{\perp} \right) \Omega_{n}^{\perp}(x; \mu^{2}) =0,
\label{eq:3rgeq}
\end{equation}
where $\gamma_{n}^{\perp}$ is the one-loop anomalous dimension
of the operator $\Omega_{n}^{\perp}$.
To establish a formal connection with the results given in the 
literature,
it is convenient to consider the case where $\Omega_{n}^{\perp}$
is diagonal in quark flavour corresponding to the flavour matrices
$\lambda_{3}, \lambda_{8}$,
and to take the forward matrix element of (\ref{eq:3rgeq})
over the nucleon state $|N(P, S)\rangle$.
Because the total derivatives drop out in this matrix element,
(\ref{eq:3rgeq}) reduces to the RG equation
for $\langle N(P, S)| \overline{\psi}(0) \sigma_{\perp}.\left(iD.
\right)^{n}
\psi(0)|N(P, S)\rangle$, which gives the $n$-th moment
of the nucleon parton distribution function $h_{1}(x, \mu^{2})$
of (\ref{eq:ftda}). 
By matching with the results for renormalization of 
$h_{1}(x, \mu^{2})$ \cite{AM90}, we obtain $\gamma_{n}^{\perp}$ as
\begin{equation}
\gamma_{n}^{\perp} = 4C_{F}
\left( \psi(n+1)+ \gamma_{E} - \frac{3}{4} + \frac{1}{n+1}
\right),
\label{eq:3anomdp}
\end{equation}
where $\psi(n+1) = \sum_{k=1}^{n}1/k - \gamma_{E}$ is
the digamma function, $\gamma_{E}$ is the Euler constant,
and $C_{F}=(N_{c}^{2}-1)/2N_{c}$.
{}From (\ref{eq:3anp2})--(\ref{eq:3anomdp}), we obtain
\begin{equation}
\left(f_{\rho}^{T} a_{n}^{\perp} \right)(Q^{2})
= L^{\gamma_{n}^{\perp}/b}
\left(f_{\rho}^{T} a_{n}^{\perp} \right)(\mu^{2}),
\label{eq:3scalep}
\end{equation}
where $L \equiv \alpha_{s}(Q^{2})/\alpha_{s}(\mu^{2})$ and
$b=(11N_{c} - 2N_{f})/3$. Combined with
(\ref{eq:3ceh}), (\ref{eq:3cee}) and (\ref{eq:3hnww}), this result 
gives the $\mu^{2}$-dependence of
$\httWW(u, \mu^{2})$ and $\hsWW(u, \mu^{2})$ and
also determines evolution of the twist~2 distribution amplitude 
$\phi_{\perp}(u, \mu^{2})$ of (\ref{eq:3ceph}).
We note that (\ref{eq:3scalep}) for $n=0$ gives
the scale-dependence of the tensor decay constant $f_{\rho}^{T}$
as
\begin{equation}
f_{\rho}^{T}(Q^{2}) = L^{C_{F}/b} f_{\rho}^{T}(\mu^{2}),
\label{eq:3scalep0}
\end{equation}
because $a_{0}^{\perp}=1$.

The three-particle contributions $\httg(u, \mu^{2})$
and $\hsg(u, \mu^{2})$ can be treated in a similar manner,
although the discussion becomes more complicated because one has to
deal  with a degenerate representation of the conformal
group
(see (\ref{eq:3ceh}), (\ref{eq:3cee}) and (\ref{eq:3solc})).
The relevant expansion coefficients $H_{n}^{g}$ and $h_{n}^{g}$
are expressed by $\omega_{k,l}^{T}$ in (\ref{eq:3cet})
and shown in (\ref{eq:3hng}) and (\ref{eq:3hng2}).
Thus the first step is to demonstrate
that $\omega_{k,l}^{T}$ are given by matrix elements
of the local conformal operators
derived in \cite{O82}. Using (\ref{eq:3cet}) and the
orthogonality relations (\ref{eq:appello0})
for the Appell polynomials, we obtain
\begin{equation}
\int{\cal D}\underline{\alpha}\: J_{k, n-k-2}(\alpha_{d}, \alpha_{u})
{\cal T}(\underline{\alpha})
= \frac{360(-1)^{n}}{2^{n+1}(n+1) (2n+1)!!}
\sum_{r=0}^{n-2} \omega_{r, n-r-2}^{T} W_{n-r-2,k}^{(n-1)}
\label{eq:3apor}
\end{equation}
for $k=0, 1, \ldots, n-2$, and $n= 2, 3, \ldots$
Making use of 
\begin{eqnarray}
\lefteqn{\langle 0| \ub(tz) \sigma^{\nu}. [tz, vz]
gG_{\nu}.(vz) [vz, wz] d(wz)
|\rho^{-} (P, \lambda)\rangle
=} 
\nonumber\\
&=& (p\cdot z)(e^{(\lambda)}\cdot z) f_{3\rho}^{T}m_{\rho}
\int {\cal D}\underline{\alpha}\:
e^{-ip\cdot z(t\alpha_{u} + w\alpha_{d} + v\alpha_{g})}
{\cal T}(\underline{\alpha}),
\label{eq:3defT}
\end{eqnarray}
which is equivalent to (\ref{eq:T3}),
the left-hand side of (\ref{eq:3apor}) gives
\begin{equation}
\frac{1}{f_{3\rho}^{T} m_{\rho}}
\frac{1}{(e^{(\lambda)}\cdot z) (p \cdot z)^{n-1}}
\langle 0| \Lambda_{k, n-k-2}^{T}(0) 
|\rho^{-} (P, \lambda) \rangle,
\label{eq:3apor2}
\end{equation}
with
\begin{equation}
\Lambda_{k,l}^{T} (0)
= \left. \left( i\partial.\right)^{k+l}
J_{k,l}\left( \frac{D^{y}_{\displaystyle{\cdot}}}{\partial.}, 
\frac{D^{x}_{\displaystyle{\cdot}}}{\partial.}
\right) \ub(x) \sigma^{\nu}. gG_{\nu}.(0) d(y)
\right|_{x=y=0},
\label{eq:3Tcnf}
\end{equation}
where the covariant
derivatives $D_{\mu}^{x}$ and $D_{\mu}^{y}$
act on the coordinates $x$ and $y$, respectively.
$\Lambda_{k,n-k-2}^{T}$ ($k=0, 1, \ldots, n-2$)
are the twist~3 conformal operators of spin
$j=n+3/2$, forming a degenerate basis 
for three-particle representation \cite{O82}.\footnote{One can
generate a tower of conformal operators with
a different ``third component'' of 
conformal spin by acting repeatedly with $\partial.$
on (\ref{eq:3Tcnf}).}
Inverting the matrix $W_{k',k}^{(n-1)}$ in (\ref{eq:3apor}),
we obtain
\begin{equation}
f_{3 \rho}^{T}\omega_{k, n-k-2}^{T}
= \frac{N_{n}^{T} (-1)^{k} }{90k!(n-k-2)!}
\langle 0|\Theta_{k, n-k-2}^{T}(0) |\rho^{-}(P, \lambda)
\rangle,
\label{eq:3omcon}
\end{equation}
where $N_{n}^{T}$ is the dimensionless and scale-independent constant:
\begin{equation}
N_{n}^{T}\equiv
\frac{2^{n-1} (n+1) (2n+1)!!}
{m_{\rho}(e^{(\lambda)}\cdot z) (p\cdot z)^{n-1}}.
\label{eq:3NnT}
\end{equation}
The numerical factor in front of the matrix element
(\ref{eq:3omcon})
is put for later convenience.
The operators $\Theta_{k, n-k-2}^{T}$ 
are given by  linear combinations of $\Lambda_{r, n-r-2}^{T}$
($r = 0, 1, \ldots, n-2$) and therefore have
conformal spin $j=n+3/2$.

The second step is to determine the explicit form
of the relevant conformal operator $\Theta_{k, n-k-2}^{T}$.
For the few lowest conformal spins, it is easy to express 
$\Theta_{k,n-k-2}^{T}$ as a linear combination of 
$\Lambda_{r, n-r-2}^{T}$ using (\ref{eq:3apor})--(\ref{eq:3Tcnf}),
but the procedure becomes more complicated for higher 
conformal spins.
Using orthogonality relations 
for the Appell polynomials, (\ref{eq:appello}),  
it proves possible, however, to determine 
$\Theta_{k,n-k-2}^{T}$ for general $n$ up to total derivatives.
\begin{eqnarray}
\lefteqn{\int {\cal D} \underline{\alpha}\: \alpha_{d}^{k}
\alpha_{u}^{n-k-2}
{\cal T}(\underline{\alpha}) = } \nonumber\\
&=&
\omega_{k,n-k-2}^{T} 
\frac{360 (-1)^{n} k!(n-k-2)!}{2^{n+1} (n+1)(2n+1)!!} 
+ (\mbox{terms involving $\left. \omega^{T}_{l,r-l-2}\right|_{r<n}$})
\nonumber\\
&=& 
\frac{1}{f_{3\rho}^{T} m_{\rho} (e^{(\lambda)}\cdot z) (p\cdot z)^{n-1}}
\langle 0| \ub(0) (i \overleftarrow{D}.)^{n-k-2} \sigma^{\nu}. 
gG_{\nu}.(0)(i \overrightarrow{D}.)^{k} d(0)|\rho^{-}(P, \lambda)\rangle,
\label{eq:33pmat}
\end{eqnarray}
where the last line is obtained by substituting (\ref{eq:3defT})
into the left-hand side. 
{}From (\ref{eq:3apor})--(\ref{eq:3Tcnf}) it follows that
$\omega_{k, n-k-2}^{T}$ are given by the matrix elements
of the operators of dimension $n+3$. This implies that 
``(terms involving $\omega^{T}_{l,r-l-2}|_{r<n}$)'' 
corresponds to matrix elements of operators
involving total derivatives, which are given by
linear combinations of terms  
$\sim \left(\partial.\right)^{n-r} \Lambda_{k,r-k-2}^{T}$
($k=0, 1, \ldots, r-2;\: r=2, 3, \ldots, n-1$).
Thus we conclude, from (\ref{eq:3omcon}) and (\ref{eq:33pmat}), that
\begin{equation}
\Theta_{k,n-k-2}^{T}(0) = 
\ub(0) (- i \overleftarrow{D}.)^{n-k-2} \sigma^{\nu}. 
gG_{\nu}.(0)(i \overrightarrow{D}.)^{k} d(0)
+ (\mbox{total derivatives}).
\label{eq:3ThetaT}
\end{equation}

The operators $\Theta_{k, n-k-2}^{T}$ 
($k=0, 1, \ldots, n-2$) have the same conformal spin and
may mix with each other,
although they do not mix with $\Theta_{l, r-l-2}^{T}$ for
$n \neq r$.
To write down the corresponding RG equation,
it is convenient to introduce combinations that are
even and odd under the substitution $k \rightarrow n-k-2$:
\begin{equation}
S_{n;k}^{\pm}\equiv 
\frac{\Theta_{k,n-k-2}^{T} \pm \Theta_{n-k-2,k}^{T}}{2}
\;\;\;\;\;\;\;\;\;\; (k= 0, 1, \ldots, \kappa_{n}^{\pm}),
\label{eq:3Snk}
\end{equation}
where
\begin{equation}
\kappa_{n}^{+} = \left[ \frac{n}{2} \right] -1,
\;\;\;\;\;\;\;\;\;\; 
\kappa_{n}^{-} = \left[\frac{n-1}{2} \right] -1.
\label{eq:3kappa}
\end{equation}
It is straightforward to see that
$S_{n;k}^{+}$ and $S_{n;k}^{-}$
possess opposite ``parity'' under the G-parity transformation.
Therefore, these two sets of the operators do not mix with each other.
The RG equations are given by
\begin{equation}
\left(\mu \frac{\partial}{\partial \mu}+
\beta(g) \frac{\partial}{\partial g}\right) 
S_{n;k}^{\pm}(0; \mu^{2})
= - \frac{\alpha_{s}}{2\pi} \sum_{l=0}^{\kappa_{n}^{\pm}}
\left(\Gamma^{T\pm}_{n}\right)_{k,l} S_{n;l}^{\pm}
(0; \mu^{2}),
\label{eq:3RG}
\end{equation}
where the one-loop anomalous dimension $\Gamma^{T+}_{n}$ 
($\Gamma^{T-}_{n}$)
is a
$[n/2]\times [n/2]$ ($[(n-1)/2]\times [(n-1)/2]$) 
matrix, describing the  mixing.
Note that the number of independent operators
and thus the size of the mixing matrix increase with conformal spin.

In order to determine the anomalous dimension matrices 
$\Gamma_{n}^{T\pm}$ in (\ref{eq:3RG}) we make contact with  
inclusive processes. 
Similarly to the discussion 
of the Wandzura-Wilczek part, we consider the case
where the operators $S_{n;k}^{\pm}$ are flavour-diagonal 
and take the forward matrix element
of (\ref{eq:3RG}) in between nucleon states $|N(P,S)\rangle$.
The total derivatives in (\ref{eq:3ThetaT}) drop out,
and (\ref{eq:3RG}) 
reduces to the RG equations for 
$$
\langle N(P,S)| \overline{\psi}(i\!\derright .)^{n-k-2}
\sigma^{\nu}. gG_{\nu}.(i\!\derright .)^{k}\psi
\pm (k \rightarrow n-k-2)|N(P,S)\rangle,
$$ 
familiar from studies of 
 the evolution of the $n$-th
moment of the nucleon parton distributions
$e(x, \mu^{2})$ and $h_{L}(x, \mu^{2})$ defined in (\ref{eq:fsda})
and (\ref{eq:ftda}).
It is straightforward to see that these operators
for the upper and lower sign,
which are even and odd under $k \rightarrow n-k-2$,
coincide exactly with the basis employed in renormalizing
the parton distribution functions $e(x, \mu^{2})$ \cite{KN97}
and $h_{L}(x, \mu^{2})$ \cite{KT95}, 
respectively.\footnote{For the case of
the parton distribution $h_{L}(x, \mu^{2})$, the relevant
quark-antiquark-gluon  operator has an additional
$i\gamma_{5}$ inbetween the quark fields.
The evolution of the corresponding operator is not affected by the
insertion of $i\gamma_{5}$ \cite{BBKT96}.}
Matching with these results, we obtain
\begin{eqnarray}
\left(\Gamma_{n}^{T+}\right)_{k,l}&=& - Y_{k+2, l+2}
\;\;\;\;\;\;\;\;\;\; (k,l = 0, 1, \ldots, \kappa_{n}^{+}),
\label{eq:3ADTp}\\
\left(\Gamma_{n}^{T-}\right)_{k,l}&=& - X_{k+2, l+2}
\;\;\;\;\;\;\;\;\;\; (k,l = 0, 1, \ldots, \kappa_{n}^{-}),
\label{eq:3ADTm}
\end{eqnarray}
where $Y_{i,j}$ and $X_{i,j}$ are the mixing matrices
in the notation of \cite{KN97} and \cite{KT95}, i.e.\  
are given by Eqs.~(3.12)--(3.16)
of \cite{KN97}
and Eqs.~(3.14)--(3.16) of \cite{KT95}, respectively.
By solving (\ref{eq:3RG}), we obtain 
\begin{equation}
S_{n;k}^{\pm}(0; Q^{2})
= \sum_{l=0}^{\kappa_{n}^{\pm}} 
\left( L^{\Gamma^{T\pm}_{n}/b} \right)_{k,l}
S_{n;l}^{\pm} (0; \mu^{2}),
\label{eq:3scTh}
\end{equation}
and the matrix elements of $S_{n;k}^{\pm}$
are related to $\omega_{[k, n-k-2]}^{T}$
and $\omega_{\{k, n-k-2\}}^{T}$ of (\ref{eq:3oma}) and (\ref{eq:3oms})
as (see (\ref{eq:3omcon}) and (\ref{eq:3Snk}))
\begin{eqnarray}
\left(f_{3 \rho}^{T}\omega^{T}_{[k, n-k-2]}\right)(\mu^{2})
&=&\frac{N_{n}^{T} (-1)^{n-k+1}}{180 k!(n-k-2)!}
\langle 0| S_{n;k}^{\mp}(0; \mu^{2}) |\rho^{-}(P, \lambda)
\rangle,
\label{eq:3omegao}\\
\left(f_{3 \rho}^{T}\omega^{T}_{\{k, n-k-2\}}\right)(\mu^{2})
&=&\frac{N_{n}^{T} (-1)^{n-k}}{180 k!(n-k-2)!}
\langle 0| S_{n;k}^{\pm}(0; \mu^{2}) |\rho^{-}(P, \lambda)
\rangle.
\label{eq:3omegae}
\end{eqnarray}
Here the upper (lower) superscript should be understood
for $n=2, 4, 6, \ldots$ ($n=3, 5, 7, \ldots$)
on the right-hand side.
The results (\ref{eq:3ADTp})--(\ref{eq:3omegae}),
combined with (\ref{eq:3ceh}), (\ref{eq:3cee}),
(\ref{eq:3hng}), (\ref{eq:3hng2}) and (\ref{eq:3scalep0})
give the $\mu^{2}$-dependence of the three-particle
contributions $\httg(u, \mu^{2})$ and $\hsg(u, \mu^{2})$
(recall that $\zeta_{3\rho}^{T}$ in (\ref{eq:3hng})
and (\ref{eq:3hng2}) is given by (\ref{eq:3pm})).
The evolution of the three-particle twist~3 distribution amplitude
${\cal T}(\underline{\alpha})$ of (\ref{eq:3cet}) is thus 
specified completely. 
For the first few moments we obtain 
from (\ref{eq:3ADTp})--(\ref{eq:3omegae}):\footnote{It is worth
noting that the evolution of 
$f_{3\rho}^{T}\omega^{T}_{\{k,n-k-2\}}$
coincides exactly with that of the coefficients, appearing
in the conformal expansion of the 
twist~3 three-particle distribution amplitude of the pion \cite{BF90}.}
\begin{eqnarray}
f_{3 \rho}^{T}(Q^{2}) &=& L^{(23C_{F} + 6C_{G})/6b}f_{3\rho}^{T}(\mu^{2}),
\label{eq:3exam1}\\
\left(f_{3 \rho}^{T}\omega^T_{[2,0]}\right)
(Q^{2}) &=& L^{(106C_{F}+75C_{G})/30b}
\left(f_{3\rho}^{T}\omega^T_{[2,0]}\right)(\mu^{2}),
\label{eq:3exam2}
\end{eqnarray}
\begin{equation}
\left(
\begin{array}{c}
3f_{3\rho}^{T}\omega^{T}_{[0,3]}\\
f_{3\rho}^{T}\omega^{T}_{[2,1]}
\end{array}
\right)^{Q^{2}} =  L^{\Gamma^{T+}_{5}/b}
\left(
\begin{array}{c}
3f_{3\rho}^{T}\omega^{T}_{[0,3]}\\
f_{3\rho}^{T}\omega^{T}_{[2,1]}
\end{array}
\right)^{\mu^{2}};
\;\;\;\;\;
\Gamma_{5}^{T+} = \left(
\begin{array}{cc}
\frac{41}{10}C_{F} + \frac{15}{6}C_{G}&
\frac{29}{15}C_{F}-\frac{11}{6}C_{G} \\
\frac{13}{15}C_{F} - \frac{13}{12}C_{G}&
\frac{47}{10}C_{F} + \frac{29}{12}C_{G}
\end{array}
\right),
\label{eq:3exam3}
\end{equation}
corresponding to $n=3, 4, 5$.
Here $C_{G}=N_{c}$, and 
$\omega_{[0,1]}^{T} = 28/3$
is substituted in (\ref{eq:3exam1})
(see (\ref{eq:3exa1})).
Substitution of (\ref{eq:3exam1})--(\ref{eq:3exam3}) in (\ref{eq:3exa1})
determines 
the scale dependence  of $H_{n}^{g}$ ($n=3,4,5$).

The results in (\ref{eq:3ADTp})--(\ref{eq:3exam3})
illustrate a complicated mixing pattern
characteristic for the higher twist operators. In particular,
$H_{n}^{g}(Q^{2})$ for $n\ge 5$ ($h_{n}^{g}(Q^{2})$ for $n\ge 4$)
are not directly related to $H_{n}^{g}(\mu^{2})$
($h_{n}^{g}(\mu^{2})$), in contrast to (\ref{eq:3scalep})
for the twist~2 operator.

There exist, however,  two important limits, 
$N_{c}\rightarrow \infty$ and $n\rightarrow \infty$,
where the three-particle coefficients
$H_{n}^{g}$ and $h_{n}^{g}$ 
obey a simple evolution equation.
The mathematical reason for this simplification is the same 
as for the similar simplification observed for
the nucleon parton distributions
$h_{L}(x, \mu^{2})$ and $e(x, \mu^{2})$ in \cite{BBKT96,KN97,BM97}.
To show this,
it is convenient to express $H_{n}^{g}$ and $h_{n}^{g}$
directly by matrix elements of $S_{n;k}^{\pm}$, by
substituting (\ref{eq:3omegao}) and (\ref{eq:3omegae})
into (\ref{eq:3hng}) and (\ref{eq:3hng2}). We find
\begin{eqnarray}
\left(f_{\rho}^{T}H_{n}^{g}\right)(\mu^{2})
&=& - \frac{N_{n}^{T}}{m_{\rho}(n+1)!}
\sum_{k=0}^{\kappa_{n}^{-}}\left(1 - \frac{2(k+2)}{n+2}\right)
\langle 0|S_{n;k}^{-}(0; \mu^{2})
|\rho^{-}(P, \lambda)\rangle,
\label{eq:3Hnge}\\
\left(f_{\rho}^{T} h_{n}^{g}\right)(\mu^{2})
&=& - \frac{N_{n}^{T}}{m_{\rho}(n+1)!}
\sum_{k=0}^{\kappa_{n}^{+}}
\left(1 - \frac{1}{2}\delta_{k, \kappa_{n}^{+}}\right) 
\langle 0|S_{n;k}^{+}(0; \mu^{2})
|\rho^{-}(P, \lambda)\rangle,
\label{eq:3hnge}
\end{eqnarray}
for $n=2, 4, 6, \ldots$, and
\begin{eqnarray}
\left(f_{\rho}^{T}H_{n}^{g}\right)(\mu^{2})&=& 
- \frac{N_{n}^{T}}{m_{\rho}(n+1)!}
\sum_{k=0}^{\kappa_{n}^{+}}
\langle 0|S_{n;k}^{+}(0; \mu^{2})
|\rho^{-}(P, \lambda)\rangle,
\label{eq:3Hngo}\\
\left(f_{\rho}^{T}h_{n}^{g}\right)(\mu^{2})&=& 
- \frac{N_{n}^{T}}{m_{\rho}(n+1)!}
\sum_{k=0}^{\kappa_{n}^{-}}\left(1 - \frac{2(k+2)}{n+2}\right)
\langle 0|S_{n;k}^{-}(0; \mu^{2})
|\rho^{-}(P, \lambda)\rangle,
\label{eq:3hngo}
\end{eqnarray}
for $n=3, 5, 7, \ldots$
Setting $\mu^{2}=Q^{2}$ in (\ref{eq:3Hnge})--(\ref{eq:3hngo})
and substituting (\ref{eq:3scTh}) into them,
we would reproduce the complicated mixing 
discussed above. However, in the large $N_{c}$ limit,
that is, neglecting terms $O(1/N_{c}^{2})$ in the anomalous dimension
matrices $\Gamma_{n}^{T\pm}$ of (\ref{eq:3ADTp}) and (\ref{eq:3ADTm}),
the following exact relations have been derived \cite{BBKT96,KN97}:
\begin{eqnarray}
\sum_{k=0}^{\kappa_{n}^{+}}\left(\Gamma_{n}^{T+}\right)_{k,l}
&=& \gamma_{n}^{T+} 
\;\;\;\;\;\;\;\;\;\; (n=3, 5, 7, \ldots),
\label{eq:3LNpo}\\
\sum_{k=0}^{\kappa_{n}^{+}}
\left(1 - \frac{1}{2}\delta_{k, \kappa_{n}^{+}}
\right)\left(\Gamma_{n}^{T+}\right)_{k,l}
&=& \left(1-\frac{1}{2} \delta_{\kappa_{n}^{+},l}\right)\gamma_{n}^{T+} 
\;\;\;\;\;\;\;\;\;\; (n=2, 4, 6, \ldots),
\label{eq:3LNpe}
\end{eqnarray}
and
\begin{equation}
\sum_{k=0}^{\kappa_{n}^{-}}\left(1-\frac{2(k+2)}{n+2}\right)
\left(\Gamma_{n}^{T-}\right)_{k,l}
= \left(1 - \frac{2(l+2)}{n+2} \right) \gamma_{n}^{T-}
\;\;\;\;\;\;\;\;\;\;(n=3,4,5,6, \ldots),
\label{eq:3LNm}
\end{equation}
where
\begin{eqnarray}
\gamma_{n}^{T+} &=& 2N_{c} \left(\psi(n+1)+\gamma_{E} - \frac{1}{4}
- \frac{1}{2(n+1)}\right),
\label{eq:3gammap}\\
\gamma_{n}^{T-} &=& 2N_{c} \left(\psi(n+1)+\gamma_{E} - \frac{1}{4}
+ \frac{3}{2(n+1)}\right).
\label{eq:3gammam}
\end{eqnarray}
As a consequence of these relations, we obtain
\begin{eqnarray}
\left(f_{\rho}^{T}H_{n}^{g} \right)(Q^{2})&=&
L^{\gamma^{T\mp}_{n}/b}\left(f_{\rho}^{T}H_{n}^{g} \right)(\mu^{2}),
\label{eq:3LNHn}\\
\left(f_{\rho}^{T}h_{n}^{g} \right)(Q^{2})&=&
L^{\gamma^{T\pm}_{n}/b}\left(f_{\rho}^{T}h_{n}^{g} \right)(\mu^{2}),
\label{eq:3LNhn}
\end{eqnarray}
where the upper (lower) superscript should be understood for 
$n=2, 4, 6, \ldots$ ($n=3, 5, 7, \ldots$)
on the right-hand side.
Therefore, in the large $N_{c}$ limit,
$H_{n}^{g}$ and $h_{n}^{g}$ obey simple DGLAP-type evolution equations
similarly to the twist~2 case (\ref{eq:3scalep});
they are governed by the anomalous dimensions given in analytic form in
(\ref{eq:3gammap}) and (\ref{eq:3gammam}).
We note that these anomalous dimensions correspond to the lowest
eigenvalues of the mixing matrices
$\Gamma_{n}^{T\pm}$ \cite{BBKT96,KN97}.

The phenomenon leading to (\ref{eq:3LNHn}) and (\ref{eq:3LNhn})
can be stated as  decoupling of the three-particle operators,
which correspond to the higher eigenvalues 
of the mixing matrix, from the RG equation.
The same decoupling is observed 
at large $n$ for arbitrary values of $N_{c}$ \cite{BBKT96,BM97}.
In this case, we obtain (\ref{eq:3LNpo})--(\ref{eq:3LNm})
with the anomalous dimensions (\ref{eq:3gammap}) and
(\ref{eq:3gammam}) shifted by
\begin{equation}
\gamma_{n}^{T\pm} \rightarrow 
\gamma_{n}^{T\pm} +(4C_{F} - 2N_{c})
\left(\ln n + \gamma_{E} - \frac{3}{4}\right).
\label{eq:3modgam}
\end{equation}
With this modification of the anomalous dimensions,
the results (\ref{eq:3LNHn}) and (\ref{eq:3LNhn})
are valid to $O\left((1/N_{c}^{2})\ln (n)/n\right)$ accuracy.

These simplifications provide 
useful insight both into the model-building
of the distribution amplitudes, and a convenient
description of their scale-dependence:
In the large $N_{c}$-limit,
each conformal partial wave of the three-particle
contributions is described by a single nonperturbative
parameter, as demonstrated in (\ref{eq:3LNHn}) and (\ref{eq:3LNhn}).
This is remarkable because in general each conformal spin
involves many independent nonperturbative matrix elements
(see (\ref{eq:3Hnge})--(\ref{eq:3hngo})).
This point can be made stronger
with the full account for effects subleading in $N_{c}$
but for large conformal spin $j$, as shown in (\ref{eq:3modgam}).
Furthermore, (\ref{eq:3modgam}) proves the conjecture
made in \cite{BB88,BF90}
that the lowest anomalous dimension of the twist~3 three-particle
operators is increasing as $\sim \ln j$, similarly to the twist~2
case (\ref{eq:3anomdp}).
This ensures convergence of the series in the Appell
polynomials at least for high energy scales.
Combined with the fact that the distribution
amplitudes can be resolved order by order in the conformal spin,
the truncation of the conformal expansion at some low order
provides a useful and consistent approximation of the 
full amplitude.

To summarize, we have worked out  the scale dependence 
of chiral-odd twist~3 distribution amplitudes
$\htt (u, \mu^{2})$ and $\hs(u, \mu^{2})$
in the leading logarithmic approximation. 
In the two limits,
$N_{c} \rightarrow \infty$ and $n \rightarrow \infty$,
the evolution of $\htt (u, \mu^{2})$ and $\hs(u, \mu^{2})$
is drastically simplified and reduces to a DGLAP-type equation.
The discussion in this section
completes the results for the chiral-odd distribution amplitudes
up to twist~3, which can be predicted based on the QCD constraints
from equations of motion, conformal invariance, and 
renormalization group invariance.

\section{Chiral-even Distribution Amplitudes}
\setcounter{equation}{0}

The analysis in the previous section can directly be extended to 
the chiral-even distribution amplitudes.   In Sec.~4.1 we  
derive
the constraint equations for 
$g_\perp^{(v)}(u)$ and $g_\perp^{(a)}(u)$  
imposed by the QCD equations of motion, and 
identify the contribution to $g_\perp^{(v)}(u)$ and $g_\perp^{(a)}(u)$ from
the twist~2 distribution amplitudes $\phi_\parallel(u)$
and $\phi_\perp(u)$
and the three-particle distribution amplitudes
${\cal V}(\underline{\alpha})$ and 
${\cal A}(\underline{\alpha})$.
In Sec.~4.2,
we study the conformal expansion for
$\phi_\parallel(u)$, $g_\perp^{(v)}(u)$ and $g_\perp^{(a)}(u)$.
In Sec.~4.3, we  work out the renormalization of 
$g_\perp^{(v)}(u)$ and $g_\perp^{(a)}(u)$,
utilizing the fact that the conformal symmetry is preserved
at one-loop level. 
Our presentation in this section will be brief, since the methods
and the results are in parallel with those in Sec.~3.

\subsection{Equations of motion}
To derive the constraint relations among the chiral-even distribution 
amplitudes
we again use operator identities
(to  twist~3 accuracy) for the nonlocal operators in (\ref{eq:vda})
and (\ref{eq:avda})\,\cite{BB88}:
\beq
\ubar (x)\gamma_\mu[x,-x] d(-x) &=&\int_0^1\, dt\, 
{ \partial \over \partial x_\mu}\,\ubar(tx)\xslash [tx,-tx] 
d(-tx)\nonumber\\
&&{}- \int_0^1\,dt\,t\,\int_{-t}^t\,dv\,\ubar(tx)
[tx,vx]g\Gtilde_{\mu\nu}(vx)
x^\nu \xslash \gamma_5 
[vx,-tx]d(-tx)\nonumber\\
&&{}-i 
\int_0^1\,dt\,\int_{-t}^t\,dv\,v\, \ubar(tx)[tx,vx]gG_{\mu\nu}(vx)x^\nu
[vx,-tx]\xslash d(-tx)\nonumber\\
&&{}- i\epsilon_{\mu\nu\alpha\beta}\int_0^1\,dt\,t\,x^\nu\partial^\alpha
\left[ \ubar(tx)\gamma^\beta\gamma_5 
[tx,-tx]d(-tx)\right]\nonumber\\
&&{}+ (m_u -m_d) x^\nu \int_0^1\,dt\,t\,\ubar(tx)\sigma_{\nu\mu}
[tx,-tx]d(-tx),
\label{eq4.1}
\eeq
and
\beq
\ubar (x)\gamma_\mu\gamma_5[x,-x] d(-x) &=&\int_0^1\, dt\, 
{ \partial \over \partial x_\mu}\,\ubar(tx)\xslash \gamma_5[tx,-tx] 
d(-tx)\nonumber\\
& &{}- \int_0^1\,dt\,t\,\int_{-t}^t\,dv\,\ubar(tx)
[tx,vx]g\Gtilde_{\mu\nu}(vx)
x^\nu \xslash  
[vx,-tx]d(-tx)\nonumber\\
&&{}-i 
\int_0^1\,dt\,\int_{-t}^t\,dv\,v\, \ubar(tx)[tx,vx]gG_{\mu\nu}(vx)x^\nu
[vx,-tx]\xslash\gamma_5 d(-tx)\nonumber\\
&&{}- i\epsilon_{\mu\nu\alpha\beta}\int_0^1\,dt\,t\,x^\nu\partial^\alpha
\left[ \ubar(tx)\gamma^\beta 
[tx,-tx]d(-tx)\right]\nonumber\\
&&{}+ (m_u +m_d) x^\nu \int_0^1\,dt\,t\,\ubar(tx)\sigma_{\nu\mu}\gamma_5
[tx,-tx]d(-tx),
\label{eq4.2}
\eeq
where $\partial_\alpha$ 
is the total derivative defined in (\ref{eq:3tdrv}), and
the terms proportional to quark masses
originate from the use of QCD equation of motion.
By sandwiching these equations between the vacuum and the $\rho$
meson state, and taking the light-cone limit $x\to z$, one obtains
the following relations among the distribution amplitudes:
\beq
\lefteqn{\int_0^1du\,e^{i\xi pz}g_\perp^{(v)}(u)\ =\
\int^1_0dt\int_0^1du\,e^{it\xi pz}\phi_\parallel(u)
-\zeta_{3\rho}^A(pz)^2\int\,t^2dt\,\int_{-1}^1dv\,{\cal
A}(v,tpz)}\makebox[1cm]{\ }\nonumber\\
& &{} -\zeta_{3\rho}^V(pz)^2\int_0^1dt\,t^2\int_{-1}^1dv\,v\,{\cal V}(v,tpz)
-{1\over 2}(pz)^2
\left( 1 - \widetilde{\delta}_+
\right)
\int_0^1dt\,t^2\,\int_0^1du\,e^{it\xi pz}g_\perp^{(a)}(u)\nonumber\\
& &{} -i\widetilde{\delta}_-(pz)\int_0^1dt\,t
\int_0^1du\,e^{it\xi pz}\phi_\perp(u),
\label{eq4.3}
\eeq
and
\beq
\lefteqn{{1\over 2}\left(1-\widetilde{\delta}_+
\right)
\int_0^1du\,e^{i\xi pz}g_\perp^{(a)}(u)\ =\ 
\int_0^1dt\,t\int_0^1du\,e^{it\xi pz}g_\perp^{(v)}(u)}\makebox[2cm]{\
}\nonumber\\
& &{} +i\zeta_{3\rho}^V(pz)\int_0^1dt\,t^2\int_{-1}^1dv{\cal V}(v,tpz)
+i\zeta_{3\rho}^A(pz)\int_0^1dt\,t^2\int_{-1}^1dv\,v{\cal A}(v,tpz)\nonumber\\
& &{} -\widetilde{\delta}_+
\int_0^1dt\,t\int_0^1du\,e^{it\xi pz}\phi_\perp(u),
\label{eq4.4}
\eeq
where ${\cal V}(v, tpz)$ and ${\cal A}(v, tpz)$ are defined in 
(\ref{eq:short})
and we introduced the notations
\begin{equation}
\widetilde{\delta}_\pm \equiv {f_\rho^{T^2}\over f_\rho^2}\delta_\pm
={f_\rho^{T}\over f_\rho}{m_u \pm m_d \over m_\rho},\qquad
\zeta_{3\rho}^{V,A} = \frac{f^{V,A}_{3\rho}}{f_\rho m_\rho}.
\label{eq4.5}
\end{equation}
In order to solve these equations, 
we expand them in powers of $pz$ and transform them into
relations among the moments of distribution amplitudes:
\begin{eqnarray}
(n+1)M_n^{(v)} & = & M_n^\parallel +
\frac{n(n-1)}{2}\left(1-\widetilde{\delta}_+ \right)M^{(a)}_{n-2} +
\zeta_{3\rho}^{A} n(n-1) \int\limits_{-1}^1\!\! dv\, 
{\cal A}_{n-2}(v)\nonumber\\
& & {} +
\zeta_{3\rho}^{V} n(n-1) \int\limits_{-1}^1\!\! dv \,v\,
{\cal V}_{n-2}(v) -\widetilde{\delta}_-
nM_{n-1}^\perp,
\label{eq4.6}
\eeq
and
\beq
\frac{1}{2}(n+2)\left(1-\widetilde{\delta}_+ \right)M_n^{(a)} 
& = &  M_n^{(v)} + \zeta_{3\rho}^{V} n
\int\limits_{-1}^1\!\! dv\, 
{\cal V}_{n-1}(v) + \zeta_{3\rho}^{A} n
\int\limits_{-1}^1\!\! dv\, v\,{\cal A}_{n-1}(v)
- \widetilde{\delta}_+ M_n^\perp,\nonumber\\
\label{eq4.7}
\end{eqnarray}
where ${\cal V}_n(v)$ and ${\cal A}_n(v)$ are defined similarly to
(\ref{eq:3mom2}) from ${\cal V}(\underline{\alpha})$
and ${\cal A}(\underline{\alpha})$, 
and we introduced the shorthand notations 
$M_n^{(a),(v)}\equiv \int_0^1\!du\,\xi^n\,g_\perp^{(a),(v)}$.
{}From these equations, one obtains a recurrence  
relation for $M_n^{(a)}$ as
\beq
\lefteqn{\left(1-\widetilde{\delta}_+ \right)
\left( (n+2)(n+1)M_n^{(a)} - n(n-1)M_{n-2}^{(a)} \right)\ =}\makebox[2cm]{\ }
\nonumber\\
& = & 2M_n^\parallel  
+ 2 \zeta_{3\rho}^V \int_{-1}^1dv\,\left[
n(n+1){\cal V}_{n-1}(v) + n(n-1)v {\cal V}_{n-2}(v)\right]\nonumber\\
& &{}+ 2 \zeta_{3\rho}^A \int_{-1}^1dv\,\left[
n(n+1)v{\cal A}_{n-1}(v) + n(n-1) {\cal A}_{n-2}(v)\right]\nonumber\\
& &{} -2(n+1)\widetilde{\delta}_+
M_n^\perp
-2n\widetilde{\delta}_-
M_{n-1}^\perp.
\label{eq4.8}
\eeq
This equation is similar to (\ref{eq:3rec2})
for $h_\parallel^{(s)}(u)$, and can
be cast into the form of a differential equation as
\begin{equation}
\left( 1 - \widetilde{\delta}_+ \right)
 u\bar{u}(g^{(a)}_{\perp})''(u) = -\Psi(u),
\label{eq4.9}
\end{equation}
where
\beq
\Psi(u)  &=&  2\phi_\parallel(u) + \widetilde{\delta}_+
\xi\phi'_\perp(u)+
\widetilde{\delta}_-
\phi'_\perp(u)\nonumber\\
& & {}+2\zeta_{3\rho}^V\,\frac{d}{du}\,\int\limits_0^u\!\!d\alpha_d
\int\limits_0^{\bar u}\!\! d\alpha_u
\,\frac{1}{1-\alpha_d-\alpha_u}\left(\alpha_d\,\frac{d}{d\alpha_d} +
\alpha_u\,\frac{d}{d\alpha_u}\right)
{\cal V}(\underline{\alpha})\nonumber\\
&&{} +
2\zeta_{3\rho}^A\,\frac{d}{du}\,\int\limits_0^u\!\!d\alpha_d
\int\limits_0^{\bar u}\!\! d\alpha_u
\,\frac{1}{1-\alpha_d-\alpha_u}\left(\alpha_d\,\frac{d}{d\alpha_d} -
\alpha_u\,\frac{d}{d\alpha_u}\right) {\cal A}(\underline{\alpha}).
\label{eq4.10}
\end{eqnarray}
{}From this equation, one immediately obtains
the solution for $g_\perp^{(a)}(u)$ as
\begin{eqnarray}
\left( 1 - \widetilde{\delta}_+ \right)
g^{(a)}_\perp(u) & = & \bar u \int\limits_0^u\!\! dv\, \frac{1}{\bar
v}\,\Psi(v)
+ u \int\limits_u^1\!\! dv\, \frac{1}{v}\,\Psi(v).
\label{eq4.11}
\eeq
Combining this result with (\ref{eq4.7}), one obtains
the solution for $g_\perp^{(v)}(u)$ as
\begin{eqnarray}
g_\perp^{(v)}(u) & = &
\frac{1}{4}
\left[
\int\limits_0^u\!\! dv\,\frac{1}{\bar v}\,\Psi(v) + 
\int\limits_u^1\!\!
dv\,\frac{1}{v}\,\Psi(v)\right]
+\widetilde{\delta}_+
\phi_\perp(u)\nonumber\\
&&{} +
\zeta_{3\rho}^{A}\int\limits_0^u \!\! d\alpha_d\!\! \int\limits_0^{\bar
u}\!\! d\alpha_u\, \frac{1}{1-\alpha_d-\alpha_u}\,
\left( \frac{d}{d\alpha_d} +
\frac{d}{d\alpha_u} \right){\cal A}(\underline{\alpha})\nonumber\\ 
&&{}+\zeta_{3\rho}^{V}\,\frac{d}{du}\int\limits_0^u \!\! d\alpha_d\!\! 
\int\limits_0^{\bar
u}\!\! d\alpha_u\, \frac{{\cal V}(\underline{\alpha})}
{1-\alpha_d-\alpha_u}.
\label{eq4.12}
\end{eqnarray}
Eqs.~(\ref{eq4.11}) and (\ref{eq4.12}) again allow the
decomposition of $g_\perp^{(v)}(u)$ and $g_\perp^{(a)}(u)$ into 
several terms according to the various source terms:
\beq
g_\perp^{(v)}(u) &=& g_\perp^{(v)WW}(u) + g_\perp^{(v)g}(u)
+ g_\perp^{(v)m}(u),
\label{eq4.13}\\
g_\perp^{(a)}(u) &=& g_\perp^{(a)WW}(u) + g_\perp^{(a)g}(u)
+ g_\perp^{(a)m}(u),
\label{eq4.14}
\eeq
where $g_\perp^{(v)WW}(u)$ and $g_\perp^{(a)WW}(u)$ 
denotes the contribution
from the twist~2 distribution amplitudes (Wandzura-Wilczek part), 
$g_\perp^{(v)g}(u)$ and $g_\perp^{(a)g}(u)$ 
are the contribution from the three-particle distribution amplitudes
${\cal V}$ and ${\cal A}$. 
In particular, we get 
\beq
g_\perp^{(v)WW}(u)&=&{1\over 2}\left[
\int_0^udv\,{1\over \bar{v}}\phi_\parallel(v)
+ \int_u^1dv\,{1\over v}\phi_\parallel(v)
\right],
\label{eq4.15}\\
g_\perp^{(a)WW}(u)&=&2\bar{u}
\int_0^udv\,{1\over \bar{v}}\phi_\parallel(v)
+ 2u\int_u^1dv\,{1\over v}\phi_\parallel(v).
\label{eq4.16}
\eeq

\subsection{Conformal expansion}

The conformal expansion for the chiral-even distribution
amplitude can be performed similarly to 
the case for the chiral-odd ones in Sec.~3.2.  
In the following, we 
restrict ourselves to the massless case.

For completeness, we start with the expansion for the 
twist~2 distribution amplitude $\phi_\parallel(u)$.  It reads
\beq
\phi_\parallel(u) = 6u\bar{u}\sum_{n=0}^\infty a_n^\parallel
C^{3/2}_n(\xi),
\label{eq4.17}
\eeq
where $C^{3/2}_n(\xi)$ is the Gegenbauer polynomial, and each
term corresponds to the conformal spin $n+2$.
We note $a_0^\parallel=1$ due to the normalization condition 
(\ref{eq:norm}).
Because of the G-parity invariance of 
the $\rho$ meson
distribution amplitude (likewise for
other mesons such as  $\omega$, $\phi$), it follows
that $a_n^\parallel=0$ for $n=1,3,5,\ldots$ in (\ref{eq4.17}).
In Sec.~5, however, we shall treat an
application to the $K^*$ distribution amplitude
with explicit 
SU(3) breaking due to the quark masses. 
With this in mind, 
we keep all terms $a_n^\parallel$ in (\ref{eq4.17}).  
For the same reason, 
we shall work out the conformal expansion for
$g_\perp^{(v)}$, $g_\perp^{(a)}$,
${\cal V}(\underline{\alpha})$ and
${\cal A}(\underline{\alpha})$ by
keeping both G-parity invariant and G-parity violating
terms in the following.

To carry out the conformal expansion for
$g_\perp^{(v)}(u)$ and $g_\perp^{(a)}(u)$, we again introduce
a set of auxiliary amplitudes $g^{\uparrow\downarrow}(u)$ and
$g^{\downarrow\uparrow}(u)$ \cite{ABS} defined by
\begin{eqnarray}
\langle 0| \overline{u}(z) \gamma.\gamma_\mu^\perp \gamma_{\ast}[z,-z] 
d(-z)|\rho^-(P,\lambda)\rangle
&=& -f_{\rho}m_{\rho}
e_{\perp \mu}^{(\lambda)}
\int_{0}^{1} du e^{i\xi pz} g^{\uparrow \downarrow} (u), 
\label{eq4.18} \\
\langle 0| \overline{u}(z) \gamma_{\ast} \gamma_\mu^\perp \gamma.[z,-z] 
d(-z)|\rho^-(P,\lambda)\rangle
&=& -f_{\rho}m_{\rho}
e_{\perp \mu}^{(\lambda)}
\int_{0}^{1} du e^{i\xi pz} g^{\downarrow \uparrow} (u), 
\label{eq4.19}
\end{eqnarray}
which are related to 
$g_\perp^{(v)}(u)$ and $g_\perp^{(a)}(u)$ as 
\beq
g^{\uparrow \downarrow}(u)&=&g^{(v)}_\perp (u) + {1\over 4}
{d\over du}g_\perp^{(a)}(u),
\label{eq4.20}\\
g^{\downarrow \uparrow}(u)&=&g^{(v)}_\perp (u) - {1\over 4}
{d\over du}g_\perp^{(a)}(u).
\label{eq4.21}
\eeq
The conformal expansion for 
$g^{\uparrow\downarrow}(u)$ and
$g^{\downarrow\uparrow}(u)$ is given by
\begin{eqnarray}
g^{\uparrow \downarrow} (u) &=& 2\overline{u} \sum_{n=0}^{\infty}
g_{n}^{\uparrow \downarrow} P_{n}^{(1,0)}(\xi),
\label{eq4.22}\\
g^{\downarrow \uparrow} (u) &=& 2u\sum_{n=0}^{\infty}
g_{n}^{\downarrow \uparrow} P_{n}^{(0,1)}(\xi).
\label{eq4.23}
\end{eqnarray}
Substituting these expansions in (\ref{eq4.20}) and (\ref{eq4.21}),
one obtains
\begin{eqnarray}
g_\perp^{(v)}(u) &=& \sum_{n=0, 2, 4, \ldots} \left(G_{n} - G_{n-1}\right)
C_{n}^{1/2}(\xi)
+ \sum_{n=1, 3, 5, \ldots} \left(g_{n} - 
g_{n-1}\right)
C_{n}^{1/2}(\xi),
\label{eq4.24}\\
g_\perp^{(a)}(u) &=& 8u\overline{u} \left(\sum_{n=0,2,4, \ldots} 
\frac{G_{n} - G_{n+1}}{(n+1)(n+2)}
C_{n}^{3/2}(\xi)
+\sum_{n=1,3,5, \ldots} 
\frac{g_{n} - g_{n+1}}{(n+1)(n+2)}
C_{n}^{3/2}(\xi)
\right),
\label{eq4.25}
\end{eqnarray}
where 
$G_n$ and $g_n$ represent, respectively, 
G-parity invariant and G-parity
violating components in the expansion defined by
\begin{eqnarray}
G_{n} &\equiv& \frac{ g_{n}^{\uparrow \downarrow} +
(-1)^{n}g_{n}^{\downarrow\uparrow}}{2},
\nonumber\\
g_{n} &\equiv& \frac{ g_{n}^{\uparrow \downarrow} -
(-1)^{n}g_{n}^{\downarrow\uparrow}}{2},
\label{eq4.27}
\end{eqnarray}
for $n= 0, 1, 2, \ldots$, and $G_{-1}=g_{-1}=0$ is implied.
Note the difference between 
$\{G_n, g_n\}$ and $\{H_n, h_n\}$
(see (\ref{eq:3coeff})) 
owing to the chiral-even or -odd nature of the distribution amplitudes.

The conformal expansion for the three-particle distribution amplitudes
${\cal V}(\underline{\alpha})$ and
${\cal A}(\underline{\alpha})$ can be written down similarly to
(\ref{eq:3cet}):
\beq
{\cal V}(\alpha_d,\alpha_u,1-\alpha_d-\alpha_u)
&=&360 \alpha_d\alpha_u(1-\alpha_d-\alpha_u)^2
\sum_{k,l=0}^\infty \omega^V_{k,l}J_{k,l}(\alpha_d, \alpha_u),
\nonumber\\
{\cal A}(\alpha_d,\alpha_u,1-\alpha_d-\alpha_u)
&=&360 \alpha_d\alpha_u(1-\alpha_d-\alpha_u)^2
\sum_{k,l=0}^\infty \omega^A_{k,l}J_{k,l}(\alpha_d, \alpha_u).
\label{eq4.28}
\eeq
The G-parity invariance of the three-particle distribution 
amplitudes leads to
$\omega^V_{k,l} =- \omega^V_{l,k}$ and 
$\omega^A_{k,l} = \omega^A_{l,k}$.  
As was stated at the begining of
this subsection, we shall not assume this symmetry
in the following.
We also note 
the normalization condition
in (\ref{eq:normalize}) gives $\omega^V_{[0,1]}=28/3$ 
and $\omega^A_{0,0}=1$.
The conformal spin for each term in the above expansion is
equal to $j=k+l+7/2$,
and the preservation of the conformal invariance
at one-loop level prevents mixing among
the contribution with different $n=k+l$.

Our next task is to
identify the twist~2 (Wandzura-Wilczek) and the three-particle contibutions
to $G_n$ and $g_n$.
We decompose
\begin{equation}
G_{n} =G_{n}^{WW} + G_{n}^{g}; \;\;\;\;\;\;\;
g_{n} = g_{n}^{WW} + g_{n}^{g},
\label{eq4.29}
\end{equation}
and consider the Wandzura-Wilczek contribution first. 
Substituting (\ref{eq4.15}) and (\ref{eq4.16}) 
into (\ref{eq4.20}) and (\ref{eq4.21}) and
using the formulae (\ref{eq:jg4}), (\ref{eq:jd1}), (\ref{A1})
and (\ref{A2}), we obtain
the Wandzura-Wilczek contribution
for $g^{\uparrow\downarrow}(u)$ and 
$g^{\downarrow\uparrow}(u)$
as
\begin{eqnarray}
{g^{\uparrow \downarrow}}^{WW}(u) 
&=& 
\int_{u}^{1}dv\frac{\phi_{\parallel}(v)}{v}\nonumber\\
&=&2\bar{u}\sum_{n=0}^\infty
a_n^\parallel\left[ {3(n+2)\over 2(2n+3)}P_n^{(1,0)}(\xi)
-{3(n+1)\over 2(2n+3)}P_{n+1}^{(1,0)}(\xi) \right],
\label{eq4.30}\\
{g^{\downarrow \uparrow}}^{WW}(u) 
&=& 
\int_{0}^{u}dv\frac{\phi_{\parallel}(v)}{\overline{v}}\nonumber\\
&=&2u\sum_{n=0}^\infty
a_n^\parallel\left[ {3(n+2)\over 2(2n+3)}P_n^{(0,1)}(\xi)
+{3(n+1)\over 2(2n+3)}P_{n+1}^{(0,1)}(\xi) \right].
\label{eq4.31}
\end{eqnarray}
These give rise to $G_n^{WW}$ and $g_n^{WW}$
in (\ref{eq4.29}) as
\beq
G_n^{WW}= {3(n+2)\over 2(2n+3)}a_n^\parallel,
\qquad g_n^{WW}=
{-3n\over 2(2n+1)}a_{n-1}^\parallel,\qquad (n=0,2,4\ldots)
\nonumber\\
G_n^{WW}= {-3n\over 2(2n+1)}a_{n-1}^\parallel,
\qquad g_n^{WW}=
{3(n+2)\over 2(2n+3)}a_{n}^\parallel.\qquad (n=1,3,5\ldots)
\label{eq4.32}
\eeq
For even $n$, we
find that $a_n^\parallel$ which corresponds to the
conformal spin $n+2$ in the expansion for
the twist~2 distribution amplitude gives rise to $G^{WW}_n$ and $G^{WW}_{n+1}$
which corresponds to the conformal spin $n+3/2$ and
$n+5/2$, respectively.  Likewise for odd $n$, $a_n^\parallel$
gives rise to $g_n^{WW}$ and $g_{n+1}^{WW}$.  This is the same pattern
as observed for the chiral-odd distribution amplitudes and is
discussed in App.~\ref{app:K}.

{}From the solution for $g_\perp^{(v)}(u)$ and $g_\perp^{(a)}(u)$
in (\ref{eq4.11}) and (\ref{eq4.12})
we can identify the three-particle contribution
to the auxiliary amplitude as
\beq
g^{\uparrow\downarrow g}(u)&=&\zeta_{3\rho}^V \left\{
\int_u^1\,dv\,{1\over v}H(v) + M(u) \right\}
+
\zeta_{3\rho}^A \left\{
\int_u^1\,dv\,{1\over v}L(v) + N(u) \right\},
\label{eq4.33}\\
g^{\downarrow\uparrow g}(u)&=&\zeta_{3\rho}^V \left\{
\int^u_0\,dv\,{1\over \bar{v}}H(v) + M(u) \right\}
+ \zeta_{3\rho}^A \left\{
\int^u_0\,dv\,{1\over \bar{v}}L(v) + N(u) \right\},
\label{eq4.34}
\eeq
where the
functions $H(u)$, $L(u)$, $M(u)$, $N(u)$ are defined as
\beq
H(u) &=& \frac{d}{du}\int_{0}^{u}d\alpha_{d} \int_{0}^{\overline{u}}
d\alpha_{u} \frac{1}{\alpha_{g}}
\left( \alpha_{d}\frac{d}{d \alpha_{d}}
+ \alpha_{u}\frac{d}{d \alpha_{u}}
\right) {\cal V}(\underline{\alpha}),
\label{eq4.35}\\
L(u) &=& \frac{d}{du}\int_{0}^{u}d\alpha_{d} \int_{0}^{\overline{u}}
d\alpha_{u} \frac{1}{\alpha_{g}}
\left( \alpha_{d}\frac{d}{d \alpha_{d}}
- \alpha_{u}\frac{d}{d \alpha_{u}}
\right) {\cal A}(\underline{\alpha}),
\label{eq4.36}\\
M(u) &=& \frac{d}{du}\int_{0}^{u}d\alpha_{d} \int_{0}^{\overline{u}}
d\alpha_{u} \frac{1}{\alpha_{g}}
{\cal V}(\underline{\alpha}),
\label{eq4.37}\\
N(u) &=& \int_{0}^{u}d\alpha_{d} \int_{0}^{\overline{u}}
d\alpha_{u} \frac{1}{\alpha_{g}}
\left( \frac{d}{d \alpha_{d}}
+ \frac{d}{d \alpha_{u}}
\right) {\cal A}(\underline{\alpha}),
\label{eq4.38}
\eeq
with $\alpha_g=1-\alpha_d-\alpha_u$.  
The calculation of 
(\ref{eq4.33}) and (\ref{eq4.34})
can be carried out similarly to the chiral-odd distribution amplitudes
in Sec.~3,
using the explicit form for ${\cal V}(\underline{\alpha})$
and ${\cal A}(\underline{\alpha})$ in (\ref{eq4.28}).
First $H(u)$, $L(u)$, $M(u)$ and $N(u)$ in (\ref{eq4.35})--(\ref{eq4.38})
can be easily obtained by using (\ref{eq:apint})$-$(\ref{A4}).
The integration of $H(u)$ and $L(u)$ in (\ref{eq4.33}) and (\ref{eq4.34})
can be done by (\ref{eq:jd1}).  
Application of (\ref{A1}) and (\ref{A2}) to the results of those integrations
and $N(u)$, and the use of (\ref{eq:jrec12}) and (\ref{eq:jrec13})
for $M(u)$ yields the anticipated form for the result.
One eventually obtains 
\beq
g^{\uparrow\downarrow g}(u) &=&
90(1-\xi)\sum_{n=2}^\infty P_n^{(1,0)}(\xi)
\sum_{k=0}^{n-2}{k!(n-k-2)!(-1)^k \over (n+1)!(n+1)}\nonumber\\
& &\times\left\{n-(2k+2-n)\right\}\left(-\zeta_{3\rho}^V\omega_{k,n-k-2}^V
+\zeta_{3\rho}^A\omega^A_{k,n-k-2} \right),
\label{eq4.39}\\
g^{\downarrow\uparrow g}(u) &=&
90(1+\xi)\sum_{n=2}^\infty P_n^{(0,1)}(\xi)
\sum_{k=0}^{n-2}{k!(n-k-2)!(-1)^k \over (n+1)!(n+1)}\nonumber\\
& &\times\left\{n+(2k+2-n)\right\}\left(\zeta_{3\rho}^V\omega_{k,n-k-2}^V
+\zeta_{3\rho}^A\omega^A_{k,n-k-2} \right),
\label{eq4.40}
\eeq
where we kept the form $(2k+2-n)$ in $\{\ldots\}$ which is
anti-symmetric under the interchange $k \leftrightarrow n-k-2$.
{}From this equation we get $G_0^g=G_1^g=g_0^g=g_1^g=0$, and
\beq\renewcommand{\arraystretch}{2}
G_n^g = \left\{
\begin{array}{ll}
& \displaystyle{
90 \sum_{k=0}^{n-2}
{k!(n-k-2)!(-1)^{n+k}\over (n+1)!(n+1)}
\left\{ (2k+2-n)\zeta_{3\rho}^V\omega^V_{[k,n-k-2]} +
n\zeta_{3\rho}^A\omega^A_{\{k,n-k-2\}}\right\}
},\nonumber\\
 & \qquad\qquad (n=2,4,6\ldots)\nonumber\\
& \displaystyle{
90 \sum_{k=0}^{n-2}
{k!(n-k-2)!(-1)^{n+k}\over (n+1)!(n+1)}
\left\{ n\zeta_{3\rho}^V\omega^V_{[k,n-k-2]}+
(2k+2-n)\zeta_{3\rho}^A\omega^A_{\{k,n-k-2\}} \right\}
},\nonumber\\
& \qquad\qquad (n=3,5,7\ldots)
\end{array}\renewcommand{\arraystretch}{1}\right.\\[-20pt]
\label{eq4.41}
\eeq
\beq
g_n^g = \left\{\renewcommand{\arraystretch}{2}
\begin{array}{ll}
& \displaystyle{
90 \sum_{k=0}^{n-2}
{k!(n-k-2)!(-1)^{n+k+1}\over (n+1)!(n+1)}
\left\{ n\zeta_{3\rho}^V\omega^V_{\{k,n-k-2\}}+
(2k+2-n)\zeta_{3\rho}^A\omega^A_{[k,n-k-2]} \right\}
},\nonumber\\
& \qquad\qquad(n=2,4,6,\ldots)\nonumber\\
& \displaystyle{
90 \sum_{k=0}^{n-2}
{k!(n-k-2)!(-1)^{n+k+1}\over (n+1)!(n+1)}
\left\{ (2k+2-n)\zeta_{3\rho}^V\omega^V_{\{k,n-k-2\}} +
n\zeta_{3\rho}^A\omega^A_{[k,n-k-2]}\right\}
},\nonumber\\
& \qquad\qquad(n=3,5,7,\ldots)
\end{array}\renewcommand{\arraystretch}{1}\right.\\[-20pt]
\label{eq4.42}
\eeq
where we introduced the anti-symmetric and symmetric components
of $\omega_{k,l}^{V,A}$ defined by
\begin{eqnarray}
\omega^{V,A}_{[k,l]} 
&\equiv& \frac{\omega_{k,l}^{V,A} - \omega_{l,k}^{V,A}}{2},
\label{eq4.43}\\
\omega^{V,A}_{\{k,l\}} &\equiv& \frac{\omega_{k,l}^{V,A} 
+ \omega_{l,k}^{V,A}}{2}.
\label{eq4.44}
\end{eqnarray}
One finds here again that 
$G_n$ and $g_n$ which correspond to the conformal spin
$j=3/2+n$ receives contributions from the
coefficients 
$\omega^{V,A}_{k,l}$ with a fixed $k+l=n-2$
which have the same conformal spin $j=3/2+n$.
The first terms in (\ref{eq4.41}) and (\ref{eq4.42}) read
\beq
& & G_2^g= 10 \zeta^A_{3\rho}\omega_{00}^A=
10\zeta^A_{3\rho},\nonumber\\
& & G_3^g={15\over 8}\left(-3\zeta_{3\rho}^V\omega_{[0,1]}^V +
\zeta_{3\rho}^A\omega_{\{0,1\}}^A\right)
= {15\over 8}\left(-28\zeta_{3\rho}^V +
\zeta_{3\rho}^A\omega_{\{0,1\}}^A\right),\nonumber\\
& & G_4^g={3\over 5}\left(-2\zeta_{3\rho}^V\omega_{[0,2]}^V +
4\zeta_{3\rho}^A\omega_{\{0,2\}}^A-
\zeta_{3\rho}^A\omega_{\{1,1\}}^A\right),\ldots,\nonumber\\
& & g_2^g= -10 \zeta^V_{3\rho}\omega_{00}^V,\qquad
g_3^g={15\over 8}\left(3\zeta_{3\rho}^A\omega_{[0,1]}^A -
\zeta_{3\rho}^V\omega_{\{0,1\}}^V\right),\nonumber\\
& & g_4^g={3\over 5}\left(2\zeta_{3\rho}^A\omega_{[0,2]}^A -
4\zeta_{3\rho}^V\omega_{\{0,2\}}^V +
\zeta_{3\rho}^V\omega_{\{1,1\}}^V\right),\ldots,\qquad
\eeq
where we have used the normalization condition for 
${\cal A}(\underline{\alpha})$ and 
${\cal V}(\underline{\alpha})$, $\omega_{00}^A=1$ and
$\omega_{[0,1]}^V=28/3$.
We also note that 
(\ref{eq4.32}) for $n=0,1$ together with
the conditions $a_0^\parallel=1$, $G_0^g=G_1^g=0$
determines the first two coefficients of $G_n$ as $G_0=1$, $G_1=-1/2$,
which gives consistent normalization
$\int_0^1du\, g_\perp^{(v)}(u)=1$ and
$\int_0^1du\, g_\perp^{(a)}(u)=1$ through (\ref{eq4.24}) and (\ref{eq4.25}).
We finally remind that the G-parity invariance of the three-particle 
distribution amplitudes imposes
$\omega^V_{\{k,l\}}=0$ and
$\omega^A_{[k,l]}=0$, leading to $g_n^g=0$
and $G_n^g= g^{\uparrow\downarrow g}_n$.

\subsection{Renormalization and scale-dependence}

In this subsection, we discuss the renormalization of the
chiral-even distribution amplitudes,
$\phi_\parallel(u,\mu^2)$, $g_\perp^{(v)}(u,\mu^2)$ and
$g_\perp^{(a)}(u,\mu^2)$, utilizing the conformal expansions
derived in the previous subsection.

For completeness we start our discussion with the renormalization
of the twist~2 distribution amplitude $\phi_\parallel(u)$. 
{}From (\ref{eq4.1}) and the orthogonality relations of the Gegenbauer
polynomials \cite{ER53}, one obtains
\begin{equation}
a_{n}^{\parallel}(\mu^{2}) = \frac{2(2n+3)}{3(n+1)(n+2)}
\int_{0}^{1}du\, C_{n}^{3/2}(\xi)\, \phi_{\parallel}(u, \mu^{2}).
\label{eq:4anp}
\end{equation}
Using (\ref{eq:vda})
in the right-hand side
of (\ref{eq:4anp}), we can express $a_n^\parallel$
in terms of the conformal operator:
\begin{equation}
a_{n}^{\parallel} 
(\mu^{2}) =
\frac{2(2n+3)}{3f_\rho m_\rho (n+1)(n+2)(e^{(\lambda)}\cdot z)(p\cdot z)^{n}}
\langle 0| \Omega_{n}^{\parallel}(0; \mu^{2})
| \rho^{-} (P, \lambda) \rangle
\label{eq:4anp2}
\end{equation}
with
\begin{equation}
\Omega_{n}^{\parallel}(x; \mu^{2})=
\left(i\partial. \right)^{n} \ub(x) \gamma. C_{n}^{3/2}
\left(\frac{\deriv\!.}
{\partial.}\right) d(x),
\label{eq:4omg}
\end{equation}
where the local operator in the right-hand side
is renormalized at $\mu^{2}$,
$\deriv \equiv \overrightarrow{D}-\overleftarrow{D}$,
and $\partial_{\mu}$ is the total derivative (\ref{eq:3tdrv}).
$\Omega_{n}^{\parallel}(x; \mu^{2})$ is the conformal
operator with conformal spin $j=n+2$ \cite{O82}.  
The scale dependence of $a_n^\parallel (\mu^2)$ is well 
known\,\cite{BLreport,exclusive}:
\begin{equation}
a_{n}^{\parallel} (Q^{2})
= L^{\gamma_{n}^{\parallel}/b} a_{n}^{\parallel}(\mu^{2}),
\label{eq:4scalep}
\end{equation}
where $L \equiv \alpha_{s}(Q^{2})/\alpha_{s}(\mu^{2})$, 
$b=(11N_{c} - 2N_{f})/3$ and the anomalous dimension $\gamma_n^\parallel$
for the conformal operator $\Omega_n^{\parallel}$
is given by
\beq
\gamma_{n}^{\parallel} = 4C_{F}
\left( \psi(n+2)+ \gamma_{E} - \frac{3}{4} - {1\over 2(n+1)(n+2)}
\right).
\label{eq:4anomdp}
\eeq
For $n=0$, $\Omega_0^\parallel$ is reduced to a conserved vector current,
and hence its anomalous dimension vanishes to all orders.
Combined with the normalization condition
for $\phi_\parallel(u)$, $a_0^\parallel =1$, this is consistent with the fact
that $f_\rho$ is scale independent.   We thus omitted $f_\rho$
in both sides of (\ref{eq:4scalep}) (compare with (\ref{eq:3scalep}) and 
(\ref{eq:3scalep0})).

Next we proceed to discuss the
renormalization of 
the twist~3 distribution amplitudes 
$g_\perp^{(v)}(u, \mu^{2})$ and $g_\perp^{(a)}(u, \mu^{2})$.
As we saw in the previous subsections,
they receive 
contributions from the twist~2 distribution amplitude (Wandzura-Wilczek parts),
the three-particle distribution amplitudes and the terms proportional to 
the quark masses 
(see (\ref{eq4.13})$-$(\ref{eq4.16})).  
The scale-dependence of $\phi_\parallel(u,\mu^2)$ discussed above
completely determines that
of $g_\perp^{(v)WW}(u, \mu^{2})$ and $g_\perp^{(a)WW}(u, \mu^{2})$
through the relations (\ref{eq4.24}), (\ref{eq4.25}) and (\ref{eq4.32}).

To understand the scale dependence
of the three-particle distribution amplitudes ${\cal V}(\underline{\alpha})$ 
and ${\cal A}(\underline{\alpha})$, one
needs to express $\omega^{V,A}_{k,l}$ in (\ref{eq4.28})
in terms of the local conformal operators.  
Owing to the orthogonality relation 
(\ref{eq:appello0}) of the Appell polynomials 
with different $k+l$, 
$\omega_{k,l}^{V,A}$ can be expressed in terms of 
the matrix elements of the conformal operators
with a definite
conformal spin $j=k+l+7/2$.  
As we saw in Sec.~3, it suffices to know the form of the conformal
operators up to total derivatives for the renormalization.  
For this purpose, we recall from
(\ref{eq:V3}) and (\ref{eq:A3})
\beq
& &\la 0|\bar{u}(tz)\gamma. [tz, vz]
g\widetilde{G}_{\perp}.(vz)[vz, wz]d(wz) | \rho^-(P,\lambda)\ra
\nonumber\\
& &\qquad\qquad\qquad
=i\epsilon_{_\perp\cdot\mu\nu}p_\mu e_{\perp\nu}^{(\lambda)} 
(p\cdot z) f_{3\rho}^V
\int {\cal D}\underline{\alpha}
e^{-ip\cdot z(t\alpha_u + w\alpha_d + v\alpha_g)}{\cal V}(\underline{\alpha}),
\label{eq:4V3}\\
& &\la 0|\bar{u}(tz)\gamma.i\gamma_5 [tz, vz]gG_{\perp}.(vz)[vz, wz]d(wz) 
| \rho^-(P,\lambda)\ra
\nonumber\\
& &\qquad\qquad\qquad
=i\epsilon_{_\perp\cdot\mu\nu}p_\mu e_{\perp\nu}^{(\lambda)} 
(p\cdot z) f_{3\rho}^A
\int {\cal D}\underline{\alpha}
e^{-ip\cdot z(t\alpha_u + w\alpha_d + v\alpha_g)}{\cal A}(\underline{\alpha}).
\label{eq:4A3}
\eeq
To obtain the actual form of the conformal operators, we 
apply ${\partial^{n-2}/\partial t^{n-k-2}\partial w^k}$
on both sides of (\ref{eq:4V3}) and (\ref{eq:4A3}) and set
$t=w=v=0$.   Using the integral formula for the Appell
polynomial (\ref{eq:appello}), one obtains
\beq
& & 
\lefteqn{ \la 0| \bar{u}(0)(i\derleft.)^{n-k-2}\gamma.
g\widetilde{G}_{\perp}.(0)(i\derright.)^k d(0) |\rho (P,\lambda )\ra }
\nonumber\\
& & \qquad\qquad
=if^V_{3\rho}\epsilon_{_\perp\cdot\mu\nu} e_{\perp\mu}^{(\lambda)} p_{\nu}
(p\cdot z)^{n-1}
\left[
\omega_{k,n-k-2}^V {360 (-1)^{n}k!(n-k-2)! \over 2^{n+1}(n+1)(2n+1)!!}
\right.\nonumber\\
& &\left.\qquad\qquad\qquad\qquad
+ ({\rm terms\ with}\ \omega_{l,r-l-2}^V|_{r<n})
\right],
\label{eq:Vdef}\\
& & 
\lefteqn{ \la 0| \bar{u}(0)(i\derleft.)^{n-k-2}\gamma.i\gamma_5
gG_{\perp}.(0)(i \derright .)^k d(0) |\rho (P,\lambda )\ra }
\nonumber\\
& & \qquad\qquad
=if^A_{3\rho}\epsilon_{_\perp\cdot\mu\nu} e_{\perp\mu}^{(\lambda)} p_\nu
(p\cdot z)^{n-1}
\left[
\omega_{k,n-k-2}^A {360 (-1)^{n}k!(n-k-2)! \over 2^{n+1}(n+1)(2n+1)!!}
\right.\nonumber\\
& &\left.\qquad\qquad\qquad\qquad
+ ({\rm terms\ with}\ \omega_{l,r-l-2}^A|_{r<n})
\right].
\label{eq:Adef}
\eeq
It is easy to see by induction 
that $\omega^{V,A}_{l,r-l-2}|_{r<n}$
in these equations are the matrix elements
of the total derivatives of the lower conformal operators.
Therefore we can identify the corresponding conformal operators
for $\omega_{k,n-k-2}^{V,A}$ as
\beq
\left(f^V_{3\rho}\omega_{k,n-k-2}^V \right)(\mu^2)
= { (-1)^k N_n \over 90 k!(n-k-2)!}
\la 0| \Theta^V_{k,n-k-2}(0;\mu^2) |\rho^-(P,\lambda)\ra, 
\label{eq:Vconf}\\
\left(f^A_{3\rho}\omega_{k,n-k-2}^A \right)(\mu^2)
= { (-1)^k N_n \over 90 k!(n-k-2)!}
\la 0| \Theta^A_{k,n-k-2}(0;\mu^2) |\rho^-(P,\lambda)\ra, 
\label{eq:Aconf}
\eeq
where the conformal operators are now obtained up to
total derivatives as
\beq
\Theta^V_{k,n-k-2}(0)
&\equiv& \bar{u}(0)(-i\derleft.)^{n-k-2}
\gamma. 
g\widetilde{G}_{\perp}.(0)(i\derright.)^{k}d(0)+ ({\rm total\ derivatives}),
\label{}
\label{eq:Uop}\\
\Theta^A_{k,n-k-2}(0)
&\equiv& \bar{u}(0)(-i\derleft.)^{n-k-2}
\gamma. gG_{\perp}.(0) (i \derright.)^{k}i\gamma_5d(0)+ ({\rm total\ derivatives}),
\label{eq:Vop}
\eeq
and we introduced for convenience a dimensionless and  
scale independent normalization constant $N_n$ as
\beq
N_n \equiv { 2^{n-1} (2n+1)!! (n+1)
\over i\epsilon_{_\perp\cdot\mu\nu} p_\mu 
e_{\perp\nu}^{(\lambda)} (p\cdot z)^{n-1}}. 
\label{eq:normeven}
\eeq
{}From Eqs.~(\ref{eq:Vconf}) and (\ref{eq:Aconf}), we can obtain 
the scale-dependence of $(f_{3\rho}^V\omega_{k,n-k-2}^V)(\mu^2)$ and 
$(f_{3\rho}^A\omega_{k,n-k-2}^A)(\mu^2)$
by working out the renormalization of 
$\{\Theta^V_{k,n-k-2},\Theta^A_{k,n-k-2}\}$
($k=0,\ldots,n-2$).\footnote{Since 
$\Theta^V_{k,n-k-2}$ and $\Theta^A_{k,n-k-2}$ 
have the same conformal spin, they generally mix with 
each other under renormalization, even though they
originate from different distribution amplitudes.} 
If we define 
$R^{V\pm}_{n,k} \equiv \Theta^V_{k,n-k-2}\pm \Theta^V_{n-k-2,k}$
($k=0,1,\ldots,\kappa^{\pm}_{n}$ with $\kappa^{\pm}_{n}$ defined in 
(\ref{eq:3kappa}))
and  
$R^{A\pm}_{n,k} \equiv  \Theta^A_{k,n-k-2}\mp \Theta^A_{n-k-2,k}$   
($k=0,1,\ldots,\kappa^{\mp}_{n}$), 
$\{R^{V+}_{n,k},R^{A+}_{n,k}\}$ and 
$\{R^{V-}_{n,k},R^{A-}_{n,k}\}$, respectively, have
G-parity $(-1)^{n+1}$ and $(-1)^{n}$ and thus they do not
mix with each other under renormalization.

By inserting (\ref{eq:Vconf}) and (\ref{eq:Aconf}) into
(\ref{eq4.41}) and (\ref{eq4.42}) and recalling the definition
of $\zeta_{3\rho}^{V,A}$ from (\ref{eq4.5}), 
one can express the
contribution from the three-particle distribution amplitude to 
the two-particle distribution amplitudes $g_\perp^{(v,a)}$ 
in terms of the conformal operators:
\beq
G_n^g(\mu^2) & = &
\left\{\renewcommand{\arraystretch}{2}
\begin{array}{ll}
\displaystyle{
{N_n\over f_\rho m_\rho (n+1)!(n+1)}\,\la 0|
\sum_{k=0}^{n-2}(k+1) 
R_{n,k}^-(0;\mu^2)|\rho^-(P,\lambda)\ra},
&(n=2,4,6,\ldots),\nonumber\\
\displaystyle{
 {-N_n\over f_\rho m_\rho (n+1)!(n+1)}\,\la 0|
\sum_{k=0}^{n-2}(n-k-1) 
R_{n,k}^+(0;\mu^2)|\rho^-(P,\lambda)\ra},
& (n=3,5,7,\ldots),
\end{array}\renewcommand{\arraystretch}{1}
\right.\\[-15pt]
\label{eq:Gcoeff}
\\
g_n^g(\mu^2) & =&
\left\{\renewcommand{\arraystretch}{2}
\begin{array}{ll}
\displaystyle{
{-N_n\over f_\rho m_\rho (n+1)!(n+1)}\,\la 0|
\sum_{k=0}^{n-2}(n-k-1) 
R_{n,k}^+(0;\mu^2)|\rho^-(P,\lambda)\ra},
& (n=2,4,6,\ldots),\nonumber\\
\displaystyle{
{N_n\over f_\rho m_\rho (n+1)!(n+1)}\, \la 0|
\sum_{k=0}^{n-2}(k+1) 
R_{n,k}^-(0;\mu^2)|\rho^-(P,\lambda)\ra},
& (n=3,5,7,\ldots),
\end{array}\renewcommand{\arraystretch}{1}
\right.\\[-15pt]
\label{eq:gcoeff}
\eeq
where the operators $R^\pm_{n,k}$ ($k=0,\ldots,n-2$) are defined as
\beq
R^{\pm}_{n,k} = 
\Theta^V_{k,n-k-2}\pm \Theta^V_{n-k-2,k} 
\mp \Theta^A_{k,n-k-2}+ \Theta^A_{n-k-2,k}.\qquad (k=0,1,\ldots,n-2)
\label{eq:Rop}
\eeq
Here we note that
$\{R_{n,k}^+ \}$ and $\{R_{n,k}^- \}$
have opposite parity under G-parity transformations
and constitute
another operator basis which is equivalent to 
$\{R^{V+}_{n,k},R^{A+}_{n,k}\}$ and 
$\{R^{V-}_{n,k},R^{A-}_{n,k}\}$, respectively, 
either of which has the same number of 
independent operators $\kappa^{+}_{n} +\kappa^{-}_{n} +2 =n-1$.
If we define the anomalous dimension matrix for 
$\{R_{n,k}^\pm\}$ as $\Gamma_n^\pm$, the scale dependence of $R_{n,k}^\pm$
is given by
\beq
R_{n,k}^\pm(0;Q^2) =\sum_{l=0}^{n-2} \left(L^{\Gamma^\pm_n/b}\right)_{k,l}
R_{n,l}^\pm(0;\mu^2).
\label{eq:Rrenorm}
\eeq

The renormalization for $\{ R_{n,k}^\pm\}$ can be conveniently worked out
by considering the forward 
matrix elements with respect to a spin 1/2 target (say a quark)
as was done in Sec.~3.  
In this case, the contribution from the 
total derivative terms disappear and its matrix element
$\la R_{n,k}^\pm \ra$ is reduced to 
$\la  \bar{\psi}(iD.)^{n-k-2}\gamma.g\widetilde{G}_\perp.(iD.)^k\psi
\pm (k\to n-k-2) \ra + \la 
 \bar{\psi}(iD.)^{k}\gamma.gG_\perp.(iD.)^{n-k-2}i\gamma_5\psi
\mp (k\to n-k-2)\ra$. 
By now, renormalization of $\{R_{n,k}^\pm\}$ has been solved 
by several different approaches in 
the context of the $Q^2$ evolution of the transverse spin structure
function $g_2(x,Q^2)$\,\cite{BB88,KYTU97,g2}. 
In particular, $\{R_{n,k}^+\}$ is exactly the operator basis
used in \cite{KYTU97}  for 
the renormalization of $g_2(x,Q^2)$.  
{}From the result in \cite{KYTU97}, $\Gamma_n^+$
in (\ref{eq:Rrenorm}) is obtained as
\beq
\left(\Gamma^+_n\right)_{k,l} = -X_{k+1,l+1}^{n+1},\qquad (k,l= 0,1,\ldots,
n-2)
\eeq
where $X_{k,l}^{n+1}$ is given in Eqs.~(15)--(17) 
of \cite{KYTU97} with $n\to n+1$. 
\footnote{The result in
Eqs.~(15)--(17) 
of \cite{KYTU97} is correct only for odd $n$ in their convention.
For general $n$, the term $(-1)^l/(n-l)$ proportional
to $2C_F - C_G$ in Eq. (16) should read  
$(-1)^{n-l+1}/(n-l)$.  See \cite{KNT} for the detail.}
The explicit form of 
$\Gamma^-_n$ is not available in the literature, but can be obtained
from the kernel given in \cite{BB88}. 
(Since $R^-_{n,k}$ does not contribute to
the deep-inelastic scattering which is the charge conjugation even process,
it has not been receiving attention up to now.)

By using the basis $\{ R_{n,k}^+\}$ and $\{ R_{n,k}^-\}$,
(\ref{eq:Vconf}) and (\ref{eq:Aconf}) can be rewritten as
\beq
\left(f^V_{3\rho}\omega_{[k,n-k-2]}^V 
\pm f^A_{3\rho}\omega_{\{k,n-k-2\}}^A \right)(\mu^2)
= { (-1)^k N_n \over 180 k!(n-k-2)!}
\la 0| R_{n,k}^\mp(0;\mu^2) |\rho^-(P,\lambda)\ra, 
\label{eq:Ginv}
\eeq
for the G-parity invariant components and
\beq
\left(f^V_{3\rho}\omega_{\{k,n-k-2\}}^V 
\mp f^A_{3\rho}\omega_{[k,n-k-2]}^A \right)(\mu^2)
= { (-1)^k N_n \over 180 k!(n-k-2)!}
\la 0| R_{n,k}^\pm(0;\mu^2) |\rho^-(P,\lambda)\ra, 
\label{eq:Gviol}
\eeq
for the G-parity violating components.
In (\ref{eq:Ginv}) and (\ref{eq:Gviol}), upper and lower
signs correspond to $n=2,4,6,\ldots$ and $n=3,5,7,\ldots$,
respectively.  
For illustration, we give here the explicit form of the scale
dependence of the left hand side of (\ref{eq:Ginv}) for $n=2,3,4$
from (\ref{eq:Rrenorm}):
\begin{eqnarray}
f_{3 \rho}^{A}(Q^{2}) = L^{\Gamma^-_2/b}f_{3\rho}^{A}(\mu^{2}),\qquad
\Gamma_2^-= -{1\over 3} C_F + 3C_G,
\label{eq:4exam1}
\end{eqnarray}
\beq
& & \left(
\begin{array}{c}
{28\over 3}f_{3\rho}^{V} - f_{3\rho}^{A}\omega^{A}_{\{0,1\}}\\
{28\over 3}f_{3\rho}^{V} + f_{3\rho}^{A}\omega^{A}_{\{0,1\}}
\end{array}
\right)^{Q^{2}} =  L^{\Gamma^{+}_{3}/b}
\left(
\begin{array}{c}
{28\over 3}f_{3\rho}^{V} - f_{3\rho}^{A}\omega^{A}_{\{0,1\}}\\
{28\over 3}f_{3\rho}^{V} + f_{3\rho}^{A}\omega^{A}_{\{0,1\}}
\end{array}
\right)^{\mu^{2}},\nonumber\\
& &\qquad\qquad\Gamma_{3}^{+} = \left(
\begin{array}{cc}
\frac{8}{3}C_{F} + {7\over 3}C_{G}\ \ &
\frac{2}{3}C_{F}-\frac{2}{3}C_{G} \\
\frac{5}{3}C_{F} - \frac{4}{3}C_{G}\ \ &
\frac{1}{6}C_{F} + 4C_{G}
\end{array}
\right),
\label{eq:4exam2}
\eeq
\beq
& &\left(
\begin{array}{c}
2f_{3\rho}^V \omega_{[0,2]} + 2f_{3\rho}^A\omega^A_{\{0,2\}}\\
-f_{3\rho}^A\omega^A_{\{1,1\}}\\
-2f_{3\rho}^V \omega_{[0,2]} + 2f_{3\rho}^A\omega^A_{\{0,2\}}
\end{array}
\right)^{Q^2}
= L^{\Gamma^-_4/b}
\left(
\begin{array}{c}
2f_{3\rho}^V \omega_{[0,2]} + 2f_{3\rho}^A\omega^A_{\{0,2\}}\\
-f_{3\rho}^A\omega^A_{\{1,1\}}\\
-2f_{3\rho}^V \omega_{[0,2]} + 2f_{3\rho}^A\omega^A_{\{0,2\}}
\end{array}
\right)^{\mu^2},\nonumber\\
& &\qquad\qquad
\Gamma^-_4 = \left(
\begin{array}{ccc}
{37\over 30}C_F + {25\over 6}C_G\ \  & {23\over 10}C_F -{7\over 4}C_G\ \  
&- {3\over 10}C_F \\
{1\over 6}C_F - {5\over 12}C_G\ \  & 3C_F +{41\over 12}C_G\ \  &
{7\over 6}C_F -{13\over 12}C_G \\
-C_F + {5\over 12}C_G\ \  & {7\over 2}C_F - {9\over 4}C_G\ \  &
{7\over 3}C_F  + {37\over 12}C_G
\end{array}
\right),
\label{eq:4exam3}
\eeq
where we have used $\omega_{00}^A=1$ and $\omega^V_{[0,1]}=28/3$, and
the anomalous dimensions 
$\Gamma_{2}^{-}$, $\Gamma_4^-$ 
can be obtained from the kernel 
given in \cite{BB88}.
\footnote{Eq.~(6.2) in \cite{BB88} for the kernel
contains a misprint: the delta function in the last line
should be replaced by $\delta(\alpha - u)$.}
For the complete discussion on $\Gamma^-_n$, see \cite{KNT}.

As one can see from this illustration, 
the $\mu^2$ evolution of 
$\{\omega_{[k,n-k-2]}^V,\omega_{\{k,n-k-2\}}^A\}$ and
$\{\omega_{\{k,n-k-2\}}^V,\omega_{[k,n-k-2]}^A\}$ 
becomes extremely complicated for general $n$.  
However, as was the case for the chiral-odd distribution amplitude in 
Sec.~3,
the $\mu^2$-dependence 
of the three-particle contribution to the
twist~3 two-particle distribution amplitudes, i.e. 
$G_n^g(\mu^2)$ and $g_n^g(\mu^2)$, 
is greatly simplified in the large $N_c$ limit.
It has been shown in \cite{ABH91} 
that the combination of $R^\pm_{n,k}$
in (\ref{eq:Gcoeff}) and (\ref{eq:gcoeff}) renormalize multiplicatively 
 at $N_c\to \infty$ to 
$O(1/N_c^2)$ accuracy.      
We thus get in this limit
\beq
G_n^g(Q^2)
&=&L^{\gamma_n/b}G_n^g(\mu^2),
\nonumber\\
g_n^g(Q^2)
&=&L^{\gamma_n/b}g_n^g(\mu^2),
\label{eq:evenlargeNc}
\eeq
with a common anomalous dimension
\beq
\gamma_n=2N_c\left( \psi(n+1) +\gamma_E -{1\over 4} + {1\over 2(n+1)} 
\right).
\label{eq:evenlargeNcAD}
\eeq
This $\gamma_n$ has been shown to be the lowest eigenvalue 
of $\Gamma^\pm_n$\,\cite{ABH91}.  This reduction to the simple evolution
equation (\ref{eq:evenlargeNc}) is equivalent to
the fact that the coefficients of $R_{n,k}^\pm$ in 
(\ref{eq:Gcoeff}) and (\ref{eq:gcoeff}) constitute the
left eigenvector of $\Gamma_n^\pm$ with the eigenvalue $\gamma_n$
in this limit:
\beq
& &\sum_{k=0}^{n-2}(k+1)\left(\Gamma_n^-\right)_{k,l} = (l+1)\gamma_n,
\nonumber\\
& &\sum_{k=0}^{n-2}(n-k-1)\left(\Gamma_n^+\right)_{k,l} = (n-l-1)\gamma_n,
\label{eq:evenmatrix}
\eeq
which can be compared with 
(\ref{eq:3LNpo})$-$(\ref{eq:3LNm}) for the chiral-odd case. 
The renormalization of the 
flavour singlet part of $g_\perp^{(v,a)}(u,\mu^2)$ 
(for $\omega$ and $\phi$ mesons) is complicated by
additional mixing with the purely gluonic
twist~3 distribution amplitudes.  
For this mixing, no simplification occurs
in the $N_c\to \infty$ limit.

As was discussed for the
chiral-odd distribution amplitudes in Sec.~3, 
another simplification of the evolution 
equation for $g_\perp^{(v,a)}(u,\mu^2)$ occures 
at large $n$ with arbitrary $N_c$.  In this limit, the scale dependence
of $G_n^g(\mu^2)$ and $g_n^g(\mu^2)$ is described by the same
equation (\ref{eq:evenlargeNc}) with a slightly 
shifted anomalous dimension
\beq
\gamma_n \to \gamma_n + (4C_F - 2N_c)\left( {\rm ln}\,n +\gamma_E -{3\over 4}
\right).
\eeq
Combined with this simplification at $n\to \infty$,
the result at large $N_c$ in (\ref{eq:evenlargeNc}) 
is valid to $O((1/N_c^2){\rm ln}\,(n)/n)$
accuracy as was the chiral-odd case in Sec.~3. 

To summarize this section, we have solved the renormalization of the
nonsinglet chiral-even twist~3 distribution amplitudes, $g_\perp^{(v)}(u)$
and $g_\perp^{(a)}(u)$.
We found that in the limits  $N_c\to \infty$ and $n\to\infty$,  
the scale-dependence of the three-particle contribution
to $g_\perp^{(v)}(u)$
and $g_\perp^{(a)}(u)$ is described by a simple DGLAP type evolution
equation similar to that for the
twist~2 distribution amplitude.  
Combined with the results of the previous section, this
means that this simplification for the scale-dependence is universal
for all twist~3 nonsinglet distribution amplitudes.

\section{Models for Distribution Amplitudes}
\setcounter{equation}{0}

In this section we present explicit models for the two-particle
distribution amplitudes of twist~3,
taking into account contributions of the first few conformal partial waves.
The main observation and important point is
that the QCD equations of motion are satisfied order by order in the 
conformal expansion, which guarantees the consistency of the approximation.
Our approximation thus introduces a minimum number of nonperturbative 
parameters describing matrix elements of certain local operators 
between the vacuum and the meson state, 
which we estimate using QCD sum rules \cite{SVZ,CZreport}. More
sophisticated models can be constructed in a systematic way by adding 
contributions of higher conformal partial waves when estimates of the relevant 
nonperturbative matrix elements will become available. 

Our  approach involves the implicit assumption that the conformal partial 
wave expansion is well-convergent. This can be justified rigorously 
at large scales, since the anomalous dimensions of twist~2 and twist~3 
operators increase logarithmically with the conformal spin $j$, but 
is nontrivial at relatively low scales of order $\mu \sim (1-2)\,$GeV
which we choose as reference scale. We believe that this assumption  
is natural and in fact necessary for any model of distribution amplitudes
at scales where they evolve perturbatively; the
last word, however, has to come from experiment. 

Since orthogonal polynomials of high orders are rapidly oscillating 
functions, a truncated expansion in conformal partial waves almost necessarily
is oscillatory as well. Such a behaviour is clearly unphysical,
but does not constitute a real problem because  physical observables 
are given by convolution integrals of distribution amplitudes with
smooth coefficient functions. 
A classical example for this
feature is the $\gamma\gamma^*$-meson form factor which is governed
by the quantity
$$
\int du\,\frac{1}{u}\, \phi(u) \sim \sum a_i,
$$
where the coefficients $a_i$ are exactly the ``reduced matrix elements''
in the conformal expansion.
The oscillating terms are averaged over and strongly suppressed. 
Stated otherwise: 
models of distribution amplitudes should generally be understood as
distributions (in the mathematical sense).    

\subsection{Leading twist distributions}

The twist~2 distribution amplitudes of vector mesons have received much
attention in the literature. Their study was
pioneered by Chernyak and Zhitnitsky (see
\cite{CZreport} for a comprehensive review). More recently, 
the results for $\rho$ mesons   
were critically examined in \cite{BBrho}. In this paper
we complete the update \cite{BBrho} by the reanalysis of SU(3) breaking
corrections, see App.~\ref{app:B}. 

A simple model of twist~2 distributions includes contributions of
the three lowest conformal 
spins (``S, P and D waves''):
\begin{eqnarray}
\phi_\parallel(u) & = & 6 u\ub \left[ 1 + 3 a_1^\parallel\, \xi +
a_2^\parallel\, \frac{3}{2} ( 5\xi^2  - 1 ) \right],\\
 \phi_\perp(u) & = & 6 u\ub \left[ 1 + 3 a_1^\perp\, \xi +
a_2^\perp\, \frac{3}{2} ( 5\xi^2  - 1 ) \right].
\end{eqnarray}
We recall that $\bar u=1-u$ and $\xi=2u-1$.
For the $K^*$ meson 
distribution amplitude, ``$u$'' designates the momentum fraction of a 
heavier (strange) quark.
The values of the decay constants $f_V$, $f^T_V$ and the Gegenbauer moments 
$a_1$, $a_2$ are collected in Tabs.~\ref{tab:B1} and \ref{tab:B2} for 
$V=\rho$, $K^*$ and $\phi$ mesons. The scaling laws for the
coefficients $a_n^\parallel$ and $a_n^\perp$ are given in
Eqs.~(\ref{eq:4scalep}) and (\ref{eq:3scalep}), respectively.

\begin{table}
$$
\renewcommand{\arraystretch}{1.4}
\addtolength{\arraycolsep}{3pt}
\begin{array}{|c|ccc|}
\hline
V & \rho^\pm  & K^{*\pm} & \phi\\ \hline
f_V [{\rm MeV}] & 198\pm 7  & 226 \pm 28 & 254 \pm 3\\
\hline
\end{array}
$$
\caption[]{Experimental values of couplings to the vector
current \protect{\cite{PDG}}.}\label{tab:B1}
\renewcommand{\arraystretch}{1.4}
\addtolength{\arraycolsep}{3pt}
$$
\begin{array}{|c|ccc|}
\hline
V & \rho^\pm & K^{*\pm} & \phi\\ \hline
f^T_V(1\,{\rm GeV}) [{\rm MeV}] & 160\pm 10 & 185\pm 10 & 215\pm15\\ \hline
a_1^\parallel(1\,{\rm GeV}) & 0 & 0.19\pm 0.05 & \phantom{-}0\\
a_2^\parallel(1\,{\rm GeV}) & 0.18\pm0.10 & 0.06\pm 0.06 & 0\pm0.1\\ 
a_1^\perp(1\,{\rm GeV}) & 0 & 0.20\pm 0.05 & \phantom{-}0\\
a_2^\perp(1\,{\rm GeV}) & 0.2\pm0.1 & 0.04\pm 0.04 & 0\pm0.1\\ \hline
\end{array}
$$
\renewcommand{\arraystretch}{1}
\addtolength{\arraycolsep}{-3pt}
\caption[]{The tensor couplings and lowest Gegenbauer moments of
vector mesons
from QCD sum rules, see App.~\protect{\ref{app:B}}.}\label{tab:B2}
\end{table}
\begin{figure}
\centerline{\epsffile{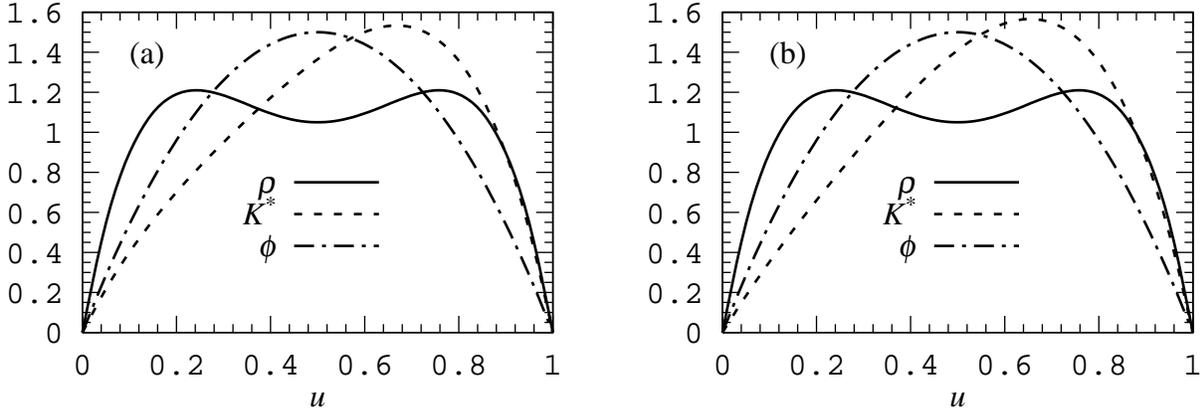}}
\caption[]{(a) Leading twist~2 distribution amplitude $\phi_\parallel(u)$ 
for a longitudinally polarized vector meson, (b) $\phi_\perp(u)$ for a
transversely polarized one. Renormalization point is $\mu=1\,$GeV.}\label{fig1}
\end{figure}

The corresponding distributions, evaluated at $\mu=1\,$GeV,
are shown in Fig.~\ref{fig1}.
Note that the leading twist distributions for longitudinally and transversely
polarized vector mesons are very similar to each other.

In the following we neglect the masses of $u$ and $d$ quarks and 
do not account for $\rho$--$\omega$ mixing. In this approximation 
the couplings and distribution amplitudes of
$\rho^{\pm}$, $\rho^0$ and $\omega$ mesons are equal if one chooses
properly normalized currents, i.e.\ $(\bar u u \pm \bar d d)/\sqrt{2}$ for the 
$\omega$ and $\rho^0$ meson, respectively, in Eqs.~(\ref{eq:tda}),
(\ref{eq:vda}), (\ref{eq:fr}) and (\ref{eq:frp}). 

The model distribution amplitudes for the $K^*$ meson given
by Chernyak and Zhitnitsky \cite{CZZsu3,CZreport} involve an additional
contribution $\sim a_3C^{3/2}_3(\xi)$.
We have not included this term since estimates of high partial waves
from QCD sum rules are not reliable. Our estimates for $a_1^\perp$ and
$a_2^\perp$ differ significantly from the results of \cite{CZZsu3,CZreport},
see Ref.~\cite{BBrho} for details. 

\subsection{Three-particle distributions of twist three}

Three-particle twist~3 distribution amplitudes were defined in Sec.~2.3
and their conformal expansion is considered in detail in Secs.~3 and 4.
Assuming  SU(3) flavour symmetry, the lowest order contributions
to the conformal expansion are  
\begin{eqnarray}
{\cal V} (\alpha_d,\alpha_u,\alpha_g) &=& 
5040\, (\alpha_d-\alpha_u)\alpha_d \alpha_u\alpha_g^2,
\label{modelV}\\
{\cal A} (\alpha_d,\alpha_u,\alpha_g) &=& 
360\, \alpha_d \alpha_u \alpha_g^2 
\Big[ 1+ \omega^A_{1,0}\frac{1}{2}(7\alpha_g-3)],
\label{modelA}\\
{\cal T} (\alpha_d,\alpha_u,\alpha_g) &=& 
5040\, (\alpha_d-\alpha_u) \alpha_d \alpha_u \alpha_g^2.
\label{modelT}
\end{eqnarray}
Our expressions for $\cal V$ and $\cal A$ agree with the corresponding
``asymptotic distributions'' in \cite{CZreport,BCZ}. The result for 
${\cal T}$ is new. An important point to note is that the contribution of 
leading conformal spin $j=7/2$ to the distribution amplitudes 
$\cal V$ and $\cal T$ vanishes by virtue of  G-parity 
invariance (in the SU(3) limit). Hence, if one takes into 
account the leading $j=7/2$ contribution to
the distribution ${\cal A}$ only, it is consistent to put
$\cal V$ and $\cal T$ to zero; the expressions given in (\ref{modelV})
and (\ref{modelT}) correspond to contributions of the next-to-leading
conformal spin $j=9/2$ and have to be taken into account together with the
{\em correction} to $\cal A$ proportional to $\omega^A_{1,0}$,
which has the same spin.

The decay constants $f_{3\rho}^V$, $f_{3\rho}^A$ and 
the few first  coefficients $\omega^A_{i,k}$, $\omega^V_{i,k}$ were estimated 
from QCD sum rules in Ref.~\cite{ZZC85}, 
in particular\footnote{The model distributions proposed in \cite{ZZC85}
include in addition the $j=11/2$ terms with coefficients 
$\omega^A_{1,1} = 11.7$, $\omega^A_{2,0}= 7$ and $\omega^V_{2,0}= -1.9$. 
We do not include these contributions for simplicity and 
because the corresponding QCD sum rules are less reliable.}           
\begin{equation}
 f_{3\rho}^A = (0.5-0.6)\cdot 10^{-2}\,\mbox{\rm GeV}^2,
 \qquad
 f_{3\rho}^V =  0.2\cdot 10^{-2}\,\mbox{\rm GeV}^2,
 \qquad 
 \omega^A_{1,0} = -2.1.
\end{equation}
We expect that the accuracy for the last two entries is of order 30-50\%. 
We have derived a new  sum rule for $f_{3\rho}^T$ (see  App.~\ref{app:B}) 
from which we obtain the estimate
\begin{equation}
 f_{3\rho}^T = (0.3\pm 0.3)\cdot 10^{-2}\,\mbox{\rm GeV}^2. 
\label{eq:5f3t}
\end{equation}
Again, the
renormalization scale is $\mu=1\,$GeV. The anomalous dimensions of the
couplings can be found in Eqs.~(\ref{eq:4exam1}), (\ref{eq:4exam2})
 and (\ref{eq:3exam1}).

Estimates of quark mass corrections are difficult to obtain from
such complicated sum rules. A detailed  study  of this point
goes beyond the scope of this paper. For the present purpose we  
neglect SU(3) corrections to three-particle distributions and
assume  that the dimensionless couplings $\zeta^{V,A,T}_3$ defined 
in (\ref{eq:3pm}) and (\ref{eq4.5}) are the same for all vector mesons. 
The most interesting effect which we miss in this ``poor-man's'' 
approximation is that the leading conformal spin contribution
$\sim 360 \alpha_d \alpha_u \alpha_g^2$  reappears in the
distribution amplitudes $\cal V$ and $\cal T$ for $K^*$ mesons.
These terms deserve a further study. Our preferred 
values for the parameters determining three-particle distributions
are collected in Tab.~\ref{tab:zetas}.

\subsection{Two-particle distributions of twist three}

As repeatedly emphasized above, the equations of motion allow the
elimination of
two-particle distribution amplitudes of higher twist 
in favour of leading twist and 
three-particle distribution amplitudes as independent
dynamical degress of freedom. 
With leading twist and three-particle 
distributions as specified above, we get the following {\em exact} expressions
(including terms up to conformal spin 9/2):
\begin{eqnarray}
\hs(u) & = & 6u\bar u \left[ 1 + a_1^\perp \xi + \left( \frac{1}{4}a_2^\perp +
\frac{35}{6}\,\zeta^T_{3} \right) (5\xi^2-1)\right]\nonumber\\
& & {}+ 3\, \delta_+\, (3 u \ub + \ub \ln \ub + u \ln u) + 3\,\delta_-\,  (\ub
\ln \ub - u \ln u),\label{eq:e}\\
\htt(u) &= & 3\xi^2+ \frac{3}{2}\,a_1^\perp \,\xi (3 \xi^2-1)
+ \frac{3}{2} a_2^\perp\, \xi^2 \,(5\xi^2-3) 
+\frac{35}{4}\zeta^T_{3}(3-30\xi^2+35\xi^4)\nonumber\\
& & {} + \frac{3}{2}\,\delta_+
\, (1 + \xi \, \ln \ub/u) + \frac{3}{2}\,\delta_- \, \xi\, ( 2
+ \ln u + \ln\ub ),\\
g_\perp^{(a)}(u) & = & 6 u \bar u \left[ 1 + a_1^\parallel \xi +
\left\{\frac{1}{4}a_2^\parallel +
\frac{5}{3}\, \zeta^A_{3} \left(1-\frac{3}{16}\,
\omega^A_{1,0}\right) + \frac{35}{4} \zeta^V_{3}\right\}
(5\xi^2-1)\right]\nonumber\\
& & {} + 6\, \widetilde{\delta}_+ \,  (3u \ub + \ub \ln \ub + u \ln u ) + 
6\, \widetilde{\delta}_- \,  (\ub \ln \ub - u \ln u),\\
 g_\perp^{(v)}(u) & = & \frac{3}{4}(1+\xi^2)
+ a_1^\parallel\,\frac{3}{2}\, \xi^3 
 + \left(\frac{3}{7} \, 
a_2^\parallel + 5 \zeta_{3}^A \right) \left(3\xi^2-1\right)
 \nonumber\\
& & {}+ \left( \frac{9}{112}\, a_2^\parallel + \frac{105}{16}\,
 \zeta_{3}^V - \frac{15}{64}\, \zeta_{3}^A \omega_{1,0}^A
 \right) \left( 3 - 30 \xi^2 + 35\xi^4\right)\nonumber\\
& & {}+\frac{3}{2}\,\widetilde{\delta}_+\,(2+\ln u + \ln\ub) +
\frac{3}{2}\,\widetilde{\delta}_-\, ( 2 \xi + \ln\ub - \ln u),\label{eq:gv}
\end{eqnarray}
where, for simplicity, we used asymptotic leading twist distribution
amplitudes in the correction terms proportional to quark masses 
$\sim \delta_{\pm}$.\footnote{For realistic parameter values 
 the correction  in $a_2$ is small and can 
safely be neglected.}

\begin{table}
\renewcommand{\arraystretch}{1.4}
\addtolength{\arraycolsep}{3pt}
$$
\begin{array}{|c|ccc|}
\hline
V & \rho^\pm & K^{*\pm} & \phi\\ \hline
\zeta_3^A & 0.032 & 0.032 & 0.032\\
\zeta_3^V & 0.013 & 0.013 & 0.013\\
\zeta_3^T & 0.024 & 0.024 & 0.024\\
\omega_{1,0}^A & -2.1 & -2.1 & -2.1 \\
\delta_+ & 0 & \phantom{-}0.24 & 0.46 \\
\delta_- & 0 & -0.24 & 0 \\
\widetilde{\delta}_+ & 0 & \phantom{-}0.16 & 0.33\\
\widetilde{\delta}_- & 0 & -0.16 & 0\\ \hline
\end{array}
$$
\renewcommand{\arraystretch}{1}
\addtolength{\arraycolsep}{-3pt}
\caption[]{Masses and couplings entering 
Eqs.~(\protect{\ref{eq:e}})--(\protect{\ref{eq:gv}}). Renormalization
point is $\mu = 1\,$GeV.
We use $m_s(1\,{\rm GeV}) = 150\,$MeV and put the $u$ and $d$ quark
mass zero.
}\label{tab:zetas}
\end{table}

\begin{figure}[t]
\centerline{\epsffile{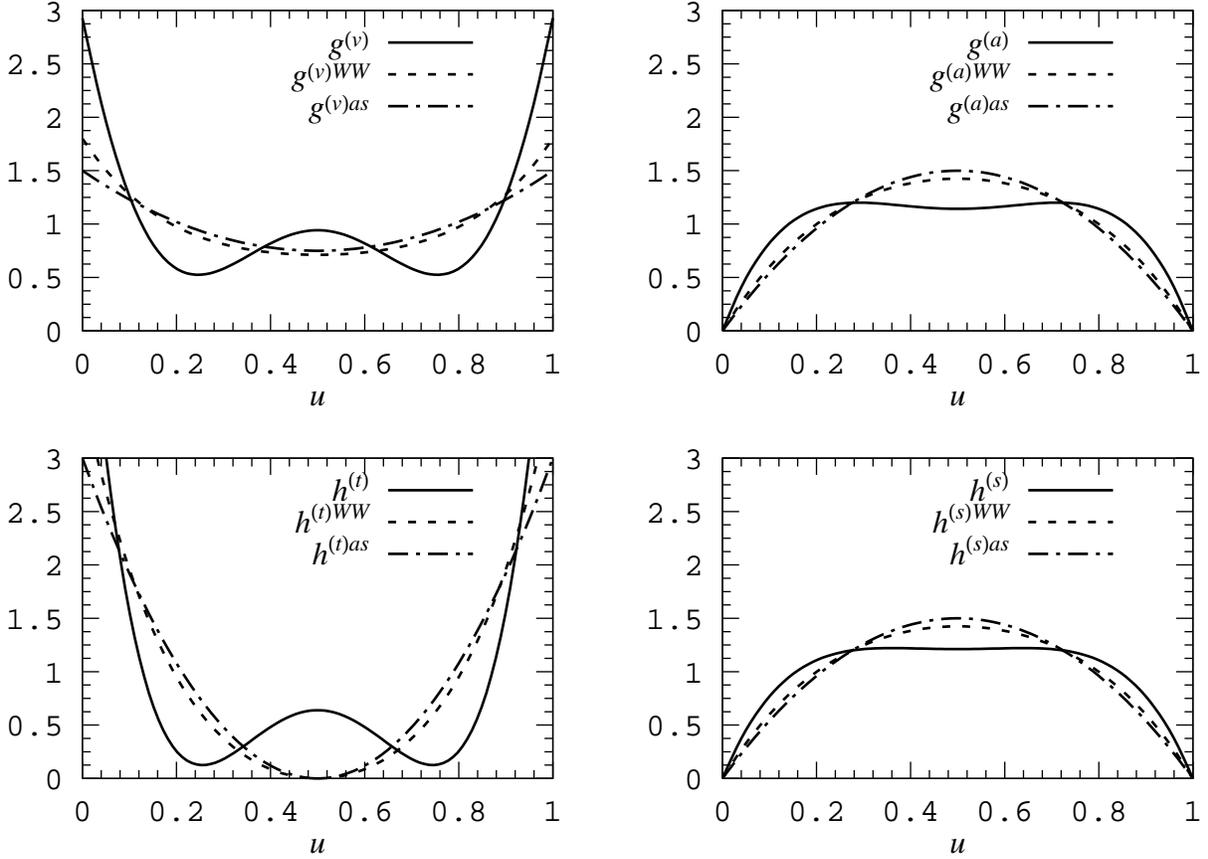}}
\caption[]{Two-particle twist~3 distribution amplitudes for the $\rho$
meson.}
\label{fig2}
\end{figure}
The resulting $\rho$ meson distributions are plotted in
 Fig.~\ref{fig2} together with the corresponding asymptotic
 distributions and with 
the distributions calculated in the Wandzura-Wilczek approximation.  
It is seen that gluon corrections of twist~3 are generally important and 
tend to broaden the distributions. For the second moments we
get (at the scale 1$\,$GeV):
\begin{eqnarray}
 \int_0^1\!du\,(2u-1)^2\, \hs(u) &=& 0.24 \,\,(0.20),\\
\int_0^1\!du\,(2u-1)^2\, \htt(u) &=& 0.63 \,\,(0.60),\\
\int_0^1\!du\,(2u-1)^2\, g_\perp^{(a)}(u) &=& 0.25 \,\,(0.20),\\
\int_0^1\!du\,(2u-1)^2\, g_\perp^{(v)}(u) &=& 0.47 \,\,(0.40),
\end{eqnarray}
where the numbers in parenthesis give the asymptotic values.
As already mentioned, the oscillatory behavior of the distributions
depicted in
Fig.~\ref{fig2}  
is an artifact of the expansion in orthogonal polynomials and
will be smoothened by contributions of higher-order partial waves.
We expect, nevertheless,  that our approximation is sufficient 
for calculating most overlap integrals that appear in
physical applications.

In Fig.~\ref{fig3} we compare the distribution amplitudes of $\rho$, 
$K^*$ and $\phi$ mesons, which differ due to the nonzero 
strange quark mass. 
Note that quark mass corrections to two-particle distributions 
in general involve logarithms of the momentum fraction and are not 
reduced to polynomials. This means that in this case the expansion in 
conformal 
partial waves does not correspond to an expansion in local operators,
which is similar to what was observed in 
\cite{CDT85,BF90} for bilinear twist~4 operators.
The quark mass effects are not large, but can result in a
logarithmic enhancement 
of  distributions close to the end-points $u\to0$ and $u\to 1$, see 
Eqs.~(\ref{eq:e})--(\ref{eq:gv}) and Fig.~\ref{fig3}.
Because of that, the calculation of SU(3) breaking effects requires caution
for physical observables which are sensitive to the end-point region 
where the linear approximation in $m_s$ breaks 
down.\footnote{For finite quark masses, 
renormalization also gets complicated  
due to the absence of conformal symmetry.  For example,  
$\Theta^T_{k,n-k-2}$ could receive additional mixing 
not only with $m_q \Omega_{n-1}^\parallel$,
but also with operators involving total derivatives
(see Eqs.~(\ref{eq:3ThetaT}) and (\ref{eq:4omg})).   
Therefore, the scale dependence of $f^T_{3}\omega^{T}_{k,n-k-2}$
may not be described by a simple extension of the anomalous dimension 
matrix involving the mixing with $m_q f_{_V} a_{n-1}^\parallel$, which 
is in contrast to the renormalization of twist~3 parton distributions.  
The complete clarification of this point is beyond the scope of this work.}

\begin{figure}[t]
\centerline{\epsffile{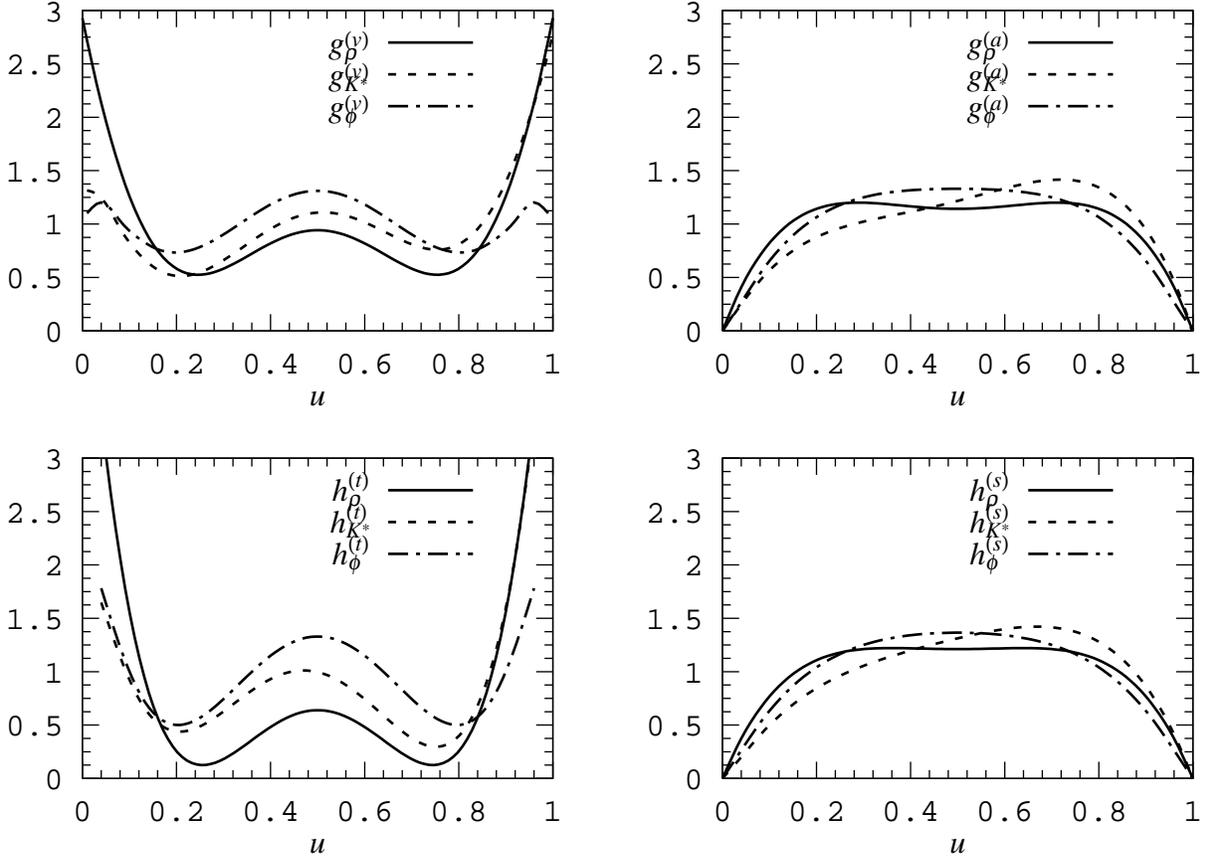}}
\caption[]{Two-particle twist~3 distribution amplitudes for the
$\rho$, $K^*$ and $\phi$ meson.}\label{fig3}
\end{figure}

\section{Summary and Conclusions}
\setcounter{equation}{0}

In the present paper, we have studied the twist three distribution amplitudes
of vector mesons in QCD and expressed them in a model-independent way
by a minimal number of nonperturbative parameters. The one key
ingredient in our approach was the use of the QCD equations of motion
which allowed us to reveal the interrelations between the different
distribution amplitudes of a given twist and to obtain exact integral
representations for distribution amplitudes that are not
dynamically independent. The other ingredient was the
use of conformal expansion which, analogously to partial wave
decomposition in quantum mechanics, allows one to separate
transverse and longitudinal variables in the wave function:
 The dependence on transverse coordinates is represented as scale-dependence
of the relevant operators and is governed by
renormalization-group equations, the dependence on the longitudinal 
momentum fraction is described in terms of irreducible
presentations of the corresponding symmetry group, the collinear
conformal group SL(2,R). The conformal partial wave expansion is
explicitly consistent with the equations of motion since the latter are 
not renormalized. The expansion thus makes maximum use of the symmetry 
of the theory in order to simplify the dynamics, which is related,
in the perturbative domain, to renormalization properties of twist three
operators. 

As it was known for some time \cite{ABH91,BBKT96}, anomalous dimensions
of twist three operators increase logarithmically with the spin.
Like in the leading twist case, this property ensures convergence 
of the conformal expansion at sufficiently large scales and suggests
that only the few lowest ``harmonics'' are important in calculations 
of physical observables. 

Based on this assumption, we have derived explicit and
consistent models for all two- and
three-particle distribution amplitudes of $\rho$, $\omega$, $K^*$ and
$\phi$ mesons of twist two and three 
including contributions up to conformal spin
$j=9/2$. The relevant nonperturbative parameters (``reduced matrix
elements'') were estimated from
QCD sum rules.  The results are immediately applicable to a range of
phenomenologically interesting processes like exclusive
semileptonic or radiative $B$ decays and hard electroproduction of
vector mesons at HERA.

Our formalism is --- in principle --- applicable to arbitrary twist,
although its realization will become technically more involved. 
The application to twist four distribution amplitudes of vector mesons
will be presented elsewhere. 

\subsection*{Acknowledgments}

 Fermilab is operated by Universities Research Association,
Inc., under contract no.\ DE--AC02--76CH03000
with the U.S.\ Department of Energy.
The work of K.T. was supported in part by the  
Grant-in-Aid for Encouragement of Young Scientists No. 09740215
from the Ministry of Education, Science, Sports and Culture.

\appendix
\renewcommand{\theequation}{\Alph{section}.\arabic{equation}}
\setcounter{table}{0}
\renewcommand{\thetable}{\Alph{table}}

\section{Formulae for Orthogonal Polynomials}
\label{app:a}
\setcounter{equation}{0}

In this appendix, we collect useful formulae for the orthogonal polynomials
which appear in the conformal expansion.

\noindent Differentiation formula for Gegenbauer polynomials:
\begin{equation}
\frac{d}{d\xi} (1-\xi^{2})C_{n}^{3/2}(\xi) = 
-(n+1)(n+2) C_{n}^{1/2}(\xi).
\label{eq:jg1}
\end{equation}
Differentiation formulae for Jacobi polynomials:
\begin{eqnarray}
\frac{d}{d\xi} P_{n}^{(\nu_{1}, \nu_{2})}(\xi)
&=& \frac{n + \nu_{1} + \nu_{2} + 1}{2} 
P_{n-1}^{(\nu_{1}+1,\nu_{2}+1)}(\xi),
\label{eq:jd2}\\
(1 + \xi) P_{n}^{(1,1)}(\xi)
&=&  2\frac{d}{d\xi}\left[
\frac{(n+1)}{(n+2)(2n+3)} P_{n+2}^{(0,0)}(\xi)\right. 
\nonumber \\
& &\left. {}+\frac{1}{n+2} P_{n+1}^{(0,0)}(\xi)
+ \frac{1}{2n+3}P_{n}^{(0,0)}(\xi)\right].
\label{eq:jd1}
\end{eqnarray}
The equation (\ref{eq:jd1}) is obtained from (\ref{eq:jd2})
combined with (\ref{eq:jrec1}) below.

\noindent
Recurrence formulae for Jacobi polynomials:
\begin{eqnarray}
(1+\xi) P_{n}^{(1,1)}(\xi) &=& \frac{(n+1)(n+3)}{(n+2)(2n+3)}
P_{n+1}^{(1,1)}(\xi) + P_{n}^{(1,1)}(\xi) + \frac{n+1}{2n+3}
P_{n-1}^{(1,1)}(\xi)
\label{eq:jrec1}\\
&=&\frac{2(n+1)}{2n+3} \left(P_{n}^{(1,0)}(\xi) + P_{n+1}^{(1,0)}(\xi)
\right),
\label{eq:jrec12}\\
(1-\xi) P_{n}^{(1,1)}(\xi) &=&
\frac{2(n+1)}{2n+3} \left(P_{n}^{(0,1)}(\xi) - P_{n+1}^{(0,1)}(\xi)
\right),
\label{eq:jrec13}\\
P_{n}^{(0,0)}(\xi) &=& \frac{n+1}{2n+1}P_{n}^{(1,0)}(\xi)
- \frac{n}{2n+1}P_{n-1}^{(1,0)}(\xi)
\nonumber \\
&=&
\frac{n+1}{2n+1}P_{n}^{(0,1)}(\xi)
+ \frac{n}{2n+1}P_{n-1}^{(0,1)}(\xi),
\label{eq:jrec2}
\end{eqnarray}
\beq
P_n^{(0,0)}(\xi) + P_{n+1}^{(0,0)}(\xi) &=&
(1+\xi)P^{(0,1)}_n(\xi),
\label{A1}\\
P_n^{(0,0)}(\xi) - P_{n+1}^{(0,0)}(\xi) &=& 
(1-\xi)P^{(1,0)}_n(\xi). 
\label{A2}
\eeq
Relations between Jacobi and Gegenbauer polynomials:
\begin{eqnarray}
(1+\xi) P_{n}^{(0,1)}(\xi) + (1-\xi)P_{n}^{(1,0)}(\xi) &=&
2 C_{n}^{1/2}(\xi),
\label{eq:jg2}\\
(1+\xi) P_{n}^{(0,1)}(\xi) - (1-\xi)P_{n}^{(1,0)}(\xi) &=&
2 C_{n+1}^{1/2}(\xi),
\label{eq:jg3}
\end{eqnarray}
\begin{equation}
(n+2)P_{n}^{(1,1)}(\xi) = 2 C_{n}^{3/2}(\xi).
\label{eq:jg4}
\end{equation}
Orthogonality relations for Appell polynomials~\cite{ER53}:
\begin{equation}
\int{\cal D}\underline{\alpha} \: \alpha_{d}\alpha_{u}\alpha_{g}^{2}
J_{k,l}(\alpha_{d}, \alpha_{u})
J_{m,n}(\alpha_{d}, \alpha_{u})
= \delta_{k+l, m+n}
\frac{(-1)^{k+l}}{2^{k+l+3}(k+l+3)(2k+2l+5)!!}
W_{k,m}^{(k+l+1)},
\label{eq:appello0}
\end{equation}
where $W_{k,m}^{(k+l+1)} \equiv 
\partial^{m+n}J_{k,l}(\alpha_{d}, \alpha_{u})/
\partial \alpha_{d}^{m} \partial \alpha_{u}^{n}$
is a $(k+l+1)\times (k+l+1)$ symmetric matrix. This result can be
obtained from the following relations:
\begin{eqnarray}
\lefteqn{\int {\cal D}
\underline{\alpha}\: \alpha_{d}^{m+1}\alpha_{u}^{n+1}
\alpha_{g}^{2} J_{k,l}(\alpha_{d}, \alpha_{u}) =}\makebox[2cm]{\ }
\nonumber \\
&=&\left\{
\begin{array}{@{\,}ll}
0& (m+n < k+l)\\
\delta_{m,k} 
\displaystyle{\frac{(-1)^{k+l}k!l!}{2^{k+l+3}(k+l+3)(2k+2l+5)!!}}&
(m+n=k+l),
\end{array}
\right.
\label{eq:appello}
\end{eqnarray}
while the integral is in general nonzero for $m+n>k+l$.

\noindent
Integral formulae for Appell polynomials:
\begin{eqnarray}
&\mbox{$ $}&\frac{d}{du}\int_{0}^{u}\!d\alpha_{d} \int_{0}^{\overline{u}}
\!d\alpha_{u}\, 
\frac{1}{1 -\alpha_{d}-\alpha_{u}}\left(
\alpha_{d}\frac{\partial}{\partial \alpha_{d}}
+\alpha_{u}\frac{\partial}{\partial \alpha_{u}}
- 1\right) \alpha_{d} \alpha_{u} (1 - \alpha_{d}-\alpha_{u})^{2}
J_{k,l}(\alpha_{d}, \alpha_{u}) =
\nonumber\\
&\mbox{$ $}&\;\;\;\;\;\;\;\;\;\;\;\;
= \frac{u\overline{u}}{2}\frac{k!l!(-1)^{k}}{(k+l+2)!}
\left(\frac{k-l}{k+l+3}P_{k+l+2}^{(1,1)}(\xi)
+ P_{k+l+1}^{(1,1)}(\xi)\right),
\label{eq:apint}
\end{eqnarray}
\begin{eqnarray}
\lefteqn{\frac{d}{du}\int_{0}^{u}d\alpha_{d} \int_{0}^{\overline{u}}
d\alpha_{u} 
\alpha_{d} \alpha_{u} (1 - \alpha_{d}-\alpha_{u})
J_{k,l}(\alpha_{d}, \alpha_{u}) =}\makebox[1cm]{\ }
\nonumber\\
&=& \frac{u\overline{u}}{2}\frac{k!l!(-1)^{k}}{(k+l+3)!}
\left(\frac{k-l}{k+l+3}P_{k+l+2}^{(1,1)}(\xi)
- P_{k+l+1}^{(1,1)}(\xi)\right),
\label{eq:apint2}
\end{eqnarray}
\beq
& &{d\over du}
\int_0^ud\alpha_d \int_0^{\bar{u}}d\alpha_u {1\over 1- \alpha_d -\alpha_u}
\left( \alpha_d {\partial \over \partial\alpha_d } 
- \alpha_u{\partial\over \partial\alpha_u}\right)
\alpha_d\alpha_u (1-\alpha_d-\alpha_u)^2 J_{k,l}(\alpha_d,\alpha_u)\ =
\nonumber\\
& &\qquad ={u\bar{u}\over 2}{k!l!(-1)^k\over (k+l+3)!} 
\left[{(k+l+2)(k+l+4)\over k+l+3}
P^{(1,1)}_{k+l+2}(\xi)
+(k-l)P^{(1,1)}_{k+l+1}(\xi) \right].
\label{A3}
\eeq
\beq
& &\int_0^ud\alpha_d \int_0^{\bar{u}}d\alpha_u {1\over 1- \alpha_d -\alpha_u}
\left( {\partial\over \partial\alpha_d } 
+ {\partial\over \partial\alpha_u}\right)
\alpha_d\alpha_u (1-\alpha_d-\alpha_u)^2 J_{k,l}(\alpha_d,\alpha_u)\ =
\nonumber\\
& &\qquad ={k!l!(-1)^k\over (k+l+3)!4} 
\left[{-k+l\over 2k+2l+5}
\left( 
P^{(0,0)}_{k+l+1}(\xi)-
P^{(0,0)}_{k+l+3}(\xi) \right)\right. \nonumber\\
& &\left.\qquad\qquad\qquad 
+{k+l+2\over 2k+2l+7}
\left( 
P^{(0,0)}_{k+l+2}(\xi)-
P^{(0,0)}_{k+l+4}(\xi) \right)\right].
\label{A4}
\eeq
The results (\ref{eq:apint})$-$(\ref{A3}) 
can be obtained by differentiating and/or
integrating 
the Appell polynomials $J_{k,l}(\alpha_{d}, \alpha_{u})$ term by term.
To obtain (\ref{A4}), it is convenient to calculate its derivative
first, which can be done similarly to (\ref{eq:apint}), (\ref{eq:apint2})
and (\ref{A3}),   
and then integrate the result with the condition that it vanishes
 at $u=0$.

\section{Conformal Expansion of Wandzura-Wilczek Contributions
         }\label{app:K}
\setcounter{equation}{0}

In this appendix we explain the structure of the conformal expansion 
of the Wandzura-Wilczek contributions to twist~3 chiral-odd two-particle
distribution amplitudes, Eq.~(\ref{eq:3hnww}).
As mentioned in Sec.~3.2, the conformal spin assignment for these 
terms seemingly does not match the expansion in Eqs.~(\ref{eq:3cehud}) and 
(\ref{eq:3cehdu}), which calls for an explanation.

The basic idea is the following: the expansion derived in 
Eqs.~(\ref{eq:3cehud}) and (\ref{eq:3cehdu}) is based on the conformal
expansion of the corresponding operators. The distribution amplitudes
$\htt$ and $\hs$ are obtained as matrix elements of these operators
between the vacuum and the longitudinally polarized $\rho$ meson state.
We call the state conformal and assign a corresponding conformal spin, if it
is annihilated by a conformal operator.
The difficulty with the Wandzura-Wilczek terms
is due to the fact that these contributions involve matrix elements
over the $\rho$ meson with a different (transverse) polarization.
Working out the Wandzura-Wilczek contributions to $\htt$, $\hs$ 
essentially corresponds to reexpressing these matrix elements in 
terms of  similar matrix elements over the longitudinally polarized meson, 
using Lorentz symmetry. In our context it is important that 
the transversely polarized state is related 
(in the $\rho$ meson rest frame) to the longitudinally
polarized state by a spin rotation which does not commute with the generators
of the collinear conformal group. Working out the necessary
commutation relations, we  reproduce the particular spin 
structure appearing in (\ref{eq:3hnww}).     

It is convenient to work in the helicity basis: $\rho$ meson states with 
$\lambda=\pm 1$ correspond to transverse polarization, $\lambda=0$
denotes the longitudinal polarization. Equation~(\ref{eq:3anp2}) reads
\begin{equation}
 f_\rho^T a_n^\perp \propto 
 \langle 0|\Omega_n^\perp|\rho(P,\lambda=\pm 1)\rangle,
\label{eq:k1} 
\end{equation}
where the proportionality factor is irrelevant for what follows.
Following Ohrndorf \cite{O82}, we define 
a set of eigenstates $|j,m\rangle$ with conformal spin $j$
and the ``third projection'' of the spin $m=j,j+1,j+2,\ldots$,
such that 
\begin{eqnarray}
 J_3|j,m\rangle = m |j,m\rangle, &\qquad& 
 J^2|j,m\rangle = j(j-1) |j,m\rangle,
\nonumber\\
 J_-|j,m\rangle = -(j-m) |j,m-1\rangle, &\qquad&
 J_+|j,m\rangle = (j+m)  |j,m+1\rangle.
\label{eq:k2}
\end{eqnarray}
The generators $J_\pm, J_3$ satisfy the canonical commutation 
relations of the algebra of the group of hyperbolic rotations, O(2,1),
and are related to the generators of the collinear conformal group
by 
\begin{equation}
 J_+ = \frac{i}{\sqrt{2}}P_.,~~~J_-= \frac{i}{\sqrt{2}}K_\ast,~~~
 J_3 =\frac{i}{2}(D+M_{\ast .}),
\label{eq:k3}
\end{equation}
where $P_\mu$ and $M_{\mu\nu}$ are the usual generators of Poincare group,
$D$ is the generator of dilatations and $K_\mu$ generates conformal 
transformations.  Note that $J_+$ and $J_-$ are just ``step-up'' and 
``step-down'' operators in this basis; the state with the 
{\em lowest} value of $m$, $m_{\rm min}=j$, is annihilated by $J_-$.

Following the discussion in Ref.~\cite{O82}, it is easy to show
that 
\begin{equation}
 J_-\wperp|0\rangle = [J_-,\wperp]|0\rangle = 0,~~~
 J_3\wperp|0\rangle =(n+2)\wperp|0\rangle,~~~
 J^2\wperp|0\rangle =(n+2)(n+1) \wperp|0\rangle, 
\label{eq:k4}
\end{equation}
so that we can identify
\begin{equation}
 |n+2,n+2\rangle \equiv \wperp|0\rangle,
\label{eq:k5}
\end{equation}
and rewrite Eq.~(\ref{eq:k1}) as
\begin{equation}
 f_\rho^T a_n^\perp \propto 
 \langle n+2,n+2|\rho(P,\lambda=\pm 1)\rangle.
\label{eq:k6} 
\end{equation}
This confirms that $a_n^\perp$ corresponds to conformal spin $j=n+2$, as 
stated in the main text.

To determine the corresponding contribution to $\htt$, $\hs$, we have 
to recast Eq.~(\ref{eq:k6}) in a different form corresponding to 
a matrix element over the longitudinally polarized meson. 
For definiteness, take $\lambda=+1$. In the $\rho$ meson rest frame the 
$\lambda=+1$ state is related to the $\lambda=0$ state by the spin
rotation
\begin{equation}
|\rho(P=0,\lambda=+1)\rangle \propto 
(M_{23}+iM_{31})|\rho(P=0,\lambda=0)\rangle,
\label{eq:k7}
\end{equation} 
where $(M_{23}+iM_{31})$ is the step-up operator  of ordinary angular 
momentum. $|\rho(P,\lambda)\rangle$ is then obtained by a Lorentz
boost in $P^3$ direction:
\begin{equation}
  |\rho(P,\lambda)\rangle = {\cal U}(\omega)|\rho(P=0,\lambda)\rangle,
\label{eq:k8}
\end{equation}
where
\begin{equation}
  {\cal U}(\omega) = e^{-i\omega M^{03}} = e^{-i\omega M_{\ast} .},~~~
  {\rm th}(\omega) = P^3/P^0.
\label{eq:k9}
\end{equation}
We can thus write 
\begin{equation}
 f_\rho^T a_n^\perp \propto 
 \langle \Psi|\rho(P,\lambda = 0)\rangle,
\label{eq:k10}  
\end{equation}
where 
\begin{equation}
|\Psi\rangle = {\cal U}(\omega)\big(M_{23}+iM_{13}\big)\,{\cal U}^{-1}(\omega)
    |n+2,n+2\rangle.
\label{eq:k11}
\end{equation}
In the following we demonstrate that $|\Psi\rangle$ is given by a 
superposition of three conformal states,
\begin{equation}
 |\Psi\rangle = C_1 |n+\frac{3}{2},n+\frac{3}{2}\rangle +
                C_2 |n+\frac{3}{2},n+\frac{5}{2}\rangle +
                C_3 |n+\frac{5}{2},n+\frac{5}{2}\rangle, 
\label{eq:k12}
\end{equation}
where the $C_k$ are C-numbers. If established, Eq.~(\ref{eq:k12}) shows that
contributions of $a_n^\perp$  to the matrix element over a 
longitudinal $\rho$ meson correspond to conformal spins $j=n+\frac{3}{2}$
and $j=n+\frac{5}{2}$, which explains the pattern appearing in 
Eq.~(\ref{eq:3hnww}).

To prove (\ref{eq:k12}), we first note that
\begin{equation}
\cu(\omega)\, (M_{23} + i M_{13})\, \cu^{-1}(\omega) = M_{23}\,{\rm
ch}(\omega) + M_{20}\, {\rm sh}(\omega) + i \{ M_{13}\, {\rm ch}(\omega) +
M_{10}\, {\rm sh}(\omega)\}.
\end{equation}
{}From this and Eq.~(\ref{eq:k11}), it follows
\begin{equation}\label{eq:k14}
|\Psi\rangle = \left[  \exp (\omega)/2\, (\opdot) -  \exp 
(-\omega)\,( \opstar) \right] |n+2,n+2\rangle
\end{equation}
with $z^\mu = (1,0,0,-1)$ and $p^\mu = (1,0,0,1)$ (note that $p\cdot
z=2$). Applying to (\ref{eq:k14}) the
commutation relations
\begin{equation}
[J_3,\opdot] = -\frac{1}{2}\:(\opdot), \qquad [J_3,\opstar] =
\frac{1}{2}\:(\opstar) 
\end{equation}
and
\begin{equation}
[J_-,\opdot] = 0, \qquad [J_-,\opstar] \propto K_2 + i K_1, \qquad [J_-,K_2 +
i K_1] = 0,
\end{equation}
it immediately follows that
\begin{eqnarray}
(\opdot) |n+2,n+2\rangle & \propto &
|n+\frac{3}{2},n+\frac{3}{2}\rangle,\nonumber \\
(\opstar) |n+2,n+2\rangle & \propto &
a_1 |n+\frac{3}{2},n+\frac{5}{2}\rangle + a_2
|n+\frac{5}{2},n+\frac{5}{2}
\rangle,
\end{eqnarray}
with C-number coefficients $a_i$, which proves Eq.~(\ref{eq:k12}).

A similar discussion also explains the mismatch observed in
Eq.~(\ref{eq4.32}) for chiral-even distribution amplitudes. 

\section{QCD Sum Rules for Expansion Coefficients of Distribution 
Amplitudes}\label{app:B}
\setcounter{equation}{0}
 
 The method of QCD sum rules in its application to distribution amplitudes
of light mesons was pioneered by Chernyak and Zhitnitsky and
is comprehensively discussed in \cite{CZreport}.
In this appendix we collect QCD sum rules and results for the twist~2
distribution amplitudes  of the vector mesons 
$\rho$, $K^*$ and $\phi$ as well as for the
twist~3 distribution amplitudes of the $\rho$ meson. 
Numerical results presented below  are obtained using the following
input parameters:
\begin{equation}
\renewcommand{\arraystretch}{1.4}
\addtolength{\arraycolsep}{2pt}
\begin{array}[b]{rcl@{\quad}rcl}
\ms(1\,{\rm GeV}) & = & (150\pm 50)\,{\rm MeV}, & \gluon & = &
(0.012\pm0.006)\,{\rm GeV}^4,\\
\quark(1\,{\rm GeV}) & = & (-240\pm 20)\,{\rm MeV}^3, & 
\squark(1\,{\rm GeV}) & = & 0.8 \quark(1\,{\rm GeV}),\\
\mixed(1\,{\rm GeV}) & = & 0.8 \quark(1\,{\rm GeV}), & 
\smixed(1\,{\rm GeV}) & = & 0.8 \mixed(1\,{\rm GeV}),\\
\multicolumn{6}{l}{\Lambda^{(3)}_{\rm QCD}\ =\ 400\,{\rm
MeV}\ \Longrightarrow\ \alpha_s(1\,{\rm GeV})\ = \ 0.56}\\[-20pt]
\end{array}
\renewcommand{\arraystretch}{1}
\addtolength{\arraycolsep}{-2pt}
\end{equation}
\makebox[2cm]{\ }\\[-7pt]
and assuming factorization of the vacuum expectation values of four-fermion
operators. The SU(3) breaking effects in the sum rules are due to explicit
corrections proportional to the quark masses, the difference in 
values of the condensates of strange and nonstrange quarks, and differences 
in the values of the continuum thresholds $s_0$ and Borel parameters 
$M^2$. Instead of fitting the continuum thresholds 
separately for each meson and for 
each sum rule, in this paper we prefer to determine 
$s_{0,\rho}$ from the simplest sum rules for vector (tensor) 
couplings and use the relations
\begin{equation}
   s_{0,K^*} - s_{0,\rho} = m^2_{K^*}-m^2_\rho,\qquad
   s_{0,\phi} - s_{0,\rho} = m^2_{\phi}-m^2_\rho,
\label{eq:split}
\end{equation}   
which are known to hold with reasonable accuracy for the spectra of 
resonances in the respective channels. Similar relations are assumed 
between the ``working windows'' in the Borel parameter.

\subsection{Distribution amplitudes of twist two}

Most of the relevant formulas were previously obtained in
\cite{CZZsu3,CZreport,BBrho}. For the $\rho$ meson, we quote
the results from \cite{BBrho}. For the other mesons we
present a new analysis which includes the radiative corrections
calculated in \cite{BBrho} and the SU(3) breaking terms calculated in
\cite{CZZsu3}. Unlike in Ref.~\cite{CZreport,CZZsu3}, where sum rules 
for the moments of distribution amplitudes were derived, we prefer to
consider the sum rules directly for the coefficients $a_n$ 
(Gegenbauer moments) in the conformal expansion, 
see \cite{BBrho} for a discussion.

QCD sum rules for even Gegenbauer moments can be derived
from the {\em diagonal} correlation functions of the conformal operators
introduced in Sec.~3.3 and Sec.~4.3:
\begin{equation}
{\rm D}_n^{\parallel\{\perp\}} = i\int\!\! d^4 y\, e^{ipy} \langle 0 | T
\Omega_n^{\parallel\{\perp\}}(y) \Omega_0^{\dagger\parallel\{\perp\}}(0) |
0 \rangle,\label{eq:CFdiag}
\end{equation}
see Eqs.~(\ref{eq:3anp})--(\ref{eq:3omg}) and (\ref{eq:4anp})--(\ref{eq:4omg})
for precise definitions.

One finds the following sum rules for vector and tensor
couplings \cite{SVZ,CZZsu3} of the $K^*$ meson:
\begin{eqnarray}
f_{K^*}^2\,e^{-m_{K^*}^2/M^2} & = & 
\frac{M^2}{4\pi^2}\,\left( 1- e^{-s_0^\parallel/M^2}\right)
\left(1+\frac{\alpha_s}{\pi}\right) + \frac{1}{12M^2}\gluon +
\frac{\ms\squark}{M^2}\nonumber\\
& & {} + 
\frac{16\pi\alpha_s}{81M^4}\,\left(\quark^2 + \squark^2 \right) -
\frac{16\pi\alpha_s}{9M^4}\, \quark \squark,
\label{eq:fKpar}\\[-30pt] \nonumber
\end{eqnarray}
\begin{eqnarray}
\lefteqn{(f_{K^*}^T(\mu))^2 e^{-m_{K^*}^2/M^2}\ =\ 
\left(\frac{\alpha_s(\mu^2)}{\alpha_s(M^2)}\right)^{\frac{2\gamma_0^\perp}{
\beta_0}}\left\{ 1 + \frac{\alpha_s(\mu^2)-\alpha_s(M^2)}{2\pi}
\,\frac{\gamma_0^\perp}{\beta_0} \left( 
\frac{\gamma_0^{\perp(1)}}{\gamma_0^\perp} -
\frac{\beta_1}{\beta_0}\right)\right\}\times}\hspace{0.5cm}\nonumber\\
& \times & {}\left[\frac{1}{4\pi^2}\int_{0}^{s_0^\perp}\!\!\! ds\, e^{-s/M^2}
\left\{1+\frac{\alpha_s}{\pi}\left(\frac{7}{9} + \frac{2}{3}\,
\ln\,\frac{s}{M^2}\right)\right\}
 - \frac{1}{12M^2}\gluon +
\frac{\ms\squark}{M^2}\right.\nonumber\\
& & {}\left. -\frac{1}{3M^4}\,\ms\smixed 
-\frac{32\pi\alpha_s}{81M^4}\,\left(\quark^2 + \squark^2\right)\right].
\makebox[1.5cm]{\ }\label{eq:fKperp}
\end{eqnarray}
In the sum rule for
$f_{K^*}^T$, 
$\beta_{0}=9, \beta_{1}=64$, and 
$\gamma_0^{\perp(1)}=310/9$ (for three running
flavours) is the two-loop anomalous dimension calculated in \cite{HKK}.

For arbitrary (even) Gegenbauer moments one obtains:
\begin{eqnarray}
\lefteqn{\frac{3(n+1)(n+2)}{2(2n+3)}\,f_{K^*}^2a_n^\parallel(\mu)
e^{-m_{K^*}^2/M^2}\ =}\nonumber\\
& = & \frac{1}{2\pi^2}\,\frac{\alpha_s}{\pi}\, M^2
\left(1-e^{-s_0^\parallel/M^2} \right) \int_0^1\!\! du\, u\bar u\,
C_n^{3/2}(2u-1)\, \ln^2\,\frac{u}{\bar u} + 
\frac{1}{2M^2}\,\ms\squark (n+1)(n+2)\nonumber\\
& & {} + \frac{1}{24M^2}\, \gluon
(n+1)(n+2) - \frac{1}{48M^4}\,\ms\smixed n(n+1)(n+2)(n+3)\nonumber\\
& & {}-\frac{8\pi\alpha_s(\mu)}{9M^4}\,\quark\squark (n+1) (n+2) +
\frac{4\pi\alpha_s}{81M^4}\,(\quark^2+\squark^2)(n+1)^2
(n+2)^2,\\[-30pt] \nonumber
\end{eqnarray}
\begin{eqnarray}
\lefteqn{\frac{3(n+1)(n+2)}{2(2n+3)}\,(f_{K^*}^T(\mu))^2 a_n^\perp(\mu)
e^{-m_{K^*}^2/M^2}\ =}\nonumber\\
& = & \frac{1}{2\pi^2}\,\frac{\alpha_s}{\pi}\, M^2
\left(1-e^{-s_0^\perp/M^2} \right) \int_0^1\!\! du\, u\bar u\,
C_n^{3/2}(2u-1) \left( \ln u + \ln \bar u + \ln^2\,\frac{u}{\bar
u}\right) 
\nonumber\\
& & {} + \frac{1}{24M^2}\, \gluon
(n^2+3n-2) - \frac{1}{48M^4}\,\ms\smixed (n+1)(n+2)(n^2+3n+8)\nonumber\\
& & {} +
\frac{4\pi\alpha_s}{81M^4}\,(\quark^2+\squark^2) (n-1)(n+1)(n+2)(n+4)+ 
\frac{1}{2M^2}\,\ms\squark (n+1)(n+2).\nonumber\\[-15pt]
\end{eqnarray}
The sum rules for $\rho$ are obtained by setting $m_s$ zero and $s$ to $q$ in
the condensates. For $\phi$, one has to replace $q$ by $s$ in the
condensates and to double the terms linear in $m_s$. On the left-hand
sides one has to insert the proper meson masses. 
All renormalization scale dependent quantities are evaluated at the scale
$\mu \sim 1$~GeV.

Gegenbauer moments with odd $n$ are nonvanishing for the $K^*$ meson only.
They can be determined most conveniently from the {\em nondiagonal} 
correlation functions\footnote{Since perturbative contributions to 
 D$^{(n)}$ vanish for odd $n$. Note also that from  ND$^{(0)}$ 
one obtains the relative sign between $f_V$ and $f_V^T$.} 
\begin{equation}
{\rm ND}_n^{\parallel\{\perp\}} = i\int\!\! d^4 y\, e^{ipy} \langle 0 | T
\Omega_n^{\parallel\{\perp\}}(y) \Omega_0^{\dagger\perp\{\parallel\}}(0) |
0 \rangle.\label{eq:CFnondiag}
\end{equation}
The sum rules read \cite{CZZsu3}:
\begin{eqnarray}
\lefteqn{\frac{3(n+1)(n+2)}{2(2n+3)}\,f_{K^*}^T(\mu)
 f_{K^*} m_{K^*} a^\parallel_n(\mu) e^{-m_{K^*}^2/M^2}\
 =}\hspace{2.5cm}\nonumber\\
& = & \frac{3}{8\pi^2}\,\ms M^2 \left(1-e^{-s_0^\parallel/M^2}\right) +
\frac{1}{2}(n+1)(n+2) (\squark-\quark)\nonumber\\
& &{} - \frac{1}{24M^2}\,
(n+1)^2(n+2)^2 (\smixed-\mixed),\\
\lefteqn{\frac{3(n+1)(n+2)}{2(2n+3)}\,f_{K^*}^T(\mu)
 f_{K^*} m_{K^*} a^\perp_n(\mu) e^{-m_{K^*}^2/M^2}\ =}\hspace{2.5cm}
\nonumber\\
& = & \frac{3}{8\pi^2}\,\ms M^2 \left(1-e^{-s_0^\perp/M^2}\right) +
\frac{1}{2}(n+1)(n+2) (\squark-\quark)\nonumber\\
& & {} - \frac{1}{48M^2}\,
(n+1)(n+2)(n^2+3n+4) (\smixed-\mixed).\hspace{2cm}
\end{eqnarray}
In these sum rules again the right-hand sides are to be
evaluated at fixed scale $\mu$ of order 1$\,$GeV.

The results are collected in Tab.~\ref{tab:B2}, 
where the quoted errors are due to uncertainties in input 
parameters and to the variation of the Borel parameter within the 
range $M^2\approx (1-2)\,{\rm GeV}^2$ (for the $\rho$ meson), with the
value of the continuum threshold $s_{0,\rho}^\parallel = 1.5$~GeV$^2$
fitted to reproduce the experimental value of the vector coupling.
As discussed in detail in \cite{BBrho}, the sum rule for the tensor
couplings contains contaminating contributions of states with the opposite
parity $1^+$, which can be effectively taken into account by using 
a lower value of the continuum threshold $s_{0,\rho}^\perp = 1.2$~GeV$^2$.
For other mesons, we assume validity of the relations (\ref{eq:split}). 
Note that in the numerical analysis of Gegenbauer moments 
we substitute the couplings on the
left-hand sides by their sum rules (\ref{eq:fKpar}) and (\ref{eq:fKperp})
rather than using the values given in the table. 
 
\subsection{Distribution amplitudes of twist three}

The vector $\fV$ and axial $\fA$ twist~3 couplings are defined as 
the local matrix elements
\begin{eqnarray}
\langle 0 | \bar d  \gamma_\mu \left[
gG_{\rho\lambda} (i\!\derright_\cdot) - (i\!\derleft_\cdot) 
gG_{\rho\lambda}\right]u|\rho^+\rangle
& \equiv & 
\langle 0 | \bar u  \gamma_\mu \left[
gG_{\rho\lambda} (i\!\derright_\cdot) - (i\!\derleft_\cdot) gG_{\rho\lambda}
\right] d|\rho^-\rangle
\nonumber\\
& = & i(pz)\,p_\mu (p_\rho e_\lambda^\perp-p_\lambda 
e_\rho^\perp) \fV + \ldots,\\
\langle 0 | \bar d  \gamma_\mu \gamma_5
g\widetilde{G}_{\lambda\rho} u|\rho^+\rangle
\ =\  \langle 0 | \bar u  \gamma_\mu \gamma_5
g\widetilde{G}_{\lambda\rho} d|\rho^-\rangle
& =& p_\mu (
p_\rho e^\perp_\lambda- p_\lambda e^\perp_\rho) \fA +\ldots ,
\end{eqnarray}
and have been estimated using the sum rule approach in Ref.~\cite{CZZfs}
together with a few matrix elements of higher dimension (and conformal spin).
The results are given in the text\footnote{Apparently the corresponding
sum rules have never been published and are not available. 
We thank V.~Chernyak for correspondence on this point.}.

The tensor coupling is defined as 
\begin{equation}
\langle 0 | \bar d  \sigma_{\mu\cdot} \left[
gG^{\mu\cdot} (i\!\derright_\cdot) - (i\!\derleft_\cdot) gG^{\mu\cdot}\right]
 u|\rho\rangle
 =  (e z) (pz)^2 m_\rho \fT
\end{equation}
and can be estimated from correlation functions of this operator
with the vector and/or tensor current. The correlation function with the 
vector current is chirality-violating and is expanded in operators with 
odd dimension. The leading contribution of the mixed quark-gluon condensate,
however, vanishes, and the first corrections comes from dimension 7
operators
whose 
vacuum expectation values are known only very poorly. For this reason
this sum rule is essentially useless.
{}From the ``diagonal'' correlation function with the tensor current, 
\begin{equation}
i\int d^4y\, e^{iqy}\,\langle 0 | T\bar d(y) \sigma_{\mu\cdot}\left[
gG^{\mu\cdot} (i\!\derright_\cdot) - (i\!\derleft_\cdot) gG^{\mu\cdot}\right]
u(y) \bar u(0) \sigma_{\mu\cdot} d(0)|0\rangle,
\end{equation}
we obtain the sum rule
\begin{equation}
e^{-m_\rho^2/M^2} m_\rho \fperp(\mu) \fT(\mu) =
\frac{\alpha_s}{720\pi^3}\,\int_0^{s_0}\!\! ds\,e^{-s/M^2} +
\frac{1}{36}\, \gluon + \frac{32}{27M^2}\, \pi\alpha_s\quark^2,
\label{eq:SRf3t}
\end{equation}
which we have studied numerically. As a general feature of the sum rules
for matrix elements of operators with high dimension, this sum rule
is dominated at small $M^2\sim$(1--2)\,GeV$^2$ by the condensates of
high dimension (four-quark operators, for the case at hand) and is 
not stable. At larger values of the Borel parameter the stability
of the sum rule is very much improved, suggesting the value 
$\fT(1\,{\rm GeV})\approx 0.3\cdot 10^{-2}\,{\rm GeV}^2$.
This number has to be considered as a rough estimate, however, since at
large values of the Borel parameter the contributions 
of higher mass resonances and of the continuum are out of control. 
Note that, similar to the case of the twist~2 tensor
coupling considered above, the sum rule (\ref{eq:SRf3t}) 
includes contributions of states with opposite (positive) parity.
Ascribing a 100\% error to this result, we arrive at the range given 
in (\ref{eq:5f3t}) as our best estimate.


\begin{thebibliography}{99}
\bibitem{BLreport} S.J.\ Brodsky and G.P.\ Lepage, in: {\em Perturbative
    Quantum Chromodynamics}, ed.\ by A.H.\ Mueller, p.~93, World
  Scientific (Singapore) 1989.
  
\bibitem{exclusive} V.L.\ Chernyak and A.R.\ Zhitnitsky, JETP Lett.\ 
  {\bf {25}} (1977) 510;
  Yad.\ Fiz.\ {\bf 31} (1980) 1053;\\
  A.V.\ Efremov and A.V.\ Radyushkin, Phys.\ Lett.\ B {\bf 94} (1980)
  245;
  Teor.\ Mat.\ Fiz.~{\bf {42}} (1980) 147;\\
  G.P.\ Lepage and S.J.\ Brodsky, Phys.\ Lett.\ B {\bf 87} (1979) 359;
  Phys.\ Rev.\ D {\bf 22} (1980) 2157;\\
  V.L.\ Chernyak, V.G.\ Serbo and A.R.\ Zhitnitsky, JETP Lett.\ {\bf
    26} (1977) 594; Sov.\ J.\ Nucl.\ Phys.\ {\bf 31} (1980) 552.
  
\bibitem{CZreport} V.L.\ Chernyak and A.R.\ Zhitnitsky, Phys.\ Rept.\ 
  {\bf {112}} (1984) 173.

\bibitem{ZZC85}
A.R.\ Zhitnitsky, I.R.\ Zhitnitsky and V.L.\ Chernyak,
Sov.\ J.\ Nucl.\ Phys.\ {\bf 41} (1985) 284.

\bibitem{BBK89} I.I.\ Balitsky, V.M.\ Braun and A.V.\ Kolesnichenko,
  Nucl.\ Phys.\ {\bf B312} (1989) 509.
 
\bibitem{G89}
    A.S.\ Gorsky, Sov.\ J.\ Nucl.\ Phys.\ {\bf 50} (1989) 708.

\bibitem{BF90} V.M.\ Braun and I.E.\ Filyanov, Z.\ Phys.\ C {\bf {48}}
  (1990) 239.

\bibitem{H97} I.\ Halperin, hep--ph/9704265.

\bibitem{BD65} J.D.\ Bjorken and S.D.\ Drell,
{\em Relativistic Quantum Fields} (McGraw-Hill, New York, 1965).

\bibitem{BB88} I.I.\ Balitsky and V.M.\ Braun, Nucl.\ Phys.\ {\bf
    B311} (1989) 541.

\bibitem{JJ92} R.L.\ Jaffe and X.\ Ji, Nucl.\ Phys.\ {\bf
    B375} (1992) 527.
  
\bibitem{BBrho} P. Ball and V.M.\ Braun, Phys.\ Rev.\ D {\bf 54}
(1996) 2182.

\bibitem{KS} J. Kogut and D. Soper, Phys.\ Rev.\ D {\bf 1} (1970)
2901.

\bibitem{WW77} S.\ Wandzura and F.\ Wilczek, 
Phys.\ Lett.\ B {\bf 72} (1977) 195.

\bibitem{B+}
S.J.\ Brodsky et al., Phys.\ Lett.\ B {\bf 91} (1980) 239;
                      Phys.\ Rev.\ D {\bf 33} (1986) 1881.
\bibitem{Makeenko}
Yu.M.\ Makeenko, Sov.\ J.\ Nucl.\ Phys. {\bf 33} (1981) 440.

\bibitem{Mueller} 
D.\ M\"uller, Phys.\ Rev.\ D {\bf 51} (1995) 3855; 
Preprint  CERN--TH/97--80 (hep--ph/9704406).

\bibitem{O82} Th.\ Ohrndorf, Nucl.\ Phys.\ {\bf {B198}} (1982) 26.

\bibitem{ER53} A.\ Erd\'{e}lyi et al., 
{\em Higher Transcendental Functions}, vol.~II (McGraw-Hill, New York, 1953).

\bibitem{ABS} A.\ Ali, V.M.\ Braun and H.\ Simma, Z.\ Phys.\ C {\bf 63}
(1994) 437.

\bibitem{KT95}
        Y. Koike and K. Tanaka, Phys.\ Rev.\ D {\bf 51} (1995) 6125.

\bibitem{KN97}
     Y. Koike and N. Nishiyama, Phys.\ Rev.\ D {\bf 55} (1997) 3068.

\bibitem{BM97} A.V.\ Belitsky and D. M\"{u}ller, Nucl.\ Phys.\ {\bf
    B503} (1997) 279.

\bibitem{AM90} X. Artru and M. Mekhfi, Z. Phys.\ C {\bf 45} (1990) 669.

\bibitem{BBKT96}
     I.I.\ Balitsky, V.M. Braun, Y. Koike and K. Tanaka,  
          Phys.\ Rev.\ Lett.\ {\bf 77} (1996) 3078.

\bibitem{ABH91}
     A. Ali, V.M.\ Braun and G. Hiller, Phys.\ Lett.\ B {\bf 266}
                               (1991) 117.

\bibitem{KYTU97} J. Kodaira, Y. Yasui, K. Tanaka and T. Uematsu,
Phys.\ Lett.\ B {\bf 387} (1996) 855.

\bibitem{g2} A.P.\ Bukhvostov, E.A.\ Kuraev and L.N.\ Lipatov, 
Sov.\ Phys.\ JETP {\bf 60} (1984) 22;\\
P.G.\ Ratcliffe, Nucl.\ Phys.\ {\bf B264} (1986) 493;\\
X. Ji and C. Chou, Phys.\ Rev.\ D {\bf 42} (1990) 3637;\\
D. M\"uller, Phys.\ Lett.\ B {\bf 407} (1997) 314.

\bibitem{KNT} Y. Koike, N. Nishiyama and K. Tanaka, hep--ph/9805460. 

\bibitem{SVZ} M.A.\ Shifman, A.I.\ Vainshtein and V.I.\ Zakharov,
Nucl.\ Phys.\ {\bf B147} (1979) 385; 448; 519.

\bibitem{PDG} R.M.\ Barnett et al.\ (PDG), Phys.\ Rev.\ D {\bf 54}
(1996) 1.

\bibitem{CZZsu3} V.L.\ Chernyak, A.R.\ Zhitnitsky and I.R.\
Zhitnitsky, Nucl.\ Phys.\ {\bf B204} (1982) 477; Erratum-ibid.\ {\bf
B214} (1983) 547; Sov.\ J.\ Nucl.\ Phys.\ {\bf 38} (1983) 775.

\bibitem{BCZ} M. Benayoun, V.L.\ Chernyak and I.R.\ Zhitnitsky,
Nucl.\ Phys.\ {\bf B348} (1991) 327.

\bibitem{CDT85} 
N.S.\ Craigie, V.K.\ Dobrev and I.T.\ Todorov,
Annals Phys.\ {\bf 159} (1985) 411.

\bibitem{HKK} A. Hayashigaki, Y. Kanazawa and Y. Koike, Phys.\ Rev.\  
D {\bf 56} (1997) 7350;\\ W. Vogelsang, Phys.\ Rev.\ D {\bf 57} (1998) 1886.

\bibitem{CZZfs} A.R.\ Zhitnitsky, I.R.\ Zhitnitsky and V.L.\ Chernyak,
Sov.\ J.\ Nucl.\ Phys.\ {\bf 41} (1985) 127.


\end{thebibliography}
\end{document}